\documentclass[aip, jcp, twocolumn, floatfix, superscriptaddress, reprint, 10pt]{revtex4-2}
\usepackage[usenames,dvipsnames]{color}
\definecolor{tangerine}{rgb}{0.944,0.522,0}
\definecolor{verde}{rgb}{0.,0.6,0}
\definecolor{rosso}{rgb}{0.9,0.0,0.2}
\definecolor{magenta}{rgb}{0.9,0.2,0.9}

\newcommand{\editor}[2]{%
  \expandafter\newcommand\csname #1note\endcsname[1]{%
    \textcolor{#2}{(\textbf{#1:} ##1)}}%
  \expandafter\newcommand\csname #1\endcsname[1]{%
    \textcolor{#2}{##1}}%
  \expandafter\newcommand\csname #1cancel\endcsname[1]{%
    \textcolor{#2}{\sout{##1}}}%
  \expandafter\newcommand\csname #1change\endcsname[2]{%
    \textcolor{#2}{\sout{##1} ##2}}%
  \newenvironment{#1text}{\color{#2}}{\color{black}}
}

\editor{FG}{verde}
\newcommand{\bba}{\Bar{\Bar{a}}}
\editor{kr}{red}

\usepackage{bm}
\usepackage{graphicx}
\usepackage[usenames]{color}
\usepackage{microtype}
\usepackage{amsmath}
\usepackage{amssymb}
\usepackage{xspace}
\usepackage{comment}
\usepackage{footnote}
\usepackage[version=4]{mhchem}
\usepackage{soul}
\usepackage{relsize}
\usepackage{svg}
\usepackage{hhline}
\usepackage{verbatim}
\usepackage[normalem]{ulem}
\usepackage[inline]{enumitem}
\usepackage{scalerel}

\newcommand{\CX}{{A} }
\newcommand{\avdos}{$\langle \text{DOS}(E) \rangle_T$}
\newcommand{\rev}[1]{{#1}}
\begin{document}

\title{
\rev{Uncertainty estimation for molecular dynamics and sampling}
}

\author{Giulio Imbalzano}
\affiliation{Laboratory of Computational Science and Modeling, IMX, \'Ecole Polytechnique F\'ed\'erale de Lausanne, 1015 Lausanne, Switzerland}

\author{Yongbin Zhuang}
\affiliation{
State Key Laboratory of Physical Chemistry of Solid Surfaces, Collaborative Innovation Center of Chemistry for Energy Materials,
Xiamen University,
Xiamen 361005 China}

\author{Venkat Kapil}
\affiliation{Laboratory of Computational Science and Modeling, IMX, \'Ecole Polytechnique F\'ed\'erale de Lausanne, 1015 Lausanne, Switzerland}
\affiliation{Department of Chemistry,  University of Cambridge,  Lensfield Road,  Cambridge,  CB2 1EW,UK}

\author{Kevin Rossi}
\affiliation{Laboratory of Computational Science and Modeling, IMX, \'Ecole Polytechnique F\'ed\'erale de Lausanne, 1015 Lausanne, Switzerland}
\affiliation{Laboratory of Nanochemistry for Energy, ISIC, \'Ecole Polytechnique F\'ed\'erale de Lausanne, 1950 Sion, Switzerland}

\author{Edgar A. Engel}
\affiliation{Laboratory of Computational Science and Modeling, IMX, \'Ecole Polytechnique F\'ed\'erale de Lausanne, 1015 Lausanne, Switzerland}

\author{Federico Grasselli}
\email{federico.grasselli@epfl.ch}
\affiliation{Laboratory of Computational Science and Modeling, IMX, \'Ecole Polytechnique F\'ed\'erale de Lausanne, 1015 Lausanne, Switzerland}

\author{Michele Ceriotti}
\email{michele.ceriotti@epfl.ch}
\affiliation{Laboratory of Computational Science and Modeling, IMX, \'Ecole Polytechnique F\'ed\'erale de Lausanne, 1015 Lausanne, Switzerland}
\begin{abstract}
Machine learning models have emerged as a very effective strategy to sidestep time-consuming electronic-structure calculations, %
enabling accurate simulations of greater size, time scale and complexity. %
Given the interpolative nature of these models, the reliability of predictions depends on the position in phase space, and it is crucial to obtain an estimate of the error that derives from the finite number of reference structures included during the training of the model.
\rev{
When using a machine-learning potential to sample a finite-temperature ensemble, the uncertainty on individual configurations translates into an error on thermodynamic averages, and provides an indication for the loss of accuracy when the simulation enters a previously unexplored region. }
Here we discuss how uncertainty quantification can be used, together with a baseline energy model, or a more robust although less accurate interatomic potential, to obtain  more resilient simulations and to support active-learning strategies. 
Furthermore, we introduce an on-the-fly reweighing scheme that makes it possible to estimate the uncertainty in the thermodynamic averages extracted from long trajectories.
We present examples covering different types of structural and thermodynamic properties, and systems as diverse as water and liquid gallium.\end{abstract}

\maketitle

\section{Introduction} %

Over the last decade, machine learning (ML) potentials~\cite{behl-parr07prl,bart+10prl,rupp+12prl} have demonstrated to be a very effective tool to improve the trade-off between accuracy and speed of atomistic simulations, allowing a quantum-level description of interatomic forces at a small fraction of the typical computational cost of ab-initio calculations.
The combination of ML potentials and traditional atomistic simulations techniques, such as molecular dynamics, has made it possible to address difficult scientific problems in chemistry~\cite{Smith_Isayev_Roitberg_2017,Devereux2020,chmi+18nc,ross+20jctc} and materials science~\cite{behl11pccp,soss+12prb,Bernstein2019,Artrith2018,Zamani2020,chen+20nature,Zeni2019,Fronzi2020}. The possibility of predicting properties beyond the potential energy -- from NMR chemical shieldings~\cite{cuny+16jctc,paru+18ncomm} to the electron density~\cite{broc+17nc,gris+19acscs} -- points to a bright future in which first-principles quality predictions of atomistic properties can be coupled with large-scale simulations~\cite{jia2020pushing} and thorough sampling of quantum statistics and dynamics~\cite{kapi+19jctc2,kapi+20jcp}. 

As ML models become ubiquitous in atomic-scale modeling, the question naturally arises of how much one can trust the predictions of a purely inductive, data-driven approach when using it on systems that are not part of the training set.
This question is particularly pressing because the regression techniques that underlie ML models are inherently interpolative, and their ability to make predictions on new systems hinges on the possibility of decomposing the target property into a sum of atom-centred contributions. Thus, a ML prediction is only reliable if all the local environments that appear in the system of interest are properly represented in the training set. 
It is therefore crucial to obtain an estimate of the error and uncertainty that derive from the finite number of reference structures, and  
many methodological frameworks have been proposed that yield a measure of the uncertainty in the prediction of a machine learning model.\cite{Tran_Neiswanger_Yoon_Zhang_Xing_Ulissi_2020}
Within Bayesian schemes, such as Gaussian process regression, the uncertainty quantification is naturally encoded in the regression algorithm -- although computing the error is substantially more demanding than evaluating the prediction.\cite{rasm05book}
Sub-sampling approaches constitute an alternative. The uncertainty is estimated on the basis of the spread of the predictions of an ensemble (committee) of independently trained ML models, e.g. by subsampling of the full training dataset \cite{Peterson_Christensen_Khorshidi_2017,behl15ijqc, Shapeev2020,poli-roma94aos,efron1979}.
These uncertainty quantification schemes provide qualitative information on the reliability of the ML predictions, and are widely used in the context of online and offline active learning, to identify regions of configuration space that need to be added to the training set
\cite{Shapeev_Gubaev_Tsymbalov_Podryabinkin_2020, shuaibi2020, ross+20jctc, Schran_Brezina_Marsalek_2020, Jinnouchi_Lahnsteiner_Karsai_Kresse_Bokdam_2019,  Vandermause_Torrisi_Batzner_Xie_Sun_Kolpak_Kozinsky_2020,Schran2020}.
When appropriately calibrated~\cite{musi+19jctc}, committee models can further provide a \emph{quantitative} assessment of the uncertainty in %
the prediction of an ML model, which can be readily propagated to estimate the error in properties that are obtained indirectly from the ML predictions such as vibrational spectra \cite{raim+19njp}.

In this paper we consider how to best exploit the availability of \rev{machine-learning models that include an error estimation} in the context of molecular dynamics simulations, and more generally in the evaluation of thermodynamic observables. 
First, we show how to construct a \emph{weighted baseline} ML scheme, in which the uncertainty is used to ensure that whenever the simulation enters an extrapolative regime, the potential falls back to a reliable (if not very accurate) baseline.
Second, we use \rev{errors computed for individual configurations} to estimate the ML uncertainty associated with static \textit{thermodynamic averages} from MD trajectories computed using a single potential. Specifically, we introduce an on-the-fly reweighting technique, which takes into account both
i) the ML uncertainty on single-configuration calculations for a given observable over a significant sample of configurations, and
ii) the distortion of the sampling probability, due to the model-dependent Boltzmann factor entering the statistical averages.
We showcase applications of these methods to several different classes of materials science and chemical systems, ranging from polypeptides, to solutions, to liquid metals, and to both structural and functional properties. 

\section{Theory}\label{sec:theory}

\rev{We consider a machine-learning model that can predict, for a structure $A$, the value of a property $y(A)$ as well as its uncertainty $\sigma^2(A)$. 
We focus our derivations on committee models, that are easy to implement and allow for straightforward error propagation. However, most of the results we derive can be applied to any scheme that provide a differentiable uncertainty estimate for each property prediction. }

\subsection{Committee model and single-point uncertainty estimation}

We start our discussion by summarizing the formulation of the uncertainty estimation scheme based on a calibrated committee of sub-sampled models, introduced in Ref.~\citenum{musi+19jctc}. In a nutshell, the full training set of $N$ input-observation pairs $(\CX, y_{\text{ref}}(\CX))$ is sub-sampled (without replacement) into $M$ training subsets of size $N_s<N$. 
$M$ models are then trained independently on this ensemble of resampled data sets, inducing a fully non-parametric estimate of the distribution $P(y|\CX)$ of the prediction $y$, given an input $\CX$. The moments of such distribution can be readily computed, so that, for instance, the first (mean value) and second (variance) moments are
\begin{align}
& \bar{y}(\CX) = \frac{1}{M} \sum_{i=1}^M y^{(i)}(\CX) \label{eq:y_committee}\\
& \sigma^2(\CX) = \frac{1}{M-1} \sum_{i=1}^M \label{eq:sigma}
\left| y^{(i)}(\CX) - \bar{y}(\CX)\right|^2.
\end{align}
Here, $y^{(i)}(\CX)$ is the prediction of the $i-$th model, while the mean value $\bar{y}(\CX)$ will be dubbed in the following as the \textit{committee} prediction.
The advantage of this machinery is that the ensemble $\lbrace y^{(i)}(\CX)\rbrace_{i=1,\ldots,M}$ of model predictions provides an immediate estimate of the single-point uncertainty $\sigma^2(\CX)$, since it fully characterises the error statistics.

The reduced size $N_s$ of the set of input-observation pairs on which the sub-sampled models are trained implies that the conditional probability distribution $P(y_{\text{ref}}(\CX)|\CX)$ may deviate from the ideal Gaussian behaviour. We assume that such deviation only affects the width of the distribution, which may be too broad or (usually) too narrow, an effect that can also be seen as a consequence of the fact that training points cannot be considered to be independent identically distributed samples. We incorporate this deviation through a linear re-scaling factor $\alpha$ of the width $\sigma$ of the distribution. We further assume that $\alpha$ is independent of $\CX$, and that any two true values $y_{\text{ref}}(\CX)$ and $y_{\text{ref}}(\CX^\prime)$ are uncorrelated if $\CX \neq \CX^\prime$, so that the predictive distribution has the following form:
\begin{equation}
\begin{split}
& P(\mathbf{y}_{\text{ref}} | \lbrace \CX \rbrace, \alpha) \\
& =
\prod_\CX \frac{1}{\sqrt{2\pi \alpha^2 \sigma^2(\CX)}} 
\exp\left[-\frac{\left| y_{\text{ref}}(\CX) - \bar{y}(\CX)\right|^2}{2\alpha^2 \sigma^2(\CX)}\right] 
\end{split}
\end{equation}
The parameter $\alpha$ is then fixed by maximizing the log-likelihood of this distribution, 
\begin{equation}
LL(\alpha) = \frac{1}{N_\mathrm{val}} \sum_{\CX\in\mathrm{val}} \log P(y_{\text{ref}}(\CX)|\CX, \alpha)
\end{equation}
over a set of $N_\mathrm{val}$ validation configurations, giving the optimal 
\begin{equation}
\alpha{^2} \equiv \frac{1}{N_\mathrm{val}} \sum_{\CX \in \mathrm{val}}
\frac{\left|y_{\text{ref}}(\CX) - \bar{y}(\CX)\right|^2}{\sigma^2(\CX)}. \label{eq:alpha}
\end{equation}
In practice, the explicit construction of a validation set can be avoided by means of a scheme where the validation points still belong to the training set, yet they are absent from a given number of sub-sampled models, as discussed in depth in Ref.~\citenum{musi+19jctc}.
Note that Eq.~\eqref{eq:alpha} is a biased estimator when the number of committee members $M$ is small. In Appendix~\ref{app:bias} we discuss the issue in more detail, and show that the bias can be corrected by computing
\begin{equation}
\alpha{^2} \equiv - \frac{1}{M} + \frac{M-3}{M-1} \frac{1}{N_\mathrm{val}} \sum_{\CX \in \mathrm{val}} 
\frac{\left|y_{\text{ref}}(\CX) - \bar{y}(\CX)\right|^2}{\sigma^2(\CX)}. \label{eq:alpha-unbiased-theory}
\end{equation}
We apply this expression in the numerical demonstrations in Section~\ref{sec:results}, but assume the asymptotic $M\rightarrow\infty$ limit in the rest of the formal derivations.%

The determination of the optimal $\alpha$ also allows us to properly re-scale the predictions of the models to be consistent with Eqs. $\eqref{eq:y_committee}$ and $\eqref{eq:sigma}$ and the optimized distribution:
\begin{align}
y^{(i)}(\CX) &\leftarrow \bar{y}(\CX) + \alpha [ y^{(i)}(\CX) - \bar{y}(\CX) ].
\end{align}
The committee prediction $\bar{y}$ is invariant under rescaling, and the spread of the predictions is adjusted according to $\sigma \leftarrow \alpha \sigma$. The rescaled predictions can be used to compute arbitrarily-complicated non-linear functions of $y$, and the mean and spread of the transformed predictions are indicative of the distribution of the target quantities. In what follows, we always assume that the committee predictions have been subject to this calibration procedure. 

\subsection{Using errors for robust sampling and active learning}

Let us consider the following \textit{baselined model}
\begin{equation}
    V^{(i)}(\CX) = V_b(\CX) + V_{\delta}^{(i)}(\CX)
\end{equation}
where the training of the $i-$th model potential $V^{(i)}_{\delta}$ is on the (set of) differences between a target, say DFT-accurate, potential $\{V_\text{ref}(\CX)\}$ and a baseline potential $\{V_b(\CX)\}$.
Splitting a potential in a cheap-to-compute but inaccurate, and an accurate-but-expensive parts has been part of the molecular dynamics toolkit for a long time~\cite{tuck+92jcp,mark-mano08cpl,kapi+16jcp}, and has proven very effective in the context of machine-learning models~\cite{rama+15jctc,bart+17sa}.
Let us define the full committee potential
\begin{equation}
    \bar{V}(\CX) = V_b(\CX) + \bar{V}_{\delta}(\CX), 
\end{equation},
the committee average of the correction potentials
\begin{equation}
    \bar{V}_{\delta}(\CX) = \frac{1}{M} \sum_{i=1}^M V_{\delta}^{(i)}(\CX), 
\end{equation}
and its uncertainty
\begin{equation}
    \sigma^2(\CX) = \frac{1}{M-1} \sum_{i=1}^{M} \left| V_{\delta}^{(i)} - \bar{V}_{\delta}(\CX) \right|^2,
\end{equation}
as in Eqs.~\eqref{eq:y_committee} and \eqref{eq:sigma}.
This uncertainty estimate, \rev{as well as any other similarly accurate and differentiable measure of the error,} can be used as an indication of the reliability of the ML predictions, and incorporated in an active-learning framework~\cite{li+15prl,smit+18jcp,jane+19cs,Schran_Brezina_Marsalek_2020}: during a molecular dynamics simulation, whenever the trajectory enters a region in which the model exhibits an extrapolative behaviour, the uncertainty $\sigma$ increases, and one can gather new configurations for an improved model~\cite{ross+20jctc}. 
Unfortunately, trajectories entering an extrapolative region often become unstable very quickly, leading to sampling of unphysical configurations or the complete failure of the simulation. 
Crucially, when using a baseline potential, one can stabilize the simulation by dynamically switching to using only $V_b$. This automatic fall-back mechanism can be realized by performing MD using the weighted-baseline potential 
\begin{equation}
\begin{split}
U(\CX) &=\left[\frac{1}{\sigma_b^2}+\frac{1}{\sigma^2(\CX)}\right]^{-1} \left[\frac{1}{\sigma_b^2} V_b(\CX) +\frac{1}{\sigma^2(\CX)} \bar{V}(\CX)\right] \\
&=V_b(\CX) + \frac{\sigma_b^2}{\sigma_b^2 + \sigma^2(\CX)} \bar{V}_{\delta}(\CX), \label{eq:U_combined}
\end{split}
\end{equation}
where the baseline uncertainty $\sigma_b$ is estimated as the variance of the difference between baseline and reference
\begin{multline}
    \sigma_b^2 \equiv \frac{1}{N-1} \left[ \sum_{\CX} \left| V_b(\CX) - V_\text{ref}(\CX) \right|^2 \right.\\
    \left.-\frac{1}{N} \left(\sum_{\CX} V_b(\CX) - V_\text{ref}(\CX)\right)^2 \right], \label{eq:sigma_b}
\end{multline} 
the sum running on the full training set, and $V_\text{ref}(\CX)$ being the target energy for configuration $\CX$. This definition explicitly takes into account the fact that the baseline and reference often differ by a huge constant.
Eq.~\eqref{eq:U_combined} corresponds to the weighted sum of the baseline potential $V_b(\CX)$ and the full committee potential $\bar{V}(\CX)$, consistent with a minimization of the combined error.
The forces (and higher derivatives) can be defined straightforwardly, paying attention to the $\CX$-dependence of $\sigma^2(\CX)$ when the derivatives of $U(\CX)$ are taken. 
Note also that in many cases -- including Behler-Parrinello neural networks~\cite{behl-parr07prl} and SOAP-GAP models~\cite{bart+10prl} -- the ML energy is computed as a sum of atom-centred contributions
\begin{equation}
\bar{V}_\delta(\CX) = \sum_{k\in \CX}
\bar{V}_\delta(\CX_k),
\end{equation}
where $\CX_k$ indicates the environment centred on the $k$-th atom in structure $\CX$. Thus, it is possible to compute uncertainty estimates at the level of individual atomic contributions, and evaluate Eq.~\eqref{eq:U_combined} as 
\begin{equation}
U(A) = V_b(\CX) + \sum_{k\in \CX} \frac{\sigma_b^2}{\sigma_b^2 + \sigma^2(\CX_k)} \bar{V}_{\delta}(\CX_k).
\end{equation}
This expression can be used even if the baseline does not entail a natural atom-centred decomposition, although in such a case one needs to re-define $\sigma_b$ so that it corresponds to the estimated error \emph{per atom}. 
This can be beneficial when the error is not spread equally across the system, e.g. when an unexpected chemical reaction occurs in an otherwise homogeneous system.

By monitoring the weight of the ML correction one can determine whether the simulation remains largely in the low-uncertainty region, or whether it enters the extrapolative regime too frequently, requiring further training. 
Finally, it is worth mentioning that a similar strategy could be used to combine multiple ML potentials with different levels of accuracy, for instance one based on short-range/two-body interactions, that is more resilient but inaccurate, and one based on a long-range and high-body-order parameterization, which is likely to be more accurate, but requires large amounts of data for training, and is therefore more likely to enter high-uncertainty regions.

\subsection{On-the-fly uncertainty of thermodynamic averages}\label{sec:ML-Uncertainty}

The machinery discussed so far paves the way for reliable estimates of the uncertainty of single-point calculations, i.e. of the value an observable quantity assumes when evaluated at a specific point in phase-space. It also allows computing the uncertainty of predictions averaged over several samples, assuming that the only source of error is that associated with the ML model of the target property~\cite{chiheb2020}.
However, the uncertainty in predictions also propagates to thermodynamic averages of target properties. 
\rev{Estimating how such uncertainty propagates is particularly straightforward in the case of a committee-based estimate.}
Computing the mean of an observable $a$ over a trajectory sampling e.g. the mean potential $\bar{V}$ from a committee of $M$ potential models (PMs) $V^{(i)}$ yields
\begin{equation}
\bba \equiv \langle \bar{a} \rangle_{\bar{V}} = \frac{1}{M'} \sum_{j=1}^{M'} \left<a^{(j)}\right>_{\bar{V}},
\label{eq:a-bbar}
\end{equation}
where  $a^{(j)}$ indicates the member of a committee of $M'$ observable models (OMs), and $\langle a\rangle_V$ the mean of an observable over the ensemble defined by the potential $V$.

When computing thermodynamic averages, one should therefore also include the uncertainty in the ensemble of configurations.
A na\"ive (but very time-consuming) way to estimate the full uncertainty relies on running $M$ simulations, each driven by the (re-scaled) force field of a specific PM, and computing the averages $\langle a ^{(j)}\rangle_{V^{(i)}}$ of the target observable $a^{(j)}$ for each OM, and finally the average 
\begin{equation}
    \tilde{a} \equiv \frac{1}{M M'} \sum_{i=1}^M \sum_{j=1}^{M'} \langle a^{(j)} \rangle_{V^{(i)}}
\label{eq:dbl-avg}
\end{equation}
and variance over both OMs and PMs. While trivially parallelizable, this strategy is inconvenient, as it prevents exploiting the considerable computational savings that can be achieved by computing multiple committee members over the same atomic configuration. %

The need for different trajectories can be avoided by employing an on-the-fly re-weighting strategy~\cite{torr-vall99jcp}. For a canonical distribution at temperature $T = 1/(\beta k_B)$,
\begin{equation}
\langle a^{(j)} \rangle_{V^{(i)}} \equiv \frac{1}{Z^{(i)}} \int  a^{(j)}(\mathbf{q}) e^{-\beta V^{(i)}(\mathbf{q})}  d\mathbf{q},\label{eq:direct}
\end{equation}
where $\mathbf{q}=(\mathbf{q}_1,\ldots,\mathbf{q}_{N_p})$ is the set of positions of the $N_p$ particles,
\begin{equation}
Z^{(i)} \equiv \int e^{-\beta V^{(i)}(\mathbf{q})} d\mathbf{q}
\end{equation}
is the configurational partition function and $V^{(i)}(\mathbf{q})$ is the potential energy of the $i-$th model.
By introducing the \textit{weights} 
\begin{equation}
w^{(i)}(\mathbf{q}) \equiv e^{-\beta [V^{(i)}(\mathbf{q}) - \bar{V}(\mathbf{q})]} ,\label{eq:weights}
\end{equation}
where $\bar{V}$ is the mean committee potential energy, we find
\begin{equation}
\langle a^{(j)} \rangle_{V^{(i)}} = \frac{\int w^{(i)}(\mathbf{q}) a^{(j)}(\mathbf{q}) e^{-\beta \bar{V}(\mathbf{q})}  d\mathbf{q}}{\int w^{(i)}(\mathbf{q}) e^{-\beta \bar{V}(\mathbf{q})} d\mathbf{q}}
\end{equation}
or, in shorthand notation,
\begin{equation}
\langle a^{(j)} \rangle_{V^{(i)}} = \frac{\left\langle w^{(i)} a^{(j)}\right\rangle_{\bar{V}}}{\left\langle w^{(i)}\right\rangle_{\bar{V}}}. \label{eq:reweighted}
\end{equation}
This means that, under the ergodic hypothesis, the re-weighting technique allows us to run \textit{a single trajectory} driven by the force field of the committee, and yet to obtain estimates for the averages as computed via the different models. 
Thus, it is possible to compute the full uncertainty, including both the error on the OMs and the PMs, by using the reweighting formula to evaluate
\begin{equation}
\tilde{\sigma}^2 \equiv \frac{1}{MM' - 1} \sum_{i=1}^M \sum_{j=1}^{M'}   \left| \langle a^{(j)} \rangle_{V^{(i)}}    -    \tilde{a} \right|^2 \label{eq:sigma_tilde}
\end{equation}
This reweighing approach has further important implications to molecular dynamics simulations: for instance, in on-the-fly learning it is customary to correct (re-train) the ML force-field from time to time along a molecular dynamics simulation so to include new configurations in the training set:\cite{csanyiPRL2003,li2015molecular} an operation which can introduce systematic errors on the estimation of canonical averages, due to the different potential-energy fields along the trajectory. 
By simply storing the model-dependent potential energies along the simulation alongside the corresponding configurations, one can at any time compute a set of weights based on the most recent value of the potential, to obtain averages that use the entire trajectory and yet are consistent with the most accurate model available.

Equation $\eqref{eq:reweighted}$ is in principle exact. However, from a computational standpoint, the efficiency in sampling the probability measure of the $i-$th model through reweighing is in general lower than what it would be by direct sampling as in Eq. $\eqref{eq:direct}$, with an error growing exponentially with the variance of $h^{(i)} \equiv-\ln w^{(i)} = \beta (V^{(i)} - \bar{V})$, that inevitably increases with system size. 
Given that we are only interested in computing an estimate of the uncertainty, we can use an approximate (but statistically more stable) expression introduced in Ref.~\citenum{ceri+12prsa}, based on a cumulant expansion.
Assuming that $ a^{(j)} $ and $h^{(i)}$ are correlated Gaussian variates %
(all with respect to the committee phase-space probability measure), we have
\begin{equation}
\begin{split}
    &\langle a^{(j)} \rangle_{V^{(i)}} \approx \langle a^{(j)} \rangle_{\bar{V}} \\ &-\beta [\langle a^{(j)} (V^{(i)} - \bar{V}) \rangle_{\bar{V}} - \langle a^{(j)} \rangle_{\bar{V}} \langle V^{(i)} - \bar{V} \rangle_{\bar{V}} ]. \label{eq:cumulant}
\end{split}
\end{equation}

\begin{figure}
\includegraphics[width=\columnwidth]{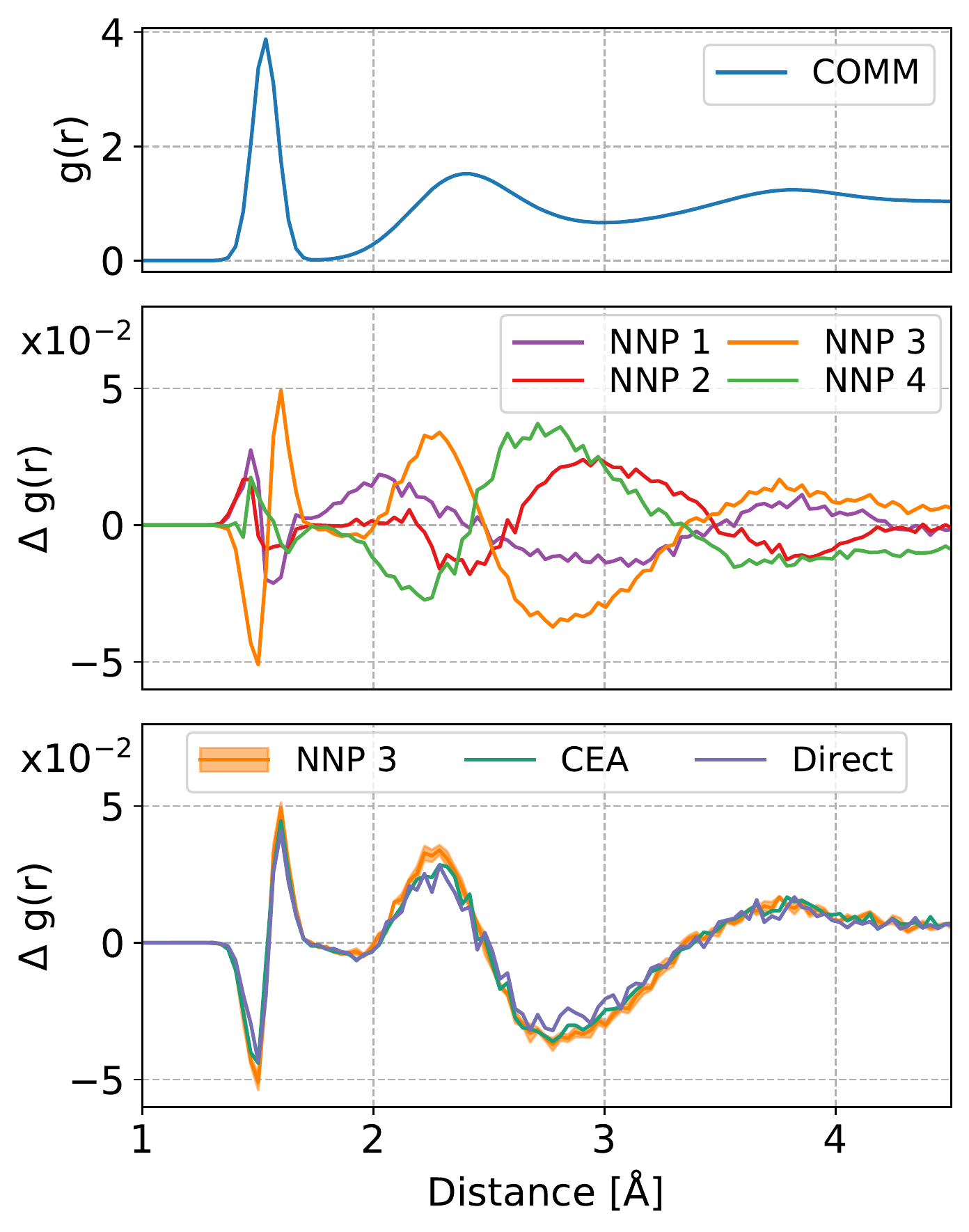} 
\caption{Hydrogen-hydrogen radial pair correlation function in water. (Top) pair distribution function computed for a simulation driven by the committee average; (middle) deviations, from the plot in the top panel, of the pair distribution functions extracted from $M=4$ independent trajectories (one for each NNP, displayed in different colours); (bottom) comparison between the result from an independent trajectory driven by NNP 3 (orange), and the pair correlation obtained from the committee-driven trajectory by direct re-weighting, Eq.~\eqref{eq:reweighted} and the cumulant expansion approximation (CEA), Eq.~\eqref{eq:cumulant}.}\label{fig:water_gofr_models}
\end{figure}

In order to compare the different definitions given so far for a physical example, we consider a simple thermodynamic average, i.e.~the radial pair correlation function $g(r)$
between H atoms in water.
We refer to Sec.~\ref{sec:results} for the specific details of the simulation. 
The top panel in Fig.~\ref{fig:water_gofr_models} displays $\bar{\bar{g}}(r)$ determined, as in Eq.~\eqref{eq:a-bbar}, by averaging over a significant number of atomic configurations sampled from a trajectory driven by a committee of $M=4$ models (neural network potentials, NNPs).
The middle panel displays the differences $\Delta g^{(i)}(r) = g^{(i)}(r) - \bar{\bar{g}}(r)$, with $g^{(i)}(r)$ obtained after sampling structures from separate trajectories driven by each NNP model.
In the bottom panel, we focus on one of the models, and we compare the deviation of the pair distribution function, with respect to $\bar{\bar{g}}(r)$, computed according to: an independent trajectory driven by NNP 3 (orange, same as in the central panel); the direct re-weighting of the sampling from the trajectory driven by the committee as in Eq.~\eqref{eq:reweighted} (purple); and within the cumulant expansion approximation (CEA), Eq.~\eqref{eq:cumulant} (dark green). 
The match between the three curves shows that the re-weighting procedure, both in its exact form and using the CEA, is capable of reproducing the result obtained from an independent trajectory generated by a specific NNP without the need of explicitly running it.

For this example, which entails a relatively small simulation cell and low discrepancy between the committee average and the individual NNPs, there is no substantial difference between the exact and CEA reweighing. 
We recommend using the CEA over the direct estimator, not only because of its improved stability and statistical efficiency, but also because the linearized form emphasizes the different sources of error associated with the single-trajectory average~\eqref{eq:a-bbar}, and has several desirable formal implications.
First, using the CEA the mean over the trajectories is consistent with the average computed over the trajectory driven by $\bar{V}$ -- whereas in general Eq.~\eqref{eq:dbl-avg} would yield a different value from~\eqref{eq:a-bbar}:
\begin{equation}
\!\tilde{a} \approx  
\bba +  \!\frac{\beta}{M}\sum_i [\langle \bar{a} (V^{(i)} - \bar{V}) \rangle_{\bar{V}} - \langle \bar{a} \rangle_{\bar{V}} \langle V^{(i)} \!-\! \bar{V} \rangle_{\bar{V}} ]= \bba.
\end{equation}
Second, one sees that 
\begin{equation}
\tilde{\sigma}^2 \approx {\frac{M(M'-1)}{MM' - 1}} \sigma^2_a + {\frac{M'(M-1)}{MM' - 1}} \sigma^2_{aV}
\underset{M,M'\rightarrow \infty}{=}\sigma^2_a+\sigma^2_{aV} \label{eq:fullsigma2_CEA}
\end{equation}
where 
\begin{equation}
    \sigma^2_a \equiv \frac{1}{M'-1} \sum_{j=1}^{M'}
    \left| \langle a^{(j)} \rangle_{\bar{V}}    -    \bba   \right|^2 \label{eq:sigma_a}
\end{equation}
indicates the uncertainty arising from the OMs, and 
\begin{equation}
\begin{split}
 \sigma_{aV}^2 & \equiv \frac{1}{M'} \sum_{j=1}^{M'} {\sigma_{aV}^{2}}^{\!(j)}, \\ 
{\sigma_{aV}^{2}}^{\!(j)} 
        &\equiv {\frac{1}{M-1}} \sum_{i=1}^M \left| \langle a^{(j)} \rangle_{V^{(i)}} - \frac{1}{M} \sum_{i=1}^M \langle a^{(j)} \rangle_{V^{(i)}}\right|^2 \\
        &\approx \frac{\beta^2}{M-1} \sum_{i=1}^M \left|\langle a^{(j)} ( V^{(i)} - \bar{V} ) \rangle_{\bar{V}}  - \langle a^{(j)} \rangle_{\bar{V}} \langle V^{(i)} - \bar{V} \rangle_{\bar{V}} \right|^2
\end{split} \label{eq:sigma_V}
\end{equation}
indicates the uncertainty that arises due to the sampling of the different PMs. 
In the general case of an uncertainty estimation that is  \emph{not} based on a committee model, where only the ``best values'', $\bar{a}(\mathbf{q})$ and $\bar{V}(\mathbf{q})$, and their uncertainties, $\sigma_{\bar{a}}(\mathbf{q})$ and $\sigma_{\bar{V}}(\mathbf{q})$, are available, the reweighting technique so far described becomes inapplicable. The error-propagation formula for the uncertainty $\tilde{\sigma}^2$ on the canonical average $\langle \bar{a} \rangle_{\bar{V}}$ cannot be straightforwardly implemented either, since it requires the off-diagonal elements of the covariance matrix, and not only $\sigma^2_{\bar{a}}(\mathbf{q})$ and $\sigma^2_{\bar{V}}(\mathbf{q})$. 
Nonetheless, as shown in Appendix \ref{app:unc-prop}, even in this case, a simple upper bound for $\tilde{\sigma}^2$ can be obtained:
\begin{equation}
\tilde{\sigma} \leq \langle \sigma_{\bar{a}} \rangle + \beta\left\langle \big|\langle \bar{a} \rangle - \bar{a} \big|\, \sigma_{\bar{V}} \right\rangle, \label{eq:sigma_gen}
\end{equation}
which corresponds, at least in spirit, to the results we obtain for the committee model, Eqs.~\eqref{eq:fullsigma2_CEA}, \eqref{eq:sigma_a}, and \eqref{eq:sigma_V}, and its implementation shows no hurdles.
\begin{figure*}
    \centering
    \includegraphics[width=18cm]{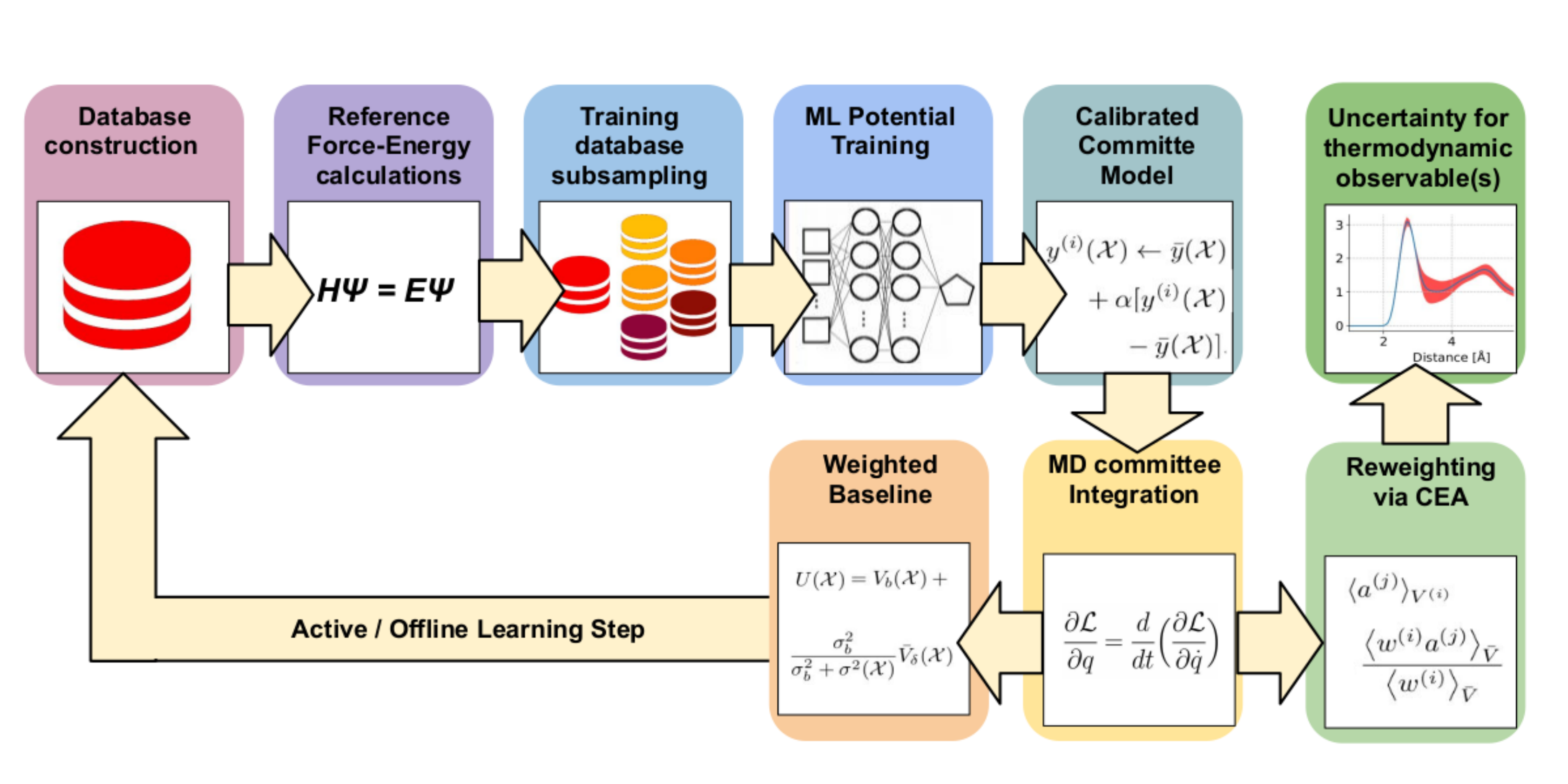}
\caption{Graphical summary concerning the essential steps of the workflow described in Sec. \ref{sec:theory} to train a committee of ML potentials (and possibly other observables), internally validate and re-calibrate it through a proper scaling factor, $\alpha$, and finally use it for uncertainty estimation in molecular dynamics and thermodynamic averages.
}
    \label{fig:workflow}
\end{figure*}

\section{Applications}
\label{sec:results}

\rev{Fig.~\ref{fig:workflow} summarizes how the weighted baseline scheme, and the on-the-fly estimation of errors for statistical averages, can be integrated with a calibrated committee model, in the context of a molecular dynamics simulation.}
After the construction of a suitable database on which reference values (say of energies and forces) are computed, the database is randomly sub-sampled into $M$ smaller training sets on which a committee of $M$ ML models are trained. 
Depending on the specific physical system/quantity analyzed we adopt two alternative but equally correct approaches to construct a validation set, in order to calibrate the uncertainty of the committee and estimate the re-scaling factor $\alpha$. The first strategy consists in extracting $N_\mathrm{val}$ decorrelated configurations from short committee MD trajectories, calculating forces and energies with the reference method, and employing these as the validation set. In the second strategy, instead, the ensemble of $N_\mathrm{val}$ validation structures was gathered by selecting, in the original training database, those structures that do not appear in at least $n$ of the training subset. 
Following the $\alpha$ calibration step, MD simulation are driven by the committee model. 
The weighted-baseline numerical integration of 
the equations of motion is based on  Eq.~\eqref{eq:U_combined}, which reduces to a non-baselined model by setting $V_b = 0$. 
During the MD simulation driven by the committee model, all the (re-scaled) model-dependent quantities of interest are stored for a significant set of (uncorrelated) configurations, eventually leading to re-weighting and, therefore, to uncertainty estimation of the chosen thermodynamic averages.
Any configuration encountered along the trajectory that is associated with an error higher than a set threshold can be used to improve the reference database, in an offline (or online) active learning scheme. 

In the next subsections we describe how we applied this routine to weighted baseline integration (Sec. \ref{sec:tripeptide}), as well as to compute thermodynamic average and the related ML uncertainty for different observables in different physico-chemical environments (Secs.~\ref{sec:PDF},  \ref{sec:FES}, \ref{sec:FT-DOS}).  
All the simulations are run with the molecular dynamics engine i-PI\cite{kapi+19cpc} interfaced with the massively parallel molecular dynamics code LAMMPS\cite{plim95jcp} with the n2p2 plugin \cite{singraber2019library} to evaluate the neural network potentials.

\begin{figure*}
    \centering
    \includegraphics[width=\textwidth]{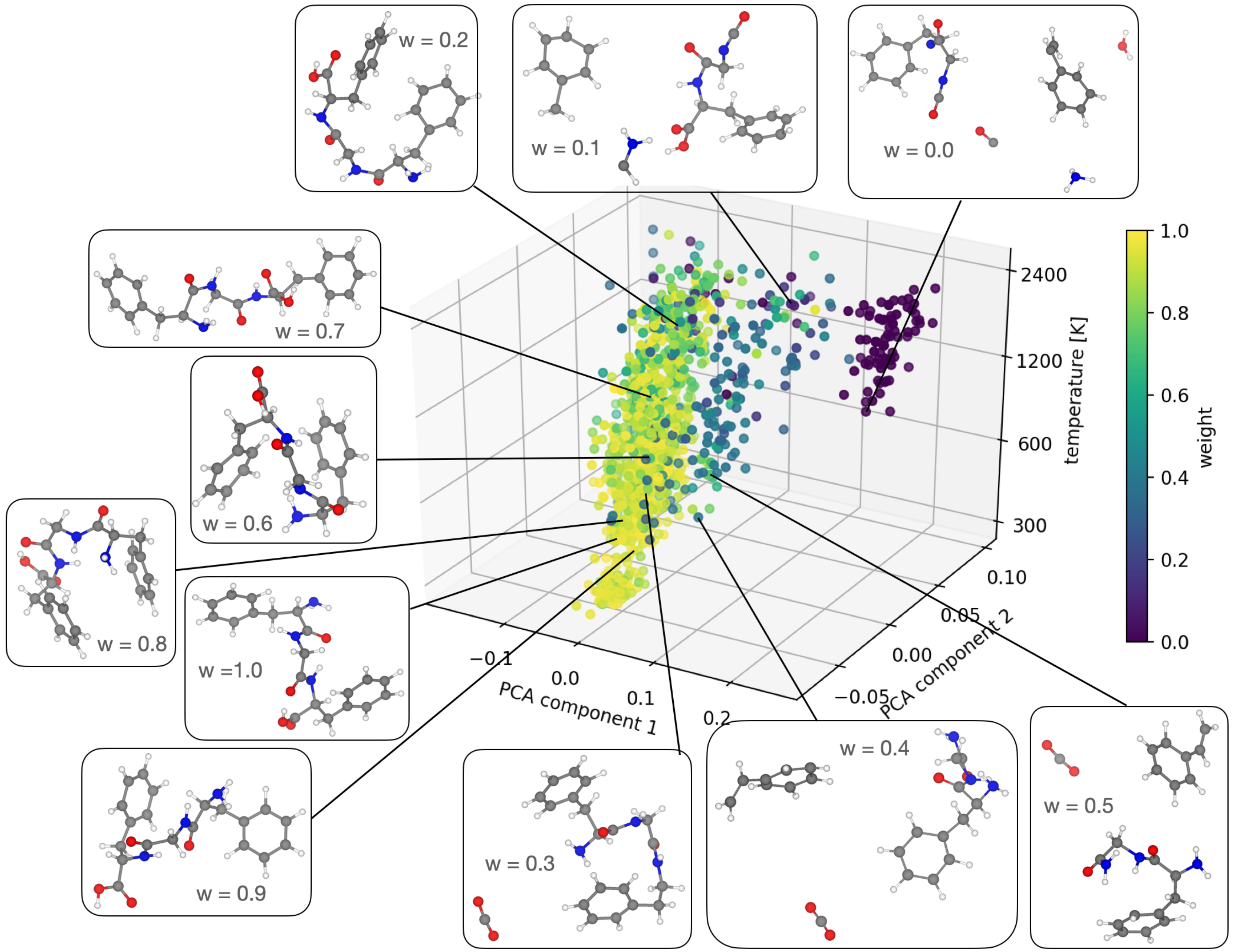}
    \caption{A visualization of the results of the replica-exchange MD simulation of the Phe-Gly-Phe tripeptide, using a weighted-baseline scheme. 
    Central scatter-plot: a set of 2,000 atomic configurations collected from all replicas is classified according to the first two principal components of their SOAP features ($x$ and $y$ axes), and the replica temperature ($z$ axis, in logarithmic scale). The  SOAP representation employs a  cut-off radius of 4 \AA, a basis of $n=6$ radial and $l=4$ angular functions, and a Gaussian width of 0.3 \AA. Each point corresponds to one configuration, colour-coded according to the weight of the ML correction to the baseline potential, see Eq.~\eqref{eq:U_combined}. Examples of typical configurations that are representative of the different temperatures and ML correction weights are displayed in the panels surrounding the scatter plot.}
    \label{fig:tripeptide}
\end{figure*}

\subsection{Weighted baseline integration}\label{sec:tripeptide}
We begin by performing and analyzing a 120\,ps temperature replica-exchange molecular dynamics (REMD)~\cite{petraglia2016beyond} simulation of the Phe-Gly-Phe tripeptide, using the weighted baseline method.
The i-PI energy and force engine~\cite{kapi+19cpc} is used to simulate 12 Langevin-thermostatted replicas with temperatures between 300\,K and 2440\,K using a time-step of 0.5\,fs. 
Baseline density-functional-based tight binding energies and forces are evaluated using the DFTB+~\cite{aradi2007dftb+} package and the DFTB3/3OB~\cite{gaus2012parametrization, gaus2014parameterization} parametrisation with a D3BJ~\cite{grimme2011effect} dispersion correction (3OB+D3BJ).
An ensemble of $M=4$ Behler-Parrinello artificial neural networks (NN)~\cite{behl-parr07prl} is then used to promote this baseline to a first-principles density-functional-theory (DFT) level of theory.
The DFT calculations are performed using the GAMESS-US~\cite{schmidt1993general, gordon2005advances} code and the PBE density functional~\cite{perdew1996generalized} with a dDsC dispersion correction~\cite{steinmann2010system, steinmann2011comprehensive, steinmann2011generalized} and the def2-TZVP basis set~\cite{Schafer1992}.
The NNs are trained to reproduce the differences between the DFTB+ baseline and the target DFT energies and forces. 
The NNs differ only in the initialisation of the NN weights and the internal cross-validation splits of the reference data into 90\% training and 10\% test data.
The reference data underlying the NNs is constructed by farthest-point sampling configurations from 1.5\,ns long REMD simulations of 26 aminoacids, each composed of 16 Langevin-thermostatted replicas with  logarithmically-spaced temperatures between 300\,K and 1000\,K. 
The resultant set of configurations is enriched with 3,380 geometry-optimised dimers from the BioFragment Database\cite{burns_biofragment_2017}.
Note that the aminoacids are simulated at less than half the maximum temperature, at which the tripeptide is simulated.
The uncertainties associated with the ensemble predictions are estimated using the scheme of Ref.~\citenum{musi+19jctc}, using a scaling correction of $\alpha = 1.0$, computed on the tripeptide validation data.
The uncertainty of the ML model is used, together with a baseline uncertainty of DFTB $\sigma_b = 7 \times 10^{-3}$\,meV/atom,  estimated according to Eq.~\eqref{eq:sigma_b}, to build a weighted baseline model following Eq.~\eqref{eq:U_combined}.

\begin{figure*}
    \centering
    \includegraphics[width=\textwidth]{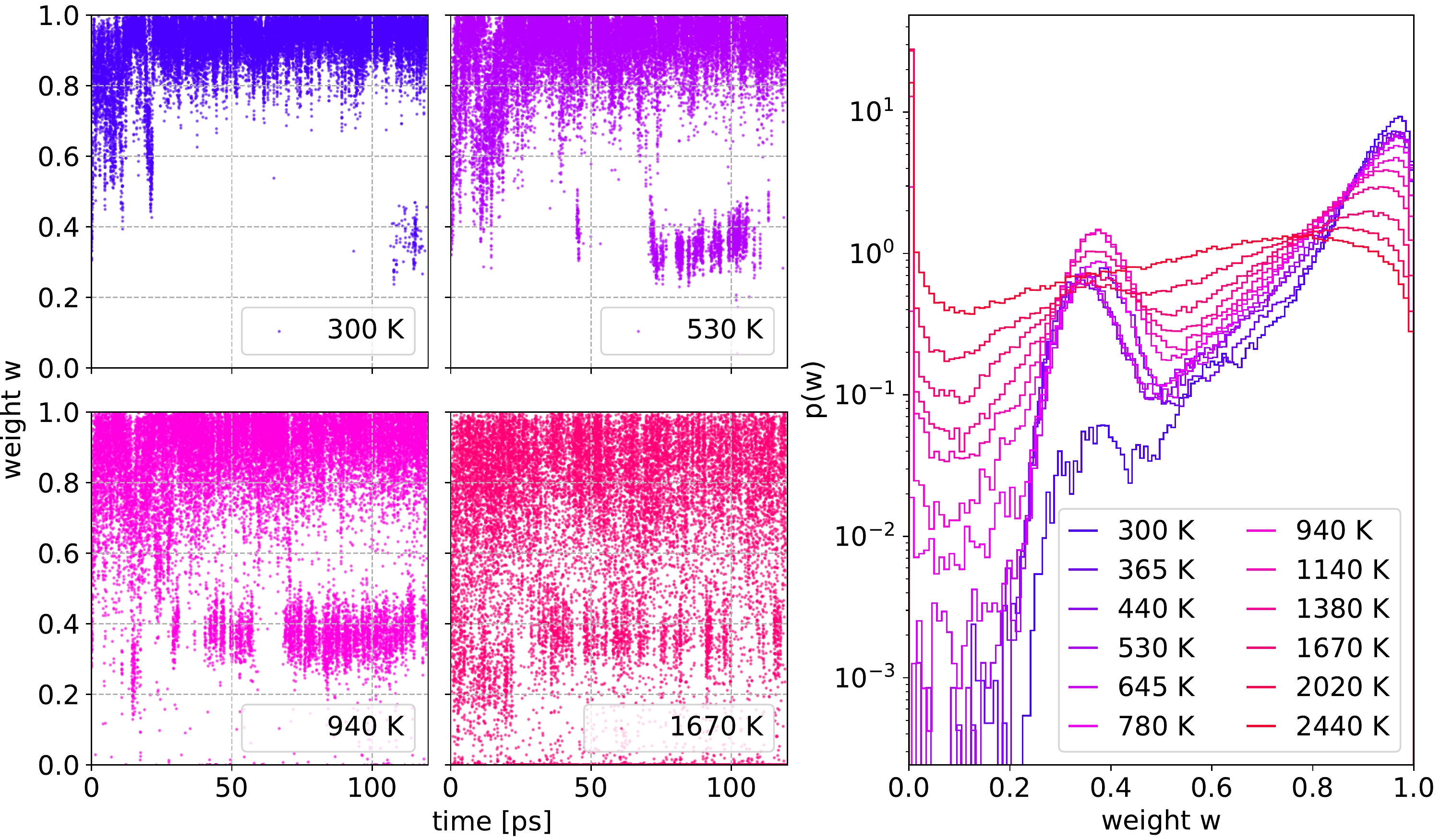}
    \caption{Weights for the ML correction in the weighted-baseline scheme for the Phe-Gly-Phe tripeptide discussed in the text. In the left panels the weights $w$ are displayed at different temperatures for a segment of the REMD trajectory. The rightmost panel shows the log-histogram of the occurrences of the weights at different temperatures.}
    \label{fig:tripep_weights}
\end{figure*}

The results of the REMD simulation of the Phe-Gly-Phe tripeptide are portrayed in Fig.~\ref{fig:tripeptide}.  
The central scatter-plot shows 2,000 atomic configurations, drawn at constant stride from all REMD target ensemble temperatures.
The configurations are classified according to the first two principal components ($x$ and $y$ axes), obtained from a principal component analysis (PCA) of their SOAP features, and temperature ($z$ axis). 
Each configuration $A$ is coloured according to the weight $w(A) = \sigma_b^2/[\sigma_b^2 + \sigma^2(A)]$ of the ML correction applied to the baseline potential during the simulation (see Eq.~\eqref{eq:U_combined}).
Examples of configurations with very low ($0\leq w \leq 0.2$), modest ($0.3\leq w \leq 0.5$), and large weights ($0.6\leq w \leq 1$) are grouped at the top, bottom and left of the scatter plot, respectively.
The figure shows that at low temperature the simulation samples exclusively different conformations of the polypeptide chain, that are well-represented in the training set and that are therefore associated with low ML uncertainty and high values of $w(A)$. 
At temperatures above $\approx 500$K, the polypeptide starts decomposing, releasing first \ce{CO2} and, at temperatures above $\approx 1000$K, \ce{NH3}, \ce{H2O}, as well as larger fragments. None of these highly energetic reactions are represented in the training set, which is reflected in the sharp decrease of the weight. 
Upon entering the extrapolative regime, the NN correction to the baseline, $\bar{V}_\delta$, is suppressed by the vanishing weight $w$, thereby ensuring numerical stability of the simulation subject to the baseline potential. 

A quantitative analysis of weight distributions is shown in Fig. \ref{fig:tripep_weights}. Higher temperatures are displayed in warmer colours. The left panels show the weights $w$ along the REMD trajectory. These values are collected in the rightmost histogram which displays, in semi-log scale, the distribution $p(w)$ of weights at different temperatures. We see that at intermediate $T$, an ``island'' at $w\approx 0.4$ -- or a peak in $p(w\approx 0.4)$ -- emerges, which corresponds to the tripeptide dissociation and the release of a \ce{CO2} molecule. At even larger temperatures the probability $p(w=0)$ grows, while the peak at $w\approx 0.4$ is levelled out by the increase in the number of low-$w$ snapshots and the $p(1)$ decreases by more than an order of magnitude due to the persistence of the extrapolative regime at $T \gg 1000$ K.   
This simulation provides a compelling example of how a weighted baseline scheme allows exploring all parts of configuration space without incurring in unphysical behaviour and instability due to extrapolations of the NNs -- which typically occur within the first 100\,ps of a similar REMD simulation using a non-weighted baseline correction.
Quite obviously, the configurations collected in the extrapolative regime do not reach the level of accuracy of the high-end electronic structure method, but only that afforded by the baseline potential. Nonetheless, simulations based on this scheme can be used whenever extrapolation occurs only over brief stretches of the trajectory, or when (as it is often the case) one is only interested in the low-temperature portion of a REMD simulation, with the high temperature replicas used only to accelerate sampling.
Furthermore, one can store configurations characterised by a large $\sigma(A)$ in order to add them to the training database, which simplifies greatly the implementation of online and offline active learning schemes.

\subsection{Pair distribution function}\label{sec:PDF}

\begin{figure*}
\includegraphics[width=2\columnwidth]{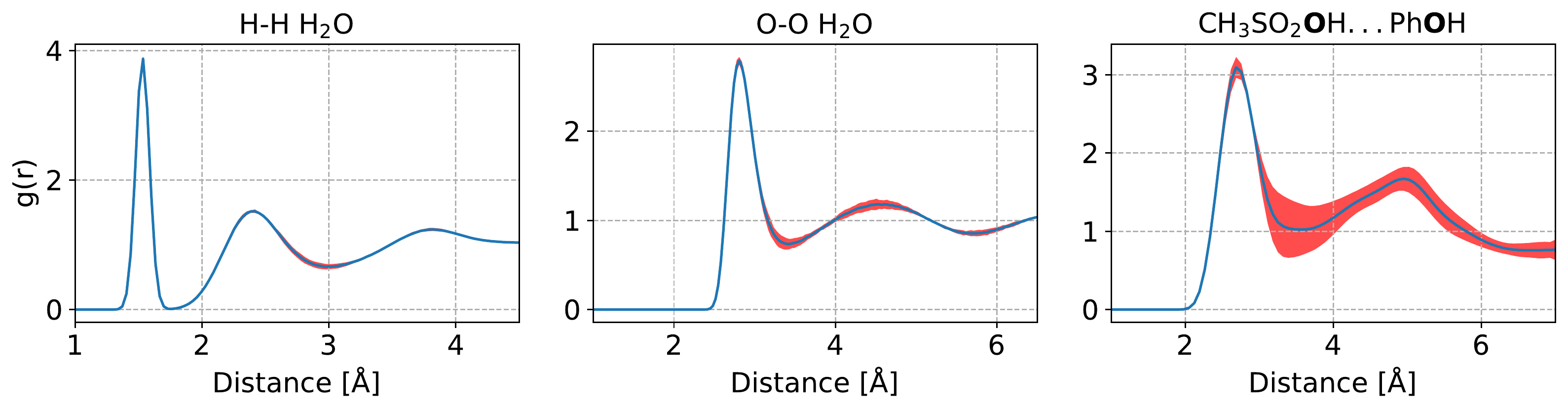} 
\caption{Pair correlation function in water (left, middle panels) and phenol-solvated methanesulphonic acid (right panel). The committee value (blue solid line) and its uncertainty (shaded red area) as estimated from Eq.~\eqref{eq:sigma_V} are displayed.}\label{fig:water_gofr}
\end{figure*}

The radial distribution function represents a simple and insightful structural observable to test the method developed in Sec.~\ref{sec:theory} to estimate the uncertainty on thermodynamic averages.
Computationally, $g(r)$ is usually determined \textit{i)} by sampling a significant number of atomic configurations from a thermodynamic ensemble; \textit{ii)} by computing the minimum image separations $\mathbf{r}_i - \mathbf{r}_j$ of all the atomic pairs, for each sampled configuration, and \textit{iii)} by sorting these separations into an histogram $h$ whose bins extend in the interval $[r,r+\delta r]$. When the reweighting procedure is considered
point \textit{i)} is performed by running a MD trajectory driven by the committee model alone, and the model-dependent phase-space sampling is accounted by the weights, Eq.~\eqref{eq:weights}.
Notice that the calculation of the radial distribution function $g^{(i)}(r)$ of the $i$-th member of the ML committee depends on $i$ through the weights alone, i.e.~through the calibrated potential energy estimate for each member.

\subsubsection{Water}\label{sec:water_gofr}
A committee of $M=4$ NNP models was trained via the n2p2 code \cite{singraber2019parallel} over a dataset of 1593 64-molecule bulk liquid water structure whose total energy and the full set of interatomic-force components were computed at the revPBE0-D3 level with CP2K~\cite{matcloud18a}. 
The atomic environments are described within a cutoff radius of $12.0~\text{a.u.}$ using the symmetry function sets for H atoms (27 functions) and O atoms (30 functions), as selected in Ref.~\onlinecite{mora+16pnas}. The hydrogen and oxygen atomic NNs consist of two hidden layers with 20 nodes each. We refer to Ref.~\onlinecite{chen+19pnas} for further details on the training set.

We run an $NVT$ MD trajectory, driven by the committee, at $T=300~\text{K}$ for $2~\text{ns}$ on a system of 64 water molecules inside an equilibrated cubic box of side $23.86~\text{\AA}$.
We obtain an unbiased estimate for the correction factor $\alpha = 2.1$, using the expression in Appendix~\ref{app:bias}. Note that without applying the correction for the estimator bias, would lead to substantial over-estimation of the correction factor, in this case $\alpha = 3.75$.
Figure \ref{fig:water_gofr} displays the hydrogen-hydrogen (left) and oxygen-oxygen (middle) pair distribution function $g(r)$. The ML uncertainty, computed as in Eq.~\eqref{eq:sigma_V}, is shown as a shaded area. The error on position and height of the first peak is minuscule, while slighlty larger uncertainty is predicted on the longer-range features for the O--O correlations. This analysis demonstrates,  with a simple post-processing of a single trajectory, that the accuracy of the NNP is sufficient to describe quantitatively the $g(r)$ -- a useful verification of the reliability of the model.

\subsubsection{Methanesulphonic acid in phenol}\label{sec:ch3so2oh-gr}

As a second example, we consider the solvation of methanesulfonic acid (CH$_3$SO$_2$OH) in phenol (C$_6$H$_6$O), a system that was studied in Ref.~\citenum{ross+20jctc} because of its relevance to the synthesis of commodity chemicals such as hydroquinone and catechol, in which methanesulfonic acid acts as a catalyst for the reaction between \ce{H2O2} and phenol. 
We use an ensemble of $M=5$ neural network (NN) machine learning potentials to simulate one acid molecule dissolved in 20 phenol molecules at $T=363$ K.
The technical details and the resulting potentials are identical to those presented in Ref.~\onlinecite{ross+20jctc}, that are available from Ref.~\citenum{matcloud20d}. Note that in the original publication the calibration factor was estimated to be $\alpha=5.8$. Using the unbiased estimator introduced here, Eq.~\eqref{eq:alpha-unbiased-theory},  yields a corrected value of $\alpha=4.1$.

An understanding of the solvation of \ce{CH3SO2OH} by phenol is a necessary preliminary step towards rationalizing the regio-selectivity of this acid in the catalytic hydroxylation of phenol to form catechol or hydroquinone.
Methanesulfonic acid acts both as a hydrogen bond acceptor through its sulfonil oxygen atoms, and as a donor through the methanesulfonic hydroxyl group.
The strength and population of hydrogen bonds can be inferred by a quantitative analysis of the pair correlation function $g(r)$ between the protonated O in the hydroxyl group of methanesulfonic acid (\ce{CH3SO2}\textbf{O}H) and the O atom in phenol (\ce{C6H5}\textbf{O}H). 
We compute the pair correlation function  from 16 independent MD simulation runs for a total of about 1.6~ns. 
A thorough discussion of the MD integration set up and the related technical details can be found in Ref.~\onlinecite{ross+20jctc}.

The uncertainty in the $g(r)$ obtained by a CEA reweighing of the committee members, as in Eq.~\eqref{eq:sigma_V}, is considerably larger than what observed for the case of water (right panel of Fig.~\ref{fig:water_gofr}), together with its uncertainty calculated as in Eq.~\eqref{eq:sigma_V} (shaded area).
This can be ascribed in part to the slightly larger test error computed for the ML potential (which is unsurprising given the considerably more complex composition), but also in part to poorer statistics due to the presence of just a single acid molecule in the simulation cell. The statistical uncertainty on the committee $g(r)$ obtained via a block analysis is indeed comparable to the one due estimated by the committee reweighing.
Similar to what we observe for the O-O $g(r)$ in water, the uncertainty is not constant, but is largest at the minimum between the first and second coordination shell. The fact that the first coordination shell is affected by a small error is reassuring, suggesting that the geometry and population of hydrogen-bonded configuration is predicted reliably. 
Overall, this example demonstrates how the estimates we introduce \rev{for the effects of the ML error on sampling} make it possible to assess the reliability of structural observables, particularly in difficult cases in which the model exhibits a substantial error, and so it is important to determine precisely whether such error does or does not (as in this case) affect the qualitative interpretation of simulations results.  %

\subsection{Free energy landscapes}\label{sec:FES}

Combining ML potentials and enhanced sampling techniques makes it possible to explore computationally free-energy landscapes that involve activated events, such as chemical reactions and phase transitions. In this Section, we show how on-the-fly reweighing can straightforwardly applied to the calculation of free-energy differences and enhanced sampling simulations.

\subsubsection{Melting point of water}

\begin{figure}
    \centering
    \includegraphics[width=\columnwidth]{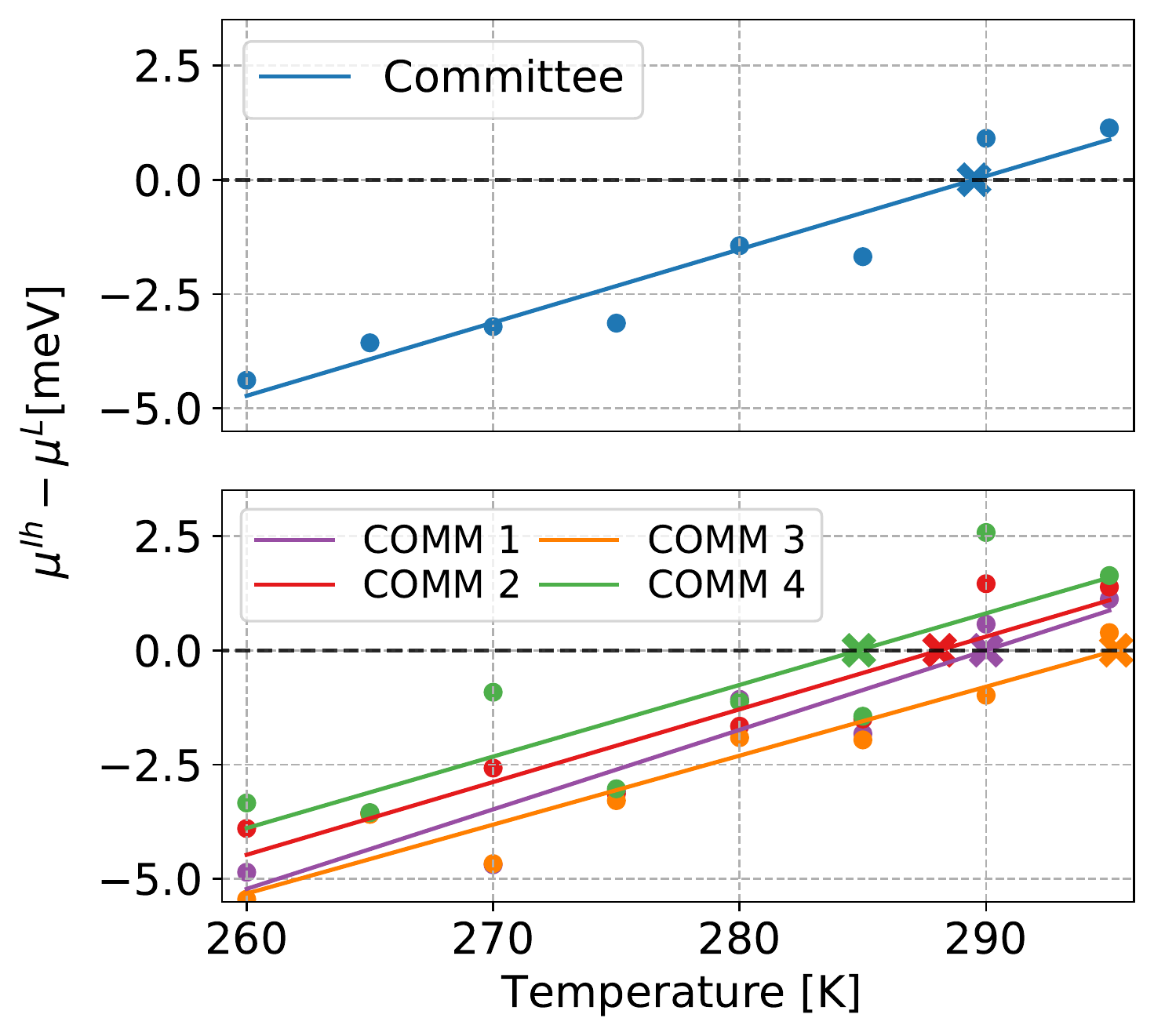}
    \caption{Chemical potential difference between hexagonal ice and liquid water as a function of temperature. Upper panel: the fit obtained for the trajectory driven by the committee mean.
    Lower panel: individual fits for each committee model. }
    \label{fig:melting_point}
\end{figure}

We begin by demonstrating the calculation of the free energy difference between hexagonal ice and liquid water, $\Delta\mu = \mu^{Ih} - \mu^{L}$, at 8 different temperatures,
using the interface pinning (IP) technique.\cite{Pedersen_Hummel_Kresse_Kahl_Dellago_2013} 
The basic idea of IP involves performing a biased simulation in which the system is forced to retain a solid-liquid interface whose position fluctuates around an average value. This is practically achieved by including an additional pinning potential
\begin{equation}
    W(A) = \frac{\kappa}{2}[Q(A) - a]^2  \label{eq:U_pinning}
\end{equation}
where $Q(A)$ is an order parameter which identifies the phase of the system (local Q6, defined as in Ref.\cite{lechner2008accurate,stei+83prb}), $\kappa$ is a spring constant dictating the amplitude of interface fluctuations, and $a$ is the reference value for the collective variable (usually taken as the value of $Q$ at which half the system is in the solid phase). 
The chemical potential difference at the simulation temperature $T$ can then be estimated by
\begin{equation}
    \Delta\mu (T) = -\kappa (\langle Q\rangle' - a) \label{eq:Deltamu_pin}
\end{equation}
where $\langle \cdot \rangle'$ indicates $NP_z\kappa T$-ensemble averages with the additional term $W$ defined in Eq. \eqref{eq:U_pinning}. 
In the present work, simulations are driven by the same committee of $M=4$ NNP models discussed in Sec.~\ref{sec:water_gofr}, using PLUMED\cite{PLUMED} to constrain the order parameter to the target value $a= 165$. A total of 336 water molecules are simulated in a supercell with an elongated side to allow probing the coexistence of the two phases, separated by the planar interface; in particular we employed an orthorhombic supercell of size $15.93 \times 13.79 \times 52.47~\text{\AA}^3$.

We compute the value of $\Delta\mu (T)$ at different temperatures, and perform a linear fit from which we determine the melting temperature $T_\text{m}$ as the intercept with the abscissa, $\Delta\mu(T_\text{m}) = 0$. We also obtain the entropy of melting per molecule, $\Delta s_\text{m} = \left.\frac{\partial \Delta \mu}{\partial T}\right|_{T_\text{m}}$, as the slope of the fit, and the latent heat of melting per molecule $\Delta h_\text{m} = T_\text{m} \Delta s_\text{m} $.
As shown in the top panel of Fig.~\ref{fig:melting_point}, even though the individual points are somewhat scattered due to statistical errors, it is possible to determine a clear linear trend resulting in $T_\text{m}=290$ K, $\Delta s_\text{m} = 0.16$ meV/K/molecule, and $\Delta h_\text{m} = 46$ meV/molecule.
It should be noted that these values deviate from those that have been computed with a similarly trained potential\cite{chen+19pnas} due to the presence of substantial finite-size effects in the present simulations, which are only meant to demonstrate the application of this uncertainty quantification approach, and not to provide size and sampling-converged values of the averages.

In order to estimate the uncertainty due to the MLPs, we combine Eq. \eqref{eq:Deltamu_pin} with the CEA, to compute the model-dependent chemical potential differences $\Delta\mu^{(i)}$ using 
\begin{equation}
    \langle Q\rangle'_{V^{(i)}} = \langle Q\rangle_{\bar{V}}' - \beta [ \langle Q (V^{(i)} - \bar{V}) \rangle'_{\bar{V}} - \langle Q \rangle'_{\bar{V}} \langle V^{(i)} - \bar{V} \rangle'_{\bar{V}}]. 
\end{equation}
In line with the uncertainty propagation framework developed in Sec. \ref{sec:theory}, we compute four different fits, one for each model, and from them four different melting temperatures $T_\text{m}^{(i)}$, indicated by the coloured crosses in the lower panel of Fig.~\ref{fig:melting_point}. By taking the average and standard deviation of the model-dependent $T_\text{m}^{(i)}$, as well as the associated $\Delta s^{(i)}_\text{m}$ and $\Delta h^{(i)}_\text{m}$, we can determine the mean values and the ML uncertainty intervals, namely $\overline{T_\text{m}} =290 \pm 5$ K, $\overline{\Delta s_\text{m}} = 0.16 \pm 0.01$ meV/K/molecule, and $\overline{\Delta h_\text{m}} = 46 \pm 3$ meV/molecule. 
In view of the linear nature of the CEA, the values of the molar entropy and latent heat of melting computed from the mean of the committee estimates match exactly those computed directly from the committee estimates. In principle, the two estimates $\overline{T_\text{m}}$ and $T_\text{m}$ differ, even if in this case they are equal within the confidence interval. Whenever a non-linear procedure is involved in the calculation of the property of interest, results may change based on the way the committee estimates are combined. Comparing different approaches is then a useful check to assess the robustness of the error estimation. 

\newcommand{\CNO}{\text{s}^{\ce{O}}}
\subsubsection{Deprotonation of methanesulfonic acid}

We use the committee model discussed in Sec.~\ref{sec:ch3so2oh-gr} and the same metadynamics protocol described in Ref.~\onlinecite{ross+20jctc} to compute the free energy profile for the deprotonation of \ce{CH3SO2OH} in phenol, a key quantity to rationalize the activity of methanesulfonic acid in catalyzing the hydroxylation of phenol.
We define the free energy as a function of the coordination, $\CNO$, of the oxygen atoms in the acid with respect to the hydrogen atoms in the system.
The free energy at $\CNO$ is by definition $kT$ times the negative of the logarithm of the population fraction $p(\CNO)$ of the configurations with a given $\CNO$.

To obtain an unbiased estimate of $p(\CNO)$ from a trajectory with a time-dependent bias $\tilde{v}(t)$, we weight the configurations by $u(A(t)) = e^{\beta(\tilde{v}(t)-c(t))}$, where the time-dependent offset $c(t)$ is computed using the Iterative Trajectory Reweighting (ITRE) algorithm.\cite{Giberti_Cheng_Tribello_Ceriotti_2020}
The population fraction for the committee, $\bar{p}(\CNO)$, is computed as the ITRE-reweighted normalized histogram of the occurrences of configurations $A$ with a given $\CNO(A)$:
\begin{equation}
\bar{p}(s) =  \langle \delta(\CNO(A)-s) u(A) \rangle_{\bar{V}}
\end{equation} 
where the average is over the metadynamics trajectory, and $\delta(\CNO(A)-s)$ selects structures with a prescribed value of the coordination number.
In turn, the model-dependent population can be readily obtained, through the CEA, as
\begin{equation}
\begin{split}
 p^{(i)}(s)  &=  \bar{p}(s) - \Delta p^{(i)}(s)\\
 \Delta p^{(i)}(s) &= \beta \langle
~ \delta(\CNO(A)-s) ~ u(A) ~ ( V^{(i)}(A) - \bar{V}(A) ) ~ \rangle_{\bar{V}}\\ & -\beta \langle
~ \delta(\CNO(A)-s) ~ u(A) ~ \rangle_{\bar{V}} \langle V^{(i)}(A) - \bar{V}(A) ~ \rangle_{\bar{V}} 
\end{split}
\end{equation}
Finally, the uncertainty in the population, $\Delta p$, is obtained as the standard deviation of $\Delta p^{(i)}$ over the $M$ models, as in Eq.~\eqref{eq:sigma_V}.
The symmetric uncertainty on the population results in a confidence range on the free energy which is \textit{asymmetric} about $-kT\log(\bar{p})$, spanning values from $-kT\log( \bar{p} + \Delta{p} ) $ to $-kT\log(\bar{p} - \Delta{p}) $.

\begin{figure}
\includegraphics[width=0.95\linewidth]{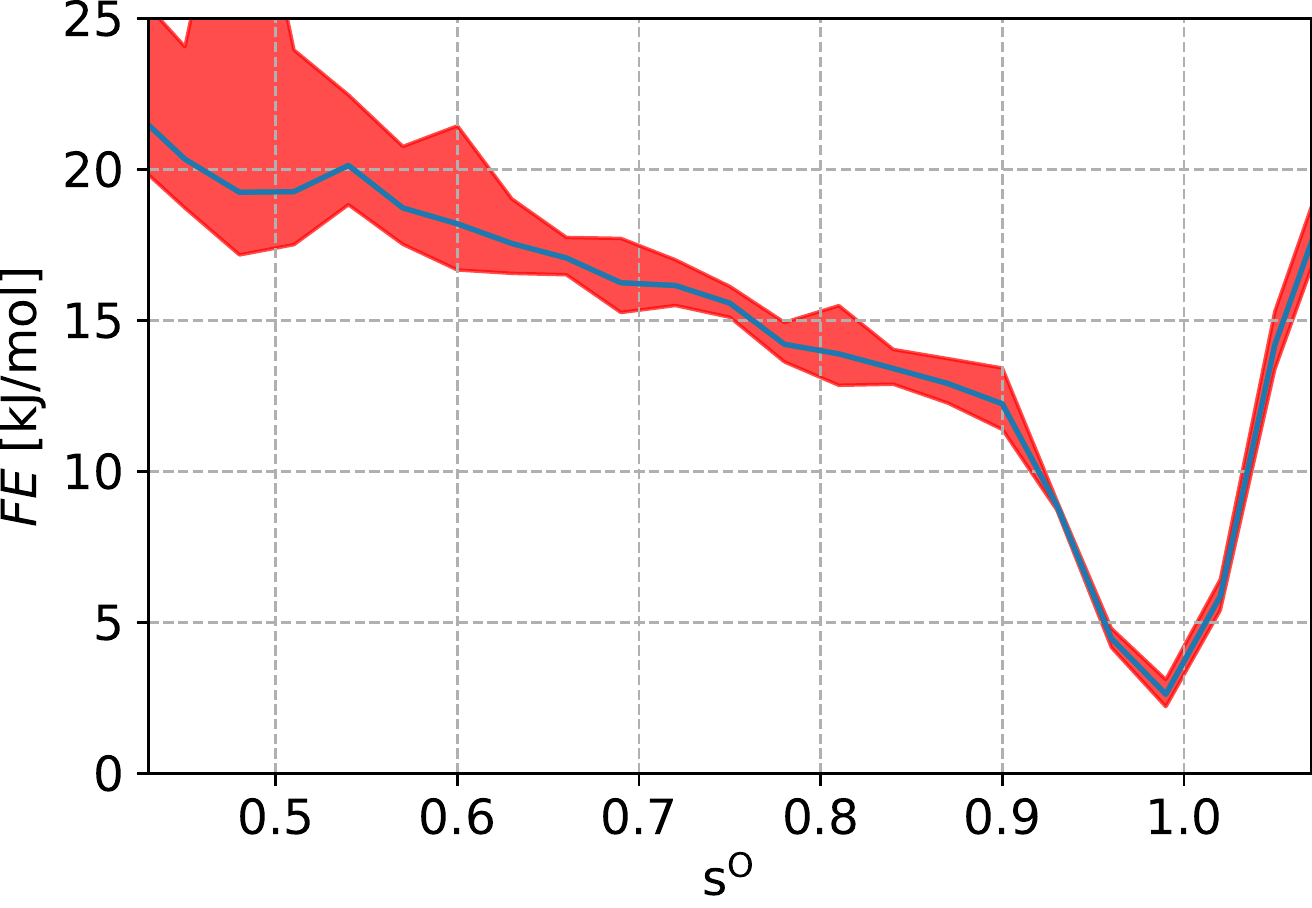}
\caption{Projection of the free energy along the proton transfer reaction $\CNO$ for a system of one methanesulfonic acid molecule dissolved in 20 phenol molecules. $\CNO \approx 1$ indicates the neutral state, while $\CNO < 1$ a deprotonated state of the acid. The shaded area represents the ML uncertainty obtained from Eq.~\eqref{eq:fullsigma2_CEA}.}\label{fig:methanesulfonic_phenol_pKa}
\end{figure}

As shown in Fig. \ref{fig:methanesulfonic_phenol_pKa}, the uncertainty between the models is very small around the minimum corresponding to the neutral state of the acid, but grows substantially in the deprotonated state -- which is consistent with the qualitative observation made in Ref.~\citenum{ross+20jctc} of the increase in the uncertainty on the NNP predictions for dissociated configurations, that are less represented in the training set. %
Interpreting the configurations with $\CNO\approx 0.5$ as the deprotonated state, the free energy cost for the dissociation of \ce{CH3SO2OH}  in phenol can be estimated to be 20{\raisebox{0.5ex}{\tiny$^{+5}_{-2}$}} kJ/mol. Even though in this specific instance other errors, e.g. those due to finite-size effects and reference energetics, are likely to be comparable with that obtained from the spread of the committee members, the substantial uncertainty computed by on-the-flight reweighting underscores the importance of error estimation when using machine learning models.

\subsection{Finite-temperature density of states}\label{sec:FT-DOS}

As a last example, that we use to highlight the interplay between sampling and model uncertainties, we consider the finite-temperature density of states (DOS) of gallium in its metallic liquid phase. 
The sampling of configurations is performed through MD simulations driven by a committee of $M=4$ NNPs, based on the potential introduced in Ref.~\citenum{zama+20am}, that is available from Ref.~\citenum{matcloud20a}. 
We consider a system of 384 Ga atoms in the NVT ensemble, sampled at a temperature $T=1800$ K using a combination of a generalized Langevin\cite{ceri+10jctc} and stochastic velocity rescaling thermostats,\cite{bussi2007canonical} as implemented in i-PI. We employ a timestep of 4 fs to integrate the equations of motion for a total of 400 ps. 
The DOS model is based on the framework developed in Ref.~\citenum{chiheb2020}, which we briefly summarise. For a given configuration $A$, the DOS is defined as
\begin{equation}
    \text{DOS}(E, A) = \frac{2}{N_b N_\mathbf{k}} \sum_n \sum_{\mathbf{k}} \delta(E- E_n(\mathbf{k}, A)) ,
\end{equation}
where $E_n(\mathbf{k})$ is the energy for the (doubly-degenerate) $n$-th band and wavevector $\mathbf{k}$. The DOS is normalized to the number of electronic states, $N_b N_\mathbf{k}$, where $N_b$ and $N_\mathbf{k}$ are the number of bands and $\mathbf{k}$-points considered, respectively. 
We adopt a ML approach based on a local-environments decomposition to predict $\text{DOS}(E,A)$, and train a committee of observable models (OM, see Sec. \ref{sec:theory}), in order to estimate a ML uncertainty. The predicted DOS of a given structure $A$, and the $j$-th model reads
\begin{equation}
    \text{DOS}^{(j)}(E, A) = \sum_{k\in A} \text{LDOS}^{(j)}(E, A_k) , \quad j=1, \ldots, M'.
\end{equation}
The training set for each OM is represented by \rev{150} random structures extracted from a total of \rev{274 Ga training configurations, including mostly liquid structures at various temperatures and pressures and a few solid ones}. For this training set, we compute reference DFT calculations for the $\text{DOS}_\text{ref}(E,A)$ as the convolution of the Kohn-Sham eigenvalues $E_{\text{ref},n}(\mathbf{k})$ with a Gaussian smearing of width 0.5 eV.\cite{chiheb2020} 
The reference DFT calculations are performed at the level of the PBE functional~\cite{perd+96prl} via the \textsc{Quantum ESPRESSO} code,\cite{quantum-espresso-1,quantum-espresso-2} with a Monkhorst-Pack $k$-point grid that ensures a density of at least 6.5 $\mathbf{k}$-points \AA.
In order to compare DOS belonging to the different structures of the training set, we align the DOS at the Fermi level. The latter, $E_F(A, T)$, is defined as the solution of the charge-neutrality constraint $N_e = \sum_E f(E,E_F,T) \text{DOS}(E, A) $, where $f(E,E_F,T)$ is the Fermi-Dirac distribution and $N_e=2$ due to spin degeneracy. 
The featurization is done using a SOAP kernel with $n=12,~l=9,~g_s=0.5,~r_c=6\text{\AA},~c=1,~m=5,~r_0=6.0$ (the parameters follow the notation in Ref.~\citenum{chiheb2020}). Given the small train set size, and that commitee predictions for sparse kernel models add negligible overhead on top of a single prediction, we use a large committee with $M'=64$ members.
According to Eqs.~\eqref{eq:fullsigma2_CEA}, \eqref{eq:sigma_a}, and \eqref{eq:sigma_V}, the total ML uncertainty $\sigma$ on \avdos{} derives from both the uncertainty on individual DOS predictions, $\sigma_a$ and the uncertainty on the phase space sampling associated with the committee of MLPs driving the dynamics, $\sigma_{aV}$.

\begin{figure}
    \centering
    \includegraphics[width=\columnwidth]{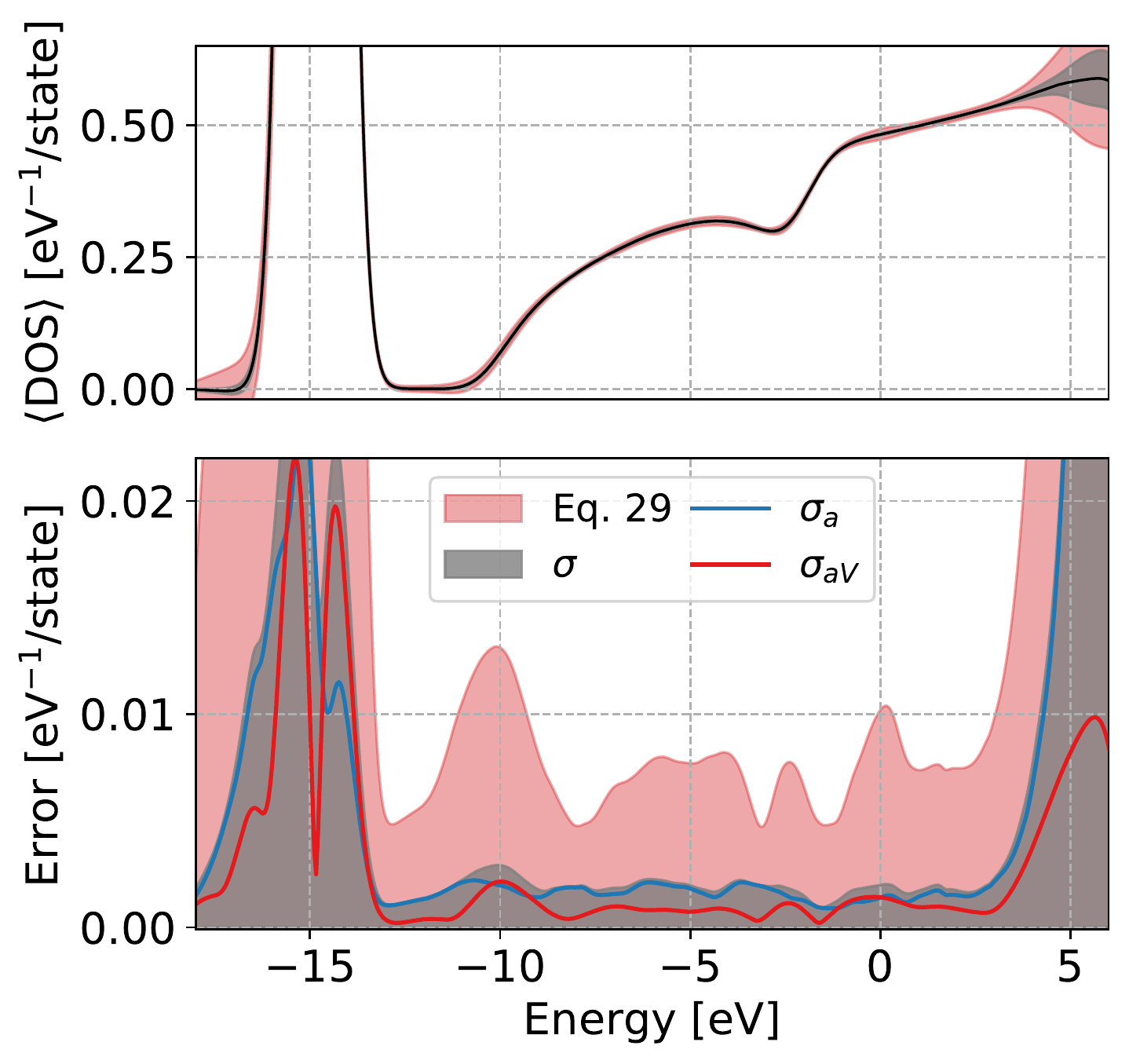}
    \caption{Machine-learned average density of states, \avdos, computed for a simulation of liquid gallium at $T=1800$ K. The zero is set at the Fermi energy, to which the single-configuration DOS entering the average were align. The average \avdos (solid line) is reported together with its statistical uncertainty (shaded \rev{gray} area).\rev{The red shaded area represents the upper bound of the uncertainty, computed as in Eq.~\eqref{eq:sigma_gen}}}
    \label{fig:DOS}
\end{figure}

The results of these calculations are displayed in Fig. \ref{fig:DOS}: in the upper panel the average \avdos{} is reported together with its total ML uncertainty, $\sigma$, as computed by in Eq. \eqref{eq:fullsigma2_CEA}. 
In the lower panel we show the individual contributions of the uncertainty on the property, $\sigma_a$, and that associated with sampling, $\sigma_{aV}$, to the total $\sigma$\rev{, together with the upper bound estimate of the uncertainty}. 
The absolute error on the DOS is small, and is dominated by $\sigma_a$.
The contribution $\sigma_{aV}$ associated with sampling is sizeable, and in some energy range it dominates the uncertainty. The coupling between the potential energy and the observable property cannot be neglected. %
\rev{Notice that the upper bound given by Eq.~\ref{eq:sigma_gen} (shaded red area) largely overestimates the uncertainty based on the committee model, where we have access to the single-configuration deviations with respect to the best (i.e. the committee) values for $\mathrm{DOS}(E,A)$ and $V(A)$, and not only to estimates of their absolute values.}

\section{Conclusions}

\rev{This work demonstrates how to use the uncertainty estimation of the machine-learning prediction of individual atomistic structures in the context of molecular dynamics simulations. 
We focus in particular on a recently-introduced calibrated committee model to account for the uncertainty stemming from the finite number of reference structures employed in the training of the model, for which the uncertainty propagation procedure is particularly natural, but our approaches can be readily adapted to any error estimation scheme.} 
First we develop an uncertainty-weighted baselined ML potential scheme, that achieves robust sampling by using the error estimator to  
interpolate smoothly between a reference baseline potential (e.g., a semiempirical electronic structure method, such as DFTB) %
and a ML correction that promotes it to a higher level of theory.
Whenever the ML correction enters an extrapolative regime, the potential reverts to the less accurate but more robust reference, guaranteeing stable trajectories, and simplifying greatly the practical implementation of ML potentials for all cases in which a reasonable baseline potential is available. 
This scheme has an obvious, straightforward application to online/offline active learning. 
Even though we demonstrate this strategy for the overall potential, a local implementation in terms of individual atomic contributions is possible even when the baseline does not offer a natural atom-centered decomposition. 
This would be particularly beneficial whenever the ML error is not equally distributed among the atoms of the system, affecting instead a rather small number of atoms, e.g., those involved into a chemical reaction. 

We also show how to obtain a quantitative estimate of the machine learning uncertainty of static thermodynamic averages of physical observables, by taking into account both the ML uncertainty on single-configuration calculations, and the distortion of the sampling probability due to the Boltzmann factor entering canonical averages.
We circumvent the poor computational efficiency of statistical reweighing using a cumulant expansion approximation (CEA), consisting in a linearized version of the reweighed average which proves to be statistically more stable.\cite{ceri+12prsa} 
This ML uncertainty propagation scheme proves to be applicable to several physical observables and condensed-phase systems, ranging from the pair distribution function and free-energy calculations on liquid water and methanesulfonic acid in phenol, to the finite-temperature electronic density of states of liquid gallium.

Depending on the application and the target property, the uncertainty that can be ascribed to the ML-driven sampling, or to the ML property models, can be substantial. 
A scheme such as the one we propose here, that allows to achieve uncertainty quantification with an affordable computational cost, should be applied across the board to all simulation based on data-driven schemes. 
\section{Data availability}
Data supporting the findings in this paper are available from public repositories as referenced, or upon reasonable request to the authors. \rev{An open-source implementation of the methods discussed in the present paper is available in i-PI\cite{ipicode}.}

\section{Acknowledgements}
We thank F\'elix Musil for insightful discussions and Chiheb Ben Mahmoud for technical assistance on machine learning the electronic density of states. Training data for the oligopeptides model was kindly provided by Alberto Fabrizio, Raimon Fabregat and Clemence Corminboeuf. 
GI, MC, VK and EAE acknowledge support by the NCCR MARVEL, funded by the Swiss National Science Foundation (SNSF).
FG and MC acknowledge funding by the Swiss National Science Foundation (Project No. 200021-182057). YZ acknowledges support for a research visit at EPFL by the Graduate School of Xiamen University.
KR was supported by an industrial grant with Solvay. 

\appendix
\section{Unbiased estimation of the calibration constant}\label{app:bias}

For a finite value of the number of models $M$, Eq.~\eqref{eq:alpha} is a biased estimator of the true scaling factor $\alpha$. To see this, and to derive a correction for this bias, let us consider the test value $y_n$ as extracted from a normal distribution whose true standard deviation is scaled by a factor $\alpha_{tr}$ with respect to the distribution of the committee models $y^{(i)}(\CX)$ (we assume the same distribution $\forall i$). 
Nonetheless, by assuming that the reference value $y_n$ and the committee predictions $y^{(i)}(\CX)$ are uncorrelated, and that the mean values of $y_{\text{ref}}(\CX)$ and $y^{(i)}(\CX)$ coincide (without loss of generality, we can set them equal to zero), we can write
\begin{equation}
\begin{split}
\alpha_{M}^2 = \lim_{N_{val}\to\infty}
\frac{1}{N_\mathrm{val}} \sum_{\CX \in \mathrm{val}} 
\frac{( y_{\text{ref}}(\CX) - \bar{y}(\CX) )^2}{\sigma^2(\CX)} = \frac{\alpha^2_{tr}}{s_M^2} + b_M^2, \label{eq:alpha_ob}
\end{split}
\end{equation}
where
\begin{equation}
\frac{1}{s_M^2} \equiv \mathbb{E} \left[ {\sigma^2} \right] \mathbb{E} \left[ \frac{1}{\sigma^2} \right]
\end{equation}
and
\begin{equation}
b_M^2 = \mathbb{E} \left[ \frac{\bar{y}^2}{\sigma^2} \right] = \mathbb{E} \left[ \bar{y}^2 \right] 
\mathbb{E} \left[\frac{1}{\sigma^2} \right] = \frac{1}{M s_M^2}, \label{eq:b2}
\end{equation}
where the factorization in the second step is justified by the independence of the sample mean and the sample variance, an application of Basu's theorem \cite{boos1998Basu}.
The expectation values are taken on a $N_{val} \to \infty$ number of validation points, for each of which there is one $\sigma$ and one $\bar{y}$, obtained by averaging over the $M$ members of the committee. 
The values $s_M$ and $b_M$ only depend on $M$ and are independent of the standard deviation 
of the original normal distribution. 
We therefore computed them for a standard normal distribution, by noticing that the $M-$sample variance is distributed, up to a factor $M-1$, according to a chi-square distribution of $M-1$ degrees of freedom, i.e. $(M-1) \varsigma^2 \sim \chi^2_{M-1}$.
Therefore 
\begin{equation}
    \mathbb{E}\left[\frac{1}{\varsigma^2}\right] = 
    \int_0^\infty \frac{M-1}{x} \frac{x^{\frac{M-1}{2} -1} e^{-\frac{x}{2}}}{2^{\frac{M-1}{2}} \Gamma(\frac{M-1}{2}) }  dx = \frac{M-1}{M-3}. \label{eq:E_isig2}
\end{equation}
From this expression it is evident that at least four models are needed to achieve a meaningful estimate of $\alpha$, and that using only 4 models would lead to an overestimation of the renormalization constant by a factor of about 2. 
We checked these results numerically by extracting points from a standard normal distribution and comparing the sampled to the expected $\alpha_M$, obtained from Eq. \eqref{eq:alpha_ob}, and Eqs. \eqref{eq:b2} and \eqref{eq:E_isig2}. 
By inverting Eq. \eqref{eq:alpha_ob}, an unbiased estimate of the calibration constant is readily obtained as
\begin{equation}
\alpha^2 = \frac{M-3}{M-1} \alpha_M^2 - \frac{1}{M},
\label{eq:alpha-unbiased}
\end{equation} 
where $\alpha_M^2$ in Eq. \eqref{eq:alpha}.

\begin{figure}
    \centering
    \includegraphics[width=\columnwidth]{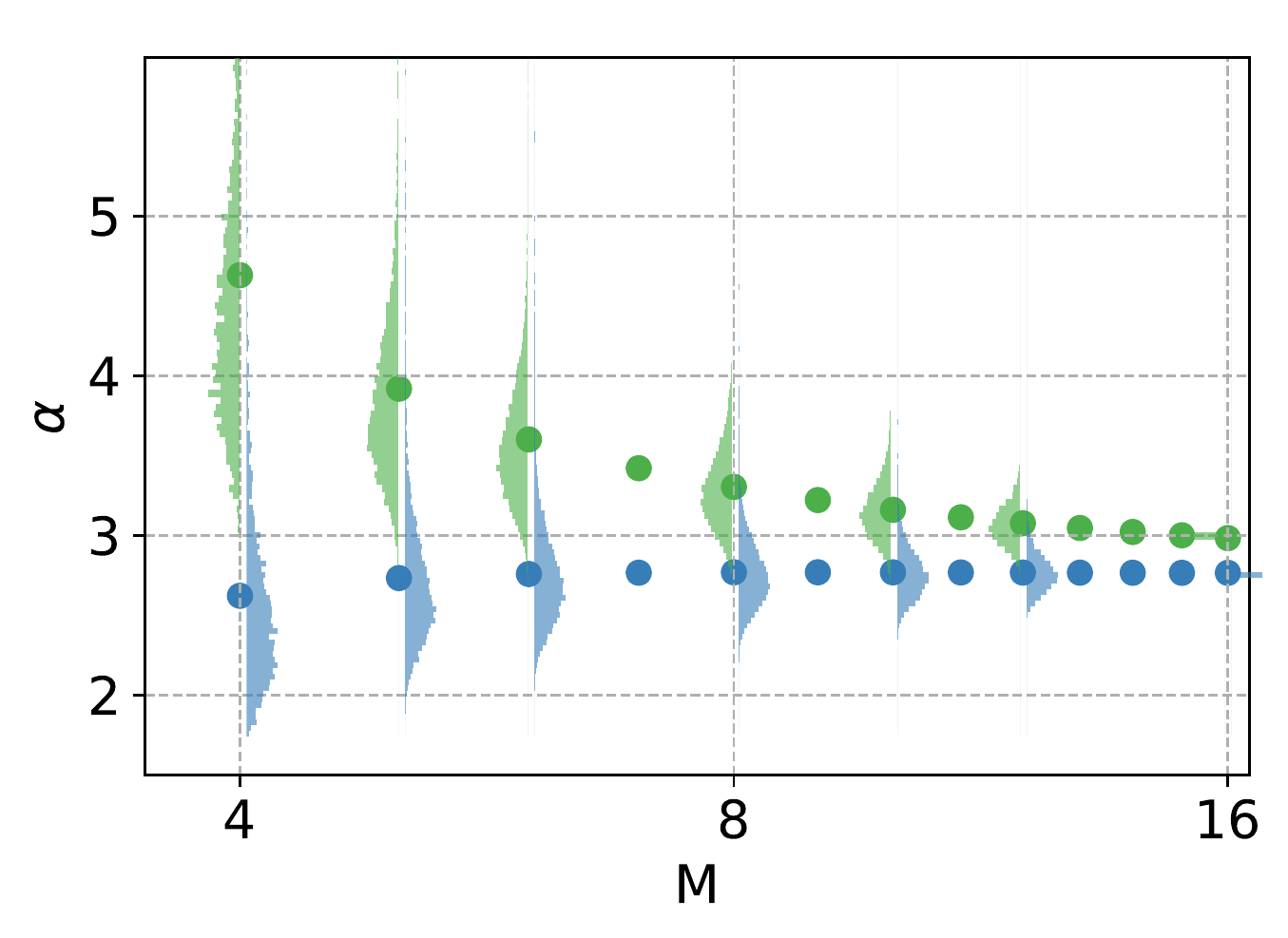}
    \caption{Violin plot of biased (green) and unbiased (blue) estimators for the correction factor $\alpha$, as a function of the number of models in the committee, $M$.}
    \label{fig:biasedVSunbiased}
\end{figure}

Figure~\ref{fig:biasedVSunbiased} displays, for the case of water (see Sec.~\ref{sec:water_gofr}), the violin plot of the biased (green) and unbiased (blue) distributions for the calibration constant $\alpha$, as a function of the number of models in the committee, $M$. The sample distributions were generated by sampling $M$ model out of a fix maximum number of trained models, $M_{\max} = 16$ in all the $\binom{M_{\max}}{M}$ possible ways for a given $M$. 
The bullets indicate the sample means. As expected, for small $M$, the biased estimator largely overestimates the unbiased value, slowly approaching the asymptotic value for $M\rightarrow \infty$.
The estimator in Eq.~\eqref{eq:alpha-unbiased}, instead, is already within 10\%{} of the asymptotic value for $M=4$. 

Two important remarks should be made: first, Eq.~\eqref{eq:alpha-unbiased} is only unbiased when the distribution of the predictions is Gaussian, which is usually only approximately true. 
Second, this correction yields a calibration factor that would provide unbiased estimates of the uncertainty in the committee predictions, but that when using the committee members in a non-linear combination to propagate uncertainty, similar biases may appear, and it is recommended, whenever possible, to increase $M$ to at least 6 members.

\rev{
\section{Uncertainty propagation on canonical averages}\label{app:unc-prop}
We aim at estimating the uncertainty $\sigma_{\langle a \rangle}$ on the canonical average  over a space of generalised coordinates $x$
\begin{equation}
    \langle a \rangle = \frac{\int a(x) e^{-\beta V(x)} dx}{\int e^{-\beta V(x)} dx},
\end{equation}
of the observable $a(x)$ under the potential energy $V(x)$ whenever we can estimate the uncertainty $\sigma_a(x)$ on $a(x)$ and $\sigma_V(x)$ on $V(x)$.
We perform a functional-derivative Taylor expansion
\begin{equation}
    \delta \langle a \rangle = \int dy \left[ \frac{\delta \langle a \rangle}{\delta a(y)} \delta a(y) + \frac{\delta \langle a \rangle}{\delta V(y)} \delta V(y) \right] .
\end{equation}
Since
\begin{align}
    &\frac{\delta \langle a \rangle}{\delta a(y)} = \frac{e^{-\beta V(y)}}{\int e^{-\beta V(x)} dx} \equiv P(y) \\
    & \frac{1}{Z} \frac{\delta \int a(x) e^{-\beta V(x)} dx}{\delta V(y)} =  -\beta a(y) P(y) \\
    &\frac{\delta [\int e^{-\beta V(x)} dx]^{-1}}{\delta V(y)} = \beta Z^{-1} P(y),
\end{align}
where $Z = \int e^{-\beta V(x)} dx$, we obtain
\begin{equation}
\begin{split}
\delta{\langle a \rangle} &= \int P(y) \delta a(y) \, dy \\
&+ \beta \int \left[ \langle a \rangle - a(y) \right] P(y) \delta V(y) \, dy.
\end{split}
\end{equation}
We have, from standard error-propagation theory,  
\begin{equation}
\begin{split}
\sigma^2_{\langle a \rangle} &= \iint P(x) P(y) \{ \rho_{aa}(x,y) \sigma_{a}(x) \sigma_{a}(y) \\
&+ \beta^2 \left[ \langle a \rangle - a(x) \right] \left[\langle a \rangle -a(y) \right] \, \rho_{VV}(x,y) \, \sigma_{V}(x)\sigma_{V}(y)\\
&+ 2 \beta \, [\langle a \rangle - a(y) ]\, \rho_{aV}(x,y) \, \sigma_{a}(x)\sigma_{V}(y)\} \, dx \,dy 
\end{split}
\end{equation}
where $\rho_{\mu\nu}(x,y)$ is the correlation coefficient. Since $|\rho_{\mu\nu}(x,y)|\leq 1$, and by taking the modulus of non positive-defined quantities, we have the following inequality:
\begin{equation}
\begin{split}
\sigma^2_{\langle a \rangle} &\leq \left(\int P(x)\, \sigma_{a}(x) \, dx \right)^2 + \\
&\beta^2 \left(\int |\langle a \rangle -a(x) | \, P(x)  \, \sigma_{V}(x) \right)^2 \\
&+ 2 \beta \int P(x) \, \sigma_{a}(x)\, dx  \int  |\langle a \rangle - a(y) | \, P(y) \, \sigma_{V}(y) \, dy  \\
&=  \left(\langle \sigma_a \rangle + \beta\left\langle |\langle a \rangle - a |\, \sigma_V \right\rangle \right)^2,
\end{split}
\end{equation}
from which Eq.~\eqref{eq:sigma_gen} follows. 
The last step is exactly what we would get from a functional generalization of the mean absolute error. 
}

%

%aipnum4-2.bst 2019-01-14 (MD) hand-edited version of apsrev4-1.bst
%Control: key (0)
%Control: author (8) initials jnrlst
%Control: editor formatted (1) identically to author
%Control: production of article title (0) allowed
%Control: page (1) range
%Control: year (1) truncated
%Control: production of eprint (0) enabled
%


\begin{thebibliography}{85}%
\makeatletter
\providecommand \@ifxundefined [1]{%
 \@ifx{#1\undefined}
}%
\providecommand \@ifnum [1]{%
 \ifnum #1\expandafter \@firstoftwo
 \else \expandafter \@secondoftwo
 \fi
}%
\providecommand \@ifx [1]{%
 \ifx #1\expandafter \@firstoftwo
 \else \expandafter \@secondoftwo
 \fi
}%
\providecommand \natexlab [1]{#1}%
\providecommand \enquote  [1]{``#1''}%
\providecommand \bibnamefont  [1]{#1}%
\providecommand \bibfnamefont [1]{#1}%
\providecommand \citenamefont [1]{#1}%
\providecommand \href@noop [0]{\@secondoftwo}%
\providecommand \href [0]{\begingroup \@sanitize@url \@href}%
\providecommand \@href[1]{\@@startlink{#1}\@@href}%
\providecommand \@@href[1]{\endgroup#1\@@endlink}%
\providecommand \@sanitize@url [0]{\catcode `\\12\catcode `\$12\catcode
  `\&12\catcode `\#12\catcode `\^12\catcode `\_12\catcode `\%12\relax}%
\providecommand \@@startlink[1]{}%
\providecommand \@@endlink[0]{}%
\providecommand \url  [0]{\begingroup\@sanitize@url \@url }%
\providecommand \@url [1]{\endgroup\@href {#1}{\urlprefix }}%
\providecommand \urlprefix  [0]{URL }%
\providecommand \Eprint [0]{\href }%
\providecommand \doibase [0]{https://doi.org/}%
\providecommand \selectlanguage [0]{\@gobble}%
\providecommand \bibinfo  [0]{\@secondoftwo}%
\providecommand \bibfield  [0]{\@secondoftwo}%
\providecommand \translation [1]{[#1]}%
\providecommand \BibitemOpen [0]{}%
\providecommand \bibitemStop [0]{}%
\providecommand \bibitemNoStop [0]{.\EOS\space}%
\providecommand \EOS [0]{\spacefactor3000\relax}%
\providecommand \BibitemShut  [1]{\csname bibitem#1\endcsname}%
\let\auto@bib@innerbib\@empty
%</preamble>
\bibitem [{\citenamefont {Behler}\ and\ \citenamefont
  {Parrinello}(2007)}]{behl-parr07prl}%
  \BibitemOpen
  \bibfield  {author} {\bibinfo {author} {\bibfnamefont {J.}~\bibnamefont
  {Behler}}\ and\ \bibinfo {author} {\bibfnamefont {M.}~\bibnamefont
  {Parrinello}},\ }\bibfield  {title} {\enquote {\bibinfo {title} {Generalized
  {{Neural}}-{{Network Representation}} of {{High}}-{{Dimensional
  Potential}}-{{Energy Surfaces}}},}\ }\href
  {https://doi.org/10.1103/PhysRevLett.98.146401} {\bibfield  {journal}
  {\bibinfo  {journal} {Phys. Rev. Lett.}\ }\textbf {\bibinfo {volume} {98}},\
  \bibinfo {pages} {146401} (\bibinfo {year} {2007})}\BibitemShut {NoStop}%
\bibitem [{\citenamefont {Bart{\'o}k}\ \emph {et~al.}(2010)\citenamefont
  {Bart{\'o}k}, \citenamefont {Payne}, \citenamefont {Kondor},\ and\
  \citenamefont {Cs{\'a}nyi}}]{bart+10prl}%
  \BibitemOpen
  \bibfield  {author} {\bibinfo {author} {\bibfnamefont {A.~P.}\ \bibnamefont
  {Bart{\'o}k}}, \bibinfo {author} {\bibfnamefont {M.~C.}\ \bibnamefont
  {Payne}}, \bibinfo {author} {\bibfnamefont {R.}~\bibnamefont {Kondor}},\ and\
  \bibinfo {author} {\bibfnamefont {G.}~\bibnamefont {Cs{\'a}nyi}},\ }\bibfield
   {title} {\enquote {\bibinfo {title} {Gaussian {{Approximation Potentials}}:
  {{The Accuracy}} of {{Quantum Mechanics}}, without the {{Electrons}}},}\
  }\href {https://doi.org/10.1103/PhysRevLett.104.136403} {\bibfield  {journal}
  {\bibinfo  {journal} {Phys. Rev. Lett.}\ }\textbf {\bibinfo {volume} {104}},\
  \bibinfo {pages} {136403} (\bibinfo {year} {2010})}\BibitemShut {NoStop}%
\bibitem [{\citenamefont {Rupp}\ \emph {et~al.}(2012)\citenamefont {Rupp},
  \citenamefont {Tkatchenko}, \citenamefont {M{\"u}ller},\ and\ \citenamefont
  {{von Lilienfeld}}}]{rupp+12prl}%
  \BibitemOpen
  \bibfield  {author} {\bibinfo {author} {\bibfnamefont {M.}~\bibnamefont
  {Rupp}}, \bibinfo {author} {\bibfnamefont {A.}~\bibnamefont {Tkatchenko}},
  \bibinfo {author} {\bibfnamefont {K.-R.}\ \bibnamefont {M{\"u}ller}},\ and\
  \bibinfo {author} {\bibfnamefont {O.~A.}\ \bibnamefont {{von Lilienfeld}}},\
  }\bibfield  {title} {\enquote {\bibinfo {title} {Fast and {{Accurate
  Modeling}} of {{Molecular Atomization Energies}} with {{Machine
  Learning}}},}\ }\href {https://doi.org/10.1103/PhysRevLett.108.058301}
  {\bibfield  {journal} {\bibinfo  {journal} {Phys. Rev. Lett.}\ }\textbf
  {\bibinfo {volume} {108}},\ \bibinfo {pages} {058301} (\bibinfo {year}
  {2012})}\BibitemShut {NoStop}%
\bibitem [{\citenamefont {Smith}, \citenamefont {Isayev},\ and\ \citenamefont
  {Roitberg}(2017)}]{Smith_Isayev_Roitberg_2017}%
  \BibitemOpen
  \bibfield  {author} {\bibinfo {author} {\bibfnamefont {J.~S.}\ \bibnamefont
  {Smith}}, \bibinfo {author} {\bibfnamefont {O.}~\bibnamefont {Isayev}},\ and\
  \bibinfo {author} {\bibfnamefont {A.~E.}\ \bibnamefont {Roitberg}},\
  }\bibfield  {title} {\enquote {\bibinfo {title} {Ani-1: an extensible neural
  network potential with dft accuracy at force field computational cost},}\
  }\href {https://doi.org/10.1039/C6SC05720A} {\bibfield  {journal} {\bibinfo
  {journal} {Chemical Science}\ }\textbf {\bibinfo {volume} {8}},\ \bibinfo
  {pages} {3192–3203} (\bibinfo {year} {2017})}\BibitemShut {NoStop}%
\bibitem [{\citenamefont {Devereux}\ \emph {et~al.}(2020)\citenamefont
  {Devereux}, \citenamefont {Smith}, \citenamefont {Davis}, \citenamefont
  {Barros}, \citenamefont {Zubatyuk}, \citenamefont {Isayev},\ and\
  \citenamefont {Roitberg}}]{Devereux2020}%
  \BibitemOpen
  \bibfield  {author} {\bibinfo {author} {\bibfnamefont {C.}~\bibnamefont
  {Devereux}}, \bibinfo {author} {\bibfnamefont {J.~S.}\ \bibnamefont {Smith}},
  \bibinfo {author} {\bibfnamefont {K.~K.}\ \bibnamefont {Davis}}, \bibinfo
  {author} {\bibfnamefont {K.}~\bibnamefont {Barros}}, \bibinfo {author}
  {\bibfnamefont {R.}~\bibnamefont {Zubatyuk}}, \bibinfo {author}
  {\bibfnamefont {O.}~\bibnamefont {Isayev}},\ and\ \bibinfo {author}
  {\bibfnamefont {A.~E.}\ \bibnamefont {Roitberg}},\ }\bibfield  {title}
  {\enquote {\bibinfo {title} {{Extending the Applicability of the ANI Deep
  Learning Molecular Potential to Sulfur and Halogens}},}\ }\href@noop {}
  {\bibfield  {journal} {\bibinfo  {journal} {Journal of Chemical Theory and
  Computation}\ }\textbf {\bibinfo {volume} {16}} (\bibinfo {year}
  {2020})}\BibitemShut {NoStop}%
\bibitem [{\citenamefont {Chmiela}\ \emph {et~al.}(2018)\citenamefont
  {Chmiela}, \citenamefont {Sauceda}, \citenamefont {M{\"u}ller},\ and\
  \citenamefont {Tkatchenko}}]{chmi+18nc}%
  \BibitemOpen
  \bibfield  {author} {\bibinfo {author} {\bibfnamefont {S.}~\bibnamefont
  {Chmiela}}, \bibinfo {author} {\bibfnamefont {H.~E.}\ \bibnamefont
  {Sauceda}}, \bibinfo {author} {\bibfnamefont {K.-R.}\ \bibnamefont
  {M{\"u}ller}},\ and\ \bibinfo {author} {\bibfnamefont {A.}~\bibnamefont
  {Tkatchenko}},\ }\bibfield  {title} {\enquote {\bibinfo {title} {Towards
  exact molecular dynamics simulations with machine-learned force fields},}\
  }\href {https://doi.org/10.1038/s41467-018-06169-2} {\bibfield  {journal}
  {\bibinfo  {journal} {Nat Commun}\ }\textbf {\bibinfo {volume} {9}},\
  \bibinfo {pages} {3887} (\bibinfo {year} {2018})}\BibitemShut {NoStop}%
\bibitem [{\citenamefont {Rossi}\ \emph
  {et~al.}(2020{\natexlab{a}})\citenamefont {Rossi}, \citenamefont
  {Jur{\'a}skov{\'a}}, \citenamefont {Wischert}, \citenamefont {Garel},
  \citenamefont {Corminboeuf},\ and\ \citenamefont {Ceriotti}}]{ross+20jctc}%
  \BibitemOpen
  \bibfield  {author} {\bibinfo {author} {\bibfnamefont {K.}~\bibnamefont
  {Rossi}}, \bibinfo {author} {\bibfnamefont {V.}~\bibnamefont
  {Jur{\'a}skov{\'a}}}, \bibinfo {author} {\bibfnamefont {R.}~\bibnamefont
  {Wischert}}, \bibinfo {author} {\bibfnamefont {L.}~\bibnamefont {Garel}},
  \bibinfo {author} {\bibfnamefont {C.}~\bibnamefont {Corminboeuf}},\ and\
  \bibinfo {author} {\bibfnamefont {M.}~\bibnamefont {Ceriotti}},\ }\bibfield
  {title} {\enquote {\bibinfo {title} {Simulating {{Solvation}} and {{Acidity}}
  in {{Complex Mixtures}} with {{First}}-{{Principles Accuracy}}: {{The Case}}
  of {{CH}} {\textsubscript{3}} {{SO}} {\textsubscript{3}} {{H}} and {{H}}
  {\textsubscript{2}} {{O}} {\textsubscript{2}} in {{Phenol}}},}\ }\href
  {https://doi.org/10.1021/acs.jctc.0c00362} {\bibfield  {journal} {\bibinfo
  {journal} {J. Chem. Theory Comput.}\ }\textbf {\bibinfo {volume} {16}},\
  \bibinfo {pages} {5139--5149} (\bibinfo {year}
  {2020}{\natexlab{a}})}\BibitemShut {NoStop}%
\bibitem [{\citenamefont {Behler}(2011)}]{behl11pccp}%
  \BibitemOpen
  \bibfield  {author} {\bibinfo {author} {\bibfnamefont {J.}~\bibnamefont
  {Behler}},\ }\bibfield  {title} {\enquote {\bibinfo {title} {Neural network
  potential-energy surfaces in chemistry: A tool for large-scale
  simulations.}}\ }\href {https://doi.org/10.1039/c1cp21668f} {\bibfield
  {journal} {\bibinfo  {journal} {Phys. Chem. Chem. Phys. PCCP}\ }\textbf
  {\bibinfo {volume} {13}},\ \bibinfo {pages} {17930--55} (\bibinfo {year}
  {2011})}\BibitemShut {NoStop}%
\bibitem [{\citenamefont {Sosso}\ \emph {et~al.}(2012)\citenamefont {Sosso},
  \citenamefont {Miceli}, \citenamefont {Caravati}, \citenamefont {Behler},\
  and\ \citenamefont {Bernasconi}}]{soss+12prb}%
  \BibitemOpen
  \bibfield  {author} {\bibinfo {author} {\bibfnamefont {G.~C.}\ \bibnamefont
  {Sosso}}, \bibinfo {author} {\bibfnamefont {G.}~\bibnamefont {Miceli}},
  \bibinfo {author} {\bibfnamefont {S.}~\bibnamefont {Caravati}}, \bibinfo
  {author} {\bibfnamefont {J.}~\bibnamefont {Behler}},\ and\ \bibinfo {author}
  {\bibfnamefont {M.}~\bibnamefont {Bernasconi}},\ }\bibfield  {title}
  {\enquote {\bibinfo {title} {Neural network interatomic potential for the
  phase change material {{GeTe}}},}\ }\href
  {https://doi.org/10.1103/PhysRevB.85.174103} {\bibfield  {journal} {\bibinfo
  {journal} {Phys. Rev. B}\ }\textbf {\bibinfo {volume} {85}},\ \bibinfo
  {pages} {174103} (\bibinfo {year} {2012})}\BibitemShut {NoStop}%
\bibitem [{\citenamefont {Bernstein}\ \emph {et~al.}(2019)\citenamefont
  {Bernstein}, \citenamefont {Bhattarai}, \citenamefont {Cs{\'{a}}nyi},
  \citenamefont {Drabold}, \citenamefont {Elliott},\ and\ \citenamefont
  {Deringer}}]{Bernstein2019}%
  \BibitemOpen
  \bibfield  {author} {\bibinfo {author} {\bibfnamefont {N.}~\bibnamefont
  {Bernstein}}, \bibinfo {author} {\bibfnamefont {B.}~\bibnamefont
  {Bhattarai}}, \bibinfo {author} {\bibfnamefont {G.}~\bibnamefont
  {Cs{\'{a}}nyi}}, \bibinfo {author} {\bibfnamefont {D.~A.}\ \bibnamefont
  {Drabold}}, \bibinfo {author} {\bibfnamefont {S.~R.}\ \bibnamefont
  {Elliott}},\ and\ \bibinfo {author} {\bibfnamefont {V.~L.}\ \bibnamefont
  {Deringer}},\ }\bibfield  {title} {\enquote {\bibinfo {title} {{Quantifying
  Chemical Structure and Machine-Learned Atomic Energies in Amorphous and
  Liquid Silicon}},}\ }\href@noop {} {\bibfield  {journal} {\bibinfo  {journal}
  {Angewandte Chemie - International Edition}\ }\textbf {\bibinfo {volume}
  {58}} (\bibinfo {year} {2019})}\BibitemShut {NoStop}%
\bibitem [{\citenamefont {Artrith}, \citenamefont {Urban},\ and\ \citenamefont
  {Ceder}(2018)}]{Artrith2018}%
  \BibitemOpen
  \bibfield  {author} {\bibinfo {author} {\bibfnamefont {N.}~\bibnamefont
  {Artrith}}, \bibinfo {author} {\bibfnamefont {A.}~\bibnamefont {Urban}},\
  and\ \bibinfo {author} {\bibfnamefont {G.}~\bibnamefont {Ceder}},\ }\bibfield
   {title} {\enquote {\bibinfo {title} {{Constructing first-principles phase
  diagrams of amorphous LixSi using machine-learning-assisted sampling with an
  evolutionary algorithm}},}\ }\href@noop {} {\bibfield  {journal} {\bibinfo
  {journal} {Journal of Chemical Physics}\ }\textbf {\bibinfo {volume} {148}}
  (\bibinfo {year} {2018})}\BibitemShut {NoStop}%
\bibitem [{\citenamefont {Zamani}\ \emph
  {et~al.}(2020{\natexlab{a}})\citenamefont {Zamani}, \citenamefont
  {Imbalzano}, \citenamefont {Tappy}, \citenamefont {Alexander}, \citenamefont
  {Mart{\'{i}}-S{\'{a}}nchez}, \citenamefont {Ghisalberti}, \citenamefont
  {Ramasse}, \citenamefont {Friedl}, \citenamefont
  {T{\"{u}}t{\"{u}}nc{\"{u}}oglu}, \citenamefont {Francaviglia}, \citenamefont
  {Bienvenue}, \citenamefont {H{\'{e}}bert}, \citenamefont {Arbiol},
  \citenamefont {Ceriotti},\ and\ \citenamefont {{Fontcuberta i
  Morral}}}]{Zamani2020}%
  \BibitemOpen
  \bibfield  {author} {\bibinfo {author} {\bibfnamefont {M.}~\bibnamefont
  {Zamani}}, \bibinfo {author} {\bibfnamefont {G.}~\bibnamefont {Imbalzano}},
  \bibinfo {author} {\bibfnamefont {N.}~\bibnamefont {Tappy}}, \bibinfo
  {author} {\bibfnamefont {D.~T.}\ \bibnamefont {Alexander}}, \bibinfo {author}
  {\bibfnamefont {S.}~\bibnamefont {Mart{\'{i}}-S{\'{a}}nchez}}, \bibinfo
  {author} {\bibfnamefont {L.}~\bibnamefont {Ghisalberti}}, \bibinfo {author}
  {\bibfnamefont {Q.~M.}\ \bibnamefont {Ramasse}}, \bibinfo {author}
  {\bibfnamefont {M.}~\bibnamefont {Friedl}}, \bibinfo {author} {\bibfnamefont
  {G.}~\bibnamefont {T{\"{u}}t{\"{u}}nc{\"{u}}oglu}}, \bibinfo {author}
  {\bibfnamefont {L.}~\bibnamefont {Francaviglia}}, \bibinfo {author}
  {\bibfnamefont {S.}~\bibnamefont {Bienvenue}}, \bibinfo {author}
  {\bibfnamefont {C.}~\bibnamefont {H{\'{e}}bert}}, \bibinfo {author}
  {\bibfnamefont {J.}~\bibnamefont {Arbiol}}, \bibinfo {author} {\bibfnamefont
  {M.}~\bibnamefont {Ceriotti}},\ and\ \bibinfo {author} {\bibfnamefont
  {A.}~\bibnamefont {{Fontcuberta i Morral}}},\ }\bibfield  {title} {\enquote
  {\bibinfo {title} {{3D Ordering at the Liquid–Solid Polar Interface of
  Nanowires}},}\ }\href@noop {} {\bibfield  {journal} {\bibinfo  {journal}
  {Advanced Materials}\ }\textbf {\bibinfo {volume} {32}} (\bibinfo {year}
  {2020}{\natexlab{a}})}\BibitemShut {NoStop}%
\bibitem [{\citenamefont {Cheng}\ \emph {et~al.}(2020)\citenamefont {Cheng},
  \citenamefont {Mazzola}, \citenamefont {Pickard},\ and\ \citenamefont
  {Ceriotti}}]{chen+20nature}%
  \BibitemOpen
  \bibfield  {author} {\bibinfo {author} {\bibfnamefont {B.}~\bibnamefont
  {Cheng}}, \bibinfo {author} {\bibfnamefont {G.}~\bibnamefont {Mazzola}},
  \bibinfo {author} {\bibfnamefont {C.~J.}\ \bibnamefont {Pickard}},\ and\
  \bibinfo {author} {\bibfnamefont {M.}~\bibnamefont {Ceriotti}},\ }\bibfield
  {title} {\enquote {\bibinfo {title} {Evidence for supercritical behaviour of
  high-pressure liquid hydrogen},}\ }\href
  {https://doi.org/10.1038/s41586-020-2677-y} {\bibfield  {journal} {\bibinfo
  {journal} {Nature}\ }\textbf {\bibinfo {volume} {585}},\ \bibinfo {pages}
  {217--220} (\bibinfo {year} {2020})}\BibitemShut {NoStop}%
\bibitem [{\citenamefont {Zeni}\ \emph {et~al.}(2019)\citenamefont {Zeni},
  \citenamefont {Rossi}, \citenamefont {Glielmo},\ and\ \citenamefont
  {Baletto}}]{Zeni2019}%
  \BibitemOpen
  \bibfield  {author} {\bibinfo {author} {\bibfnamefont {C.}~\bibnamefont
  {Zeni}}, \bibinfo {author} {\bibfnamefont {K.}~\bibnamefont {Rossi}},
  \bibinfo {author} {\bibfnamefont {A.}~\bibnamefont {Glielmo}},\ and\ \bibinfo
  {author} {\bibfnamefont {F.}~\bibnamefont {Baletto}},\ }\bibfield  {title}
  {\enquote {\bibinfo {title} {{On machine learning force fields for metallic
  nanoparticles}},}\ }\href@noop {} {\bibfield  {journal} {\bibinfo  {journal}
  {Advances in Physics: X}\ }\textbf {\bibinfo {volume} {4}} (\bibinfo {year}
  {2019})}\BibitemShut {NoStop}%
\bibitem [{\citenamefont {Fronzi}\ \emph {et~al.}(2020)\citenamefont {Fronzi},
  \citenamefont {Tawfik}, \citenamefont {Ghazaleh}, \citenamefont {Isayev},
  \citenamefont {Winkler}, \citenamefont {Shapter},\ and\ \citenamefont
  {Ford}}]{Fronzi2020}%
  \BibitemOpen
  \bibfield  {author} {\bibinfo {author} {\bibfnamefont {M.}~\bibnamefont
  {Fronzi}}, \bibinfo {author} {\bibfnamefont {S.~A.}\ \bibnamefont {Tawfik}},
  \bibinfo {author} {\bibfnamefont {M.~A.}\ \bibnamefont {Ghazaleh}}, \bibinfo
  {author} {\bibfnamefont {O.}~\bibnamefont {Isayev}}, \bibinfo {author}
  {\bibfnamefont {D.~A.}\ \bibnamefont {Winkler}}, \bibinfo {author}
  {\bibfnamefont {J.}~\bibnamefont {Shapter}},\ and\ \bibinfo {author}
  {\bibfnamefont {M.~J.}\ \bibnamefont {Ford}},\ }\bibfield  {title} {\enquote
  {\bibinfo {title} {{High Throughput Screening of Millions of van der Waals
  Heterostructures for Superlubricant Applications}},}\ }\href@noop {}
  {\bibfield  {journal} {\bibinfo  {journal} {Advanced Theory and Simulations}\
  }\textbf {\bibinfo {volume} {3}},\ \bibinfo {pages} {2000029} (\bibinfo
  {year} {2020})}\BibitemShut {NoStop}%
\bibitem [{\citenamefont {Cuny}\ \emph {et~al.}(2016)\citenamefont {Cuny},
  \citenamefont {Xie}, \citenamefont {Pickard},\ and\ \citenamefont
  {Hassanali}}]{cuny+16jctc}%
  \BibitemOpen
  \bibfield  {author} {\bibinfo {author} {\bibfnamefont {J.}~\bibnamefont
  {Cuny}}, \bibinfo {author} {\bibfnamefont {Y.}~\bibnamefont {Xie}}, \bibinfo
  {author} {\bibfnamefont {C.~J.}\ \bibnamefont {Pickard}},\ and\ \bibinfo
  {author} {\bibfnamefont {A.~A.}\ \bibnamefont {Hassanali}},\ }\bibfield
  {title} {\enquote {\bibinfo {title} {Ab {{Initio Quality NMR Parameters}} in
  {{Solid}}-{{State Materials Using}} a {{High}}-{{Dimensional
  Neural}}-{{Network Representation}}},}\ }\href
  {https://doi.org/10.1021/acs.jctc.5b01006} {\bibfield  {journal} {\bibinfo
  {journal} {J. Chem. Theory Comput.}\ }\textbf {\bibinfo {volume} {12}},\
  \bibinfo {pages} {765--773} (\bibinfo {year} {2016})}\BibitemShut {NoStop}%
\bibitem [{\citenamefont {Paruzzo}\ \emph {et~al.}(2018)\citenamefont
  {Paruzzo}, \citenamefont {Hofstetter}, \citenamefont {Musil}, \citenamefont
  {De}, \citenamefont {Ceriotti},\ and\ \citenamefont {Emsley}}]{paru+18ncomm}%
  \BibitemOpen
  \bibfield  {author} {\bibinfo {author} {\bibfnamefont {F.~M.}\ \bibnamefont
  {Paruzzo}}, \bibinfo {author} {\bibfnamefont {A.}~\bibnamefont {Hofstetter}},
  \bibinfo {author} {\bibfnamefont {F.}~\bibnamefont {Musil}}, \bibinfo
  {author} {\bibfnamefont {S.}~\bibnamefont {De}}, \bibinfo {author}
  {\bibfnamefont {M.}~\bibnamefont {Ceriotti}},\ and\ \bibinfo {author}
  {\bibfnamefont {L.}~\bibnamefont {Emsley}},\ }\bibfield  {title} {\enquote
  {\bibinfo {title} {Chemical shifts in molecular solids by machine
  learning},}\ }\href {https://doi.org/10.1038/s41467-018-06972-x} {\bibfield
  {journal} {\bibinfo  {journal} {Nat. Commun.}\ }\textbf {\bibinfo {volume}
  {9}},\ \bibinfo {pages} {4501} (\bibinfo {year} {2018})}\BibitemShut
  {NoStop}%
\bibitem [{\citenamefont {Brockherde}\ \emph {et~al.}(2017)\citenamefont
  {Brockherde}, \citenamefont {Vogt}, \citenamefont {Li}, \citenamefont
  {Tuckerman}, \citenamefont {Burke},\ and\ \citenamefont
  {M{\"u}ller}}]{broc+17nc}%
  \BibitemOpen
  \bibfield  {author} {\bibinfo {author} {\bibfnamefont {F.}~\bibnamefont
  {Brockherde}}, \bibinfo {author} {\bibfnamefont {L.}~\bibnamefont {Vogt}},
  \bibinfo {author} {\bibfnamefont {L.}~\bibnamefont {Li}}, \bibinfo {author}
  {\bibfnamefont {M.~E.}\ \bibnamefont {Tuckerman}}, \bibinfo {author}
  {\bibfnamefont {K.}~\bibnamefont {Burke}},\ and\ \bibinfo {author}
  {\bibfnamefont {K.~R.}\ \bibnamefont {M{\"u}ller}},\ }\bibfield  {title}
  {\enquote {\bibinfo {title} {Bypassing the {{Kohn}}-{{Sham}} equations with
  machine learning},}\ }\href {https://doi.org/10.1038/s41467-017-00839-3}
  {\bibfield  {journal} {\bibinfo  {journal} {Nat. Commun.}\ }\textbf {\bibinfo
  {volume} {8}},\ \bibinfo {pages} {872} (\bibinfo {year} {2017})}\BibitemShut
  {NoStop}%
\bibitem [{\citenamefont {Grisafi}\ \emph {et~al.}(2019)\citenamefont
  {Grisafi}, \citenamefont {Fabrizio}, \citenamefont {Meyer}, \citenamefont
  {Wilkins}, \citenamefont {Corminboeuf},\ and\ \citenamefont
  {Ceriotti}}]{gris+19acscs}%
  \BibitemOpen
  \bibfield  {author} {\bibinfo {author} {\bibfnamefont {A.}~\bibnamefont
  {Grisafi}}, \bibinfo {author} {\bibfnamefont {A.}~\bibnamefont {Fabrizio}},
  \bibinfo {author} {\bibfnamefont {B.}~\bibnamefont {Meyer}}, \bibinfo
  {author} {\bibfnamefont {D.~M.}\ \bibnamefont {Wilkins}}, \bibinfo {author}
  {\bibfnamefont {C.}~\bibnamefont {Corminboeuf}},\ and\ \bibinfo {author}
  {\bibfnamefont {M.}~\bibnamefont {Ceriotti}},\ }\bibfield  {title} {\enquote
  {\bibinfo {title} {Transferable {{Machine}}-{{Learning Model}} of the
  {{Electron Density}}},}\ }\href {https://doi.org/10.1021/acscentsci.8b00551}
  {\bibfield  {journal} {\bibinfo  {journal} {ACS Cent. Sci.}\ }\textbf
  {\bibinfo {volume} {5}},\ \bibinfo {pages} {57--64} (\bibinfo {year}
  {2019})}\BibitemShut {NoStop}%
\bibitem [{\citenamefont {Jia}\ \emph {et~al.}(2020)\citenamefont {Jia},
  \citenamefont {Wang}, \citenamefont {Chen}, \citenamefont {Lu}, \citenamefont
  {Lin}, \citenamefont {Car}, \citenamefont {E},\ and\ \citenamefont
  {Zhang}}]{jia2020pushing}%
  \BibitemOpen
  \bibfield  {author} {\bibinfo {author} {\bibfnamefont {W.}~\bibnamefont
  {Jia}}, \bibinfo {author} {\bibfnamefont {H.}~\bibnamefont {Wang}}, \bibinfo
  {author} {\bibfnamefont {M.}~\bibnamefont {Chen}}, \bibinfo {author}
  {\bibfnamefont {D.}~\bibnamefont {Lu}}, \bibinfo {author} {\bibfnamefont
  {L.}~\bibnamefont {Lin}}, \bibinfo {author} {\bibfnamefont {R.}~\bibnamefont
  {Car}}, \bibinfo {author} {\bibfnamefont {W.}~\bibnamefont {E}},\ and\
  \bibinfo {author} {\bibfnamefont {L.}~\bibnamefont {Zhang}},\ }\href@noop {}
  {\enquote {\bibinfo {title} {Pushing the limit of molecular dynamics with ab
  initio accuracy to 100 million atoms with machine learning},}\ } (\bibinfo
  {year} {2020}),\ \Eprint {https://arxiv.org/abs/2005.00223} {arXiv:2005.00223
  [physics.comp-ph]} \BibitemShut {NoStop}%
\bibitem [{\citenamefont {Kapil}\ \emph
  {et~al.}(2019{\natexlab{a}})\citenamefont {Kapil}, \citenamefont {Engel},
  \citenamefont {Rossi},\ and\ \citenamefont {Ceriotti}}]{kapi+19jctc2}%
  \BibitemOpen
  \bibfield  {author} {\bibinfo {author} {\bibfnamefont {V.}~\bibnamefont
  {Kapil}}, \bibinfo {author} {\bibfnamefont {E.}~\bibnamefont {Engel}},
  \bibinfo {author} {\bibfnamefont {M.}~\bibnamefont {Rossi}},\ and\ \bibinfo
  {author} {\bibfnamefont {M.}~\bibnamefont {Ceriotti}},\ }\bibfield  {title}
  {\enquote {\bibinfo {title} {Assessment of {{Approximate Methods}} for
  {{Anharmonic Free Energies}}},}\ }\href
  {https://doi.org/10.1021/acs.jctc.9b00596} {\bibfield  {journal} {\bibinfo
  {journal} {J. Chem. Theory Comput.}\ }\textbf {\bibinfo {volume} {15}},\
  \bibinfo {pages} {5845--5857} (\bibinfo {year}
  {2019}{\natexlab{a}})}\BibitemShut {NoStop}%
\bibitem [{\citenamefont {Kapil}\ \emph {et~al.}(2020)\citenamefont {Kapil},
  \citenamefont {Wilkins}, \citenamefont {Lan},\ and\ \citenamefont
  {Ceriotti}}]{kapi+20jcp}%
  \BibitemOpen
  \bibfield  {author} {\bibinfo {author} {\bibfnamefont {V.}~\bibnamefont
  {Kapil}}, \bibinfo {author} {\bibfnamefont {D.~M.}\ \bibnamefont {Wilkins}},
  \bibinfo {author} {\bibfnamefont {J.}~\bibnamefont {Lan}},\ and\ \bibinfo
  {author} {\bibfnamefont {M.}~\bibnamefont {Ceriotti}},\ }\bibfield  {title}
  {\enquote {\bibinfo {title} {Inexpensive modeling of quantum dynamics using
  path integral generalized {{Langevin}} equation thermostats},}\ }\href
  {https://doi.org/10.1063/1.5141950} {\bibfield  {journal} {\bibinfo
  {journal} {J. Chem. Phys.}\ }\textbf {\bibinfo {volume} {152}},\ \bibinfo
  {pages} {124104} (\bibinfo {year} {2020})}\BibitemShut {NoStop}%
\bibitem [{\citenamefont {Tran}\ \emph {et~al.}(2020)\citenamefont {Tran},
  \citenamefont {Neiswanger}, \citenamefont {Yoon}, \citenamefont {Zhang},
  \citenamefont {Xing},\ and\ \citenamefont
  {Ulissi}}]{Tran_Neiswanger_Yoon_Zhang_Xing_Ulissi_2020}%
  \BibitemOpen
  \bibfield  {author} {\bibinfo {author} {\bibfnamefont {K.}~\bibnamefont
  {Tran}}, \bibinfo {author} {\bibfnamefont {W.}~\bibnamefont {Neiswanger}},
  \bibinfo {author} {\bibfnamefont {J.}~\bibnamefont {Yoon}}, \bibinfo {author}
  {\bibfnamefont {Q.}~\bibnamefont {Zhang}}, \bibinfo {author} {\bibfnamefont
  {E.}~\bibnamefont {Xing}},\ and\ \bibinfo {author} {\bibfnamefont {Z.~W.}\
  \bibnamefont {Ulissi}},\ }\bibfield  {title} {\enquote {\bibinfo {title}
  {Methods for comparing uncertainty quantifications for material property
  predictions},}\ }\href {https://doi.org/10.1088/2632-2153/ab7e1a} {\bibfield
  {journal} {\bibinfo  {journal} {Machine Learning: Science and Technology}\
  }\textbf {\bibinfo {volume} {1}},\ \bibinfo {pages} {025006} (\bibinfo {year}
  {2020})}\BibitemShut {NoStop}%
\bibitem [{\citenamefont {Rasmussen}\ and\ \citenamefont
  {Williams}(2005)}]{rasm05book}%
  \BibitemOpen
  \bibfield  {author} {\bibinfo {author} {\bibfnamefont {C.~E.}\ \bibnamefont
  {Rasmussen}}\ and\ \bibinfo {author} {\bibfnamefont {C.~K.~I.}\ \bibnamefont
  {Williams}},\ }\href@noop {} {\emph {\bibinfo {title} {Gaussian {{Processes}}
  for {{Machine Learning}} ({{Adaptive Computation}} and {{Machine
  Learning}})}}}\ (\bibinfo  {publisher} {{The MIT Press}},\ \bibinfo {year}
  {2005})\BibitemShut {NoStop}%
\bibitem [{\citenamefont {Peterson}, \citenamefont {Christensen},\ and\
  \citenamefont {Khorshidi}(2017)}]{Peterson_Christensen_Khorshidi_2017}%
  \BibitemOpen
  \bibfield  {author} {\bibinfo {author} {\bibfnamefont {A.~A.}\ \bibnamefont
  {Peterson}}, \bibinfo {author} {\bibfnamefont {R.}~\bibnamefont
  {Christensen}},\ and\ \bibinfo {author} {\bibfnamefont {A.}~\bibnamefont
  {Khorshidi}},\ }\bibfield  {title} {\enquote {\bibinfo {title} {Addressing
  uncertainty in atomistic machine learning},}\ }\href
  {https://doi.org/10.1039/C7CP00375G} {\bibfield  {journal} {\bibinfo
  {journal} {Physical Chemistry Chemical Physics}\ }\textbf {\bibinfo {volume}
  {19}},\ \bibinfo {pages} {10978–10985} (\bibinfo {year}
  {2017})}\BibitemShut {NoStop}%
\bibitem [{\citenamefont {Behler}(2015)}]{behl15ijqc}%
  \BibitemOpen
  \bibfield  {author} {\bibinfo {author} {\bibfnamefont {J.}~\bibnamefont
  {Behler}},\ }\bibfield  {title} {\enquote {\bibinfo {title} {Constructing
  high-dimensional neural network potentials: {{A}} tutorial review},}\ }\href
  {https://doi.org/10.1002/qua.24890} {\bibfield  {journal} {\bibinfo
  {journal} {Int. J. Quantum Chem.}\ }\textbf {\bibinfo {volume} {115}},\
  \bibinfo {pages} {1032--1050} (\bibinfo {year} {2015})}\BibitemShut {NoStop}%
\bibitem [{\citenamefont {Shapeev}\ \emph
  {et~al.}(2020{\natexlab{a}})\citenamefont {Shapeev}, \citenamefont {Gubaev},
  \citenamefont {Tsymbalov},\ and\ \citenamefont {Podryabinkin}}]{Shapeev2020}%
  \BibitemOpen
  \bibfield  {author} {\bibinfo {author} {\bibfnamefont {A.}~\bibnamefont
  {Shapeev}}, \bibinfo {author} {\bibfnamefont {K.}~\bibnamefont {Gubaev}},
  \bibinfo {author} {\bibfnamefont {E.}~\bibnamefont {Tsymbalov}},\ and\
  \bibinfo {author} {\bibfnamefont {E.}~\bibnamefont {Podryabinkin}},\
  }\enquote {\bibinfo {title} {{Active Learning and Uncertainty Estimation}},}\
  in\ \href {https://doi.org/10.1007/978-3-030-40245-7_15} {\emph {\bibinfo
  {booktitle} {Machine Learning Meets Quantum Physics}}},\ \bibinfo {editor}
  {edited by\ \bibinfo {editor} {\bibfnamefont {K.~T.}\ \bibnamefont
  {Sch{\"{u}}tt}}, \bibinfo {editor} {\bibfnamefont {S.}~\bibnamefont
  {Chmiela}}, \bibinfo {editor} {\bibfnamefont {O.~A.}\ \bibnamefont {von
  Lilienfeld}}, \bibinfo {editor} {\bibfnamefont {A.}~\bibnamefont
  {Tkatchenko}}, \bibinfo {editor} {\bibfnamefont {K.}~\bibnamefont {Tsuda}},\
  and\ \bibinfo {editor} {\bibfnamefont {K.-R.}\ \bibnamefont {M{\"{u}}ller}}}\
  (\bibinfo  {publisher} {Springer International Publishing},\ \bibinfo
  {address} {Cham},\ \bibinfo {year} {2020})\ pp.\ \bibinfo {pages}
  {309--329}\BibitemShut {NoStop}%
\bibitem [{\citenamefont {Politis}\ and\ \citenamefont
  {Romano}(1994)}]{poli-roma94aos}%
  \BibitemOpen
  \bibfield  {author} {\bibinfo {author} {\bibfnamefont {D.~N.}\ \bibnamefont
  {Politis}}\ and\ \bibinfo {author} {\bibfnamefont {J.~P.}\ \bibnamefont
  {Romano}},\ }\bibfield  {title} {\enquote {\bibinfo {title} {Large {{Sample
  Confidence Regions Based}} on {{Subsamples}} under {{Minimal
  Assumptions}}},}\ }\href {https://doi.org/10.1214/aos/1176325770} {\bibfield
  {journal} {\bibinfo  {journal} {Ann. Stat.}\ }\textbf {\bibinfo {volume}
  {22}},\ \bibinfo {pages} {2031--2050} (\bibinfo {year} {1994})}\BibitemShut
  {NoStop}%
\bibitem [{\citenamefont {Efron}(1979)}]{efron1979}%
  \BibitemOpen
  \bibfield  {author} {\bibinfo {author} {\bibfnamefont {B.}~\bibnamefont
  {Efron}},\ }\bibfield  {title} {\enquote {\bibinfo {title} {Bootstrap
  methods: Another look at the jackknife},}\ }\href
  {https://doi.org/10.1214/aos/1176344552} {\bibfield  {journal} {\bibinfo
  {journal} {Ann. Statist.}\ }\textbf {\bibinfo {volume} {7}},\ \bibinfo
  {pages} {1--26} (\bibinfo {year} {1979})}\BibitemShut {NoStop}%
\bibitem [{\citenamefont {Shapeev}\ \emph
  {et~al.}(2020{\natexlab{b}})\citenamefont {Shapeev}, \citenamefont {Gubaev},
  \citenamefont {Tsymbalov},\ and\ \citenamefont
  {Podryabinkin}}]{Shapeev_Gubaev_Tsymbalov_Podryabinkin_2020}%
  \BibitemOpen
  \bibfield  {author} {\bibinfo {author} {\bibfnamefont {A.}~\bibnamefont
  {Shapeev}}, \bibinfo {author} {\bibfnamefont {K.}~\bibnamefont {Gubaev}},
  \bibinfo {author} {\bibfnamefont {E.}~\bibnamefont {Tsymbalov}},\ and\
  \bibinfo {author} {\bibfnamefont {E.}~\bibnamefont {Podryabinkin}},\
  }\enquote {\bibinfo {title} {Active learning and uncertainty estimation},}\
  in\ \href {https://doi.org/10.1007/978-3-030-40245-7_15} {\emph {\bibinfo
  {booktitle} {Machine Learning Meets Quantum Physics}}},\ \bibinfo {series and
  number} {Lecture Notes in Physics},\ \bibinfo {editor} {edited by\ \bibinfo
  {editor} {\bibfnamefont {K.~T.}\ \bibnamefont {Schütt}}, \bibinfo {editor}
  {\bibfnamefont {S.}~\bibnamefont {Chmiela}}, \bibinfo {editor} {\bibfnamefont
  {O.~A.}\ \bibnamefont {von Lilienfeld}}, \bibinfo {editor} {\bibfnamefont
  {A.}~\bibnamefont {Tkatchenko}}, \bibinfo {editor} {\bibfnamefont
  {K.}~\bibnamefont {Tsuda}},\ and\ \bibinfo {editor} {\bibfnamefont {K.-R.}\
  \bibnamefont {Müller}}}\ (\bibinfo  {publisher} {Springer International
  Publishing},\ \bibinfo {year} {2020})\ p.\ \bibinfo {pages}
  {309–329}\BibitemShut {NoStop}%
\bibitem [{\citenamefont {Shuaibi}\ \emph {et~al.}(2020)\citenamefont
  {Shuaibi}, \citenamefont {Sivakumar}, \citenamefont {Chen},\ and\
  \citenamefont {Ulissi}}]{shuaibi2020}%
  \BibitemOpen
  \bibfield  {author} {\bibinfo {author} {\bibfnamefont {M.}~\bibnamefont
  {Shuaibi}}, \bibinfo {author} {\bibfnamefont {S.}~\bibnamefont {Sivakumar}},
  \bibinfo {author} {\bibfnamefont {R.~Q.}\ \bibnamefont {Chen}},\ and\
  \bibinfo {author} {\bibfnamefont {Z.~W.}\ \bibnamefont {Ulissi}},\ }\bibfield
   {title} {\enquote {\bibinfo {title} {Enabling robust offline active learning
  for machine learning potentials using simple physics-based priors},}\
  }\href@noop {} {\  (\bibinfo {year} {2020})},\ \Eprint
  {https://arxiv.org/abs/2008.10773} {arXiv:2008.10773 [physics.comp-ph]}
  \BibitemShut {NoStop}%
\bibitem [{\citenamefont {Schran}, \citenamefont {Brezina},\ and\ \citenamefont
  {Marsalek}(2020{\natexlab{a}})}]{Schran_Brezina_Marsalek_2020}%
  \BibitemOpen
  \bibfield  {author} {\bibinfo {author} {\bibfnamefont {C.}~\bibnamefont
  {Schran}}, \bibinfo {author} {\bibfnamefont {K.}~\bibnamefont {Brezina}},\
  and\ \bibinfo {author} {\bibfnamefont {O.}~\bibnamefont {Marsalek}},\
  }\bibfield  {title} {\enquote {\bibinfo {title} {Committee neural network
  potentials control generalization errors and enable active learning},}\
  }\href {http://arxiv.org/abs/2006.01541} {\bibfield  {journal} {\bibinfo
  {journal} {arXiv:2006.01541 [physics, stat]}\ } (\bibinfo {year}
  {2020}{\natexlab{a}})},\ \bibinfo {note} {arXiv: 2006.01541}\BibitemShut
  {NoStop}%
\bibitem [{\citenamefont {Jinnouchi}\ \emph {et~al.}(2019)\citenamefont
  {Jinnouchi}, \citenamefont {Lahnsteiner}, \citenamefont {Karsai},
  \citenamefont {Kresse},\ and\ \citenamefont
  {Bokdam}}]{Jinnouchi_Lahnsteiner_Karsai_Kresse_Bokdam_2019}%
  \BibitemOpen
  \bibfield  {author} {\bibinfo {author} {\bibfnamefont {R.}~\bibnamefont
  {Jinnouchi}}, \bibinfo {author} {\bibfnamefont {J.}~\bibnamefont
  {Lahnsteiner}}, \bibinfo {author} {\bibfnamefont {F.}~\bibnamefont {Karsai}},
  \bibinfo {author} {\bibfnamefont {G.}~\bibnamefont {Kresse}},\ and\ \bibinfo
  {author} {\bibfnamefont {M.}~\bibnamefont {Bokdam}},\ }\bibfield  {title}
  {\enquote {\bibinfo {title} {Phase transitions of hybrid perovskites
  simulated by machine-learning force fields trained on the fly with bayesian
  inference},}\ }\href {https://doi.org/10.1103/PhysRevLett.122.225701}
  {\bibfield  {journal} {\bibinfo  {journal} {Physical Review Letters}\
  }\textbf {\bibinfo {volume} {122}},\ \bibinfo {pages} {225701} (\bibinfo
  {year} {2019})}\BibitemShut {NoStop}%
\bibitem [{\citenamefont {Vandermause}\ \emph {et~al.}(2020)\citenamefont
  {Vandermause}, \citenamefont {Torrisi}, \citenamefont {Batzner},
  \citenamefont {Xie}, \citenamefont {Sun}, \citenamefont {Kolpak},\ and\
  \citenamefont
  {Kozinsky}}]{Vandermause_Torrisi_Batzner_Xie_Sun_Kolpak_Kozinsky_2020}%
  \BibitemOpen
  \bibfield  {author} {\bibinfo {author} {\bibfnamefont {J.}~\bibnamefont
  {Vandermause}}, \bibinfo {author} {\bibfnamefont {S.~B.}\ \bibnamefont
  {Torrisi}}, \bibinfo {author} {\bibfnamefont {S.}~\bibnamefont {Batzner}},
  \bibinfo {author} {\bibfnamefont {Y.}~\bibnamefont {Xie}}, \bibinfo {author}
  {\bibfnamefont {L.}~\bibnamefont {Sun}}, \bibinfo {author} {\bibfnamefont
  {A.~M.}\ \bibnamefont {Kolpak}},\ and\ \bibinfo {author} {\bibfnamefont
  {B.}~\bibnamefont {Kozinsky}},\ }\bibfield  {title} {\enquote {\bibinfo
  {title} {On-the-fly active learning of interpretable bayesian force fields
  for atomistic rare events},}\ }\href
  {https://doi.org/10.1038/s41524-020-0283-z} {\bibfield  {journal} {\bibinfo
  {journal} {npj Computational Materials}\ }\textbf {\bibinfo {volume} {6}},\
  \bibinfo {pages} {1–11} (\bibinfo {year} {2020})}\BibitemShut {NoStop}%
\bibitem [{\citenamefont {Schran}, \citenamefont {Brezina},\ and\ \citenamefont
  {Marsalek}(2020{\natexlab{b}})}]{Schran2020}%
  \BibitemOpen
  \bibfield  {author} {\bibinfo {author} {\bibfnamefont {C.}~\bibnamefont
  {Schran}}, \bibinfo {author} {\bibfnamefont {K.}~\bibnamefont {Brezina}},\
  and\ \bibinfo {author} {\bibfnamefont {O.}~\bibnamefont {Marsalek}},\
  }\bibfield  {title} {\enquote {\bibinfo {title} {Committee neural network
  potentials control generalization errors and enable active learning},}\
  }\href {https://doi.org/10.1063/5.0016004} {\bibfield  {journal} {\bibinfo
  {journal} {The Journal of Chemical Physics}\ }\textbf {\bibinfo {volume}
  {153}},\ \bibinfo {pages} {104105} (\bibinfo {year}
  {2020}{\natexlab{b}})}\BibitemShut {NoStop}%
\bibitem [{\citenamefont {Musil}\ \emph {et~al.}(2019)\citenamefont {Musil},
  \citenamefont {Willatt}, \citenamefont {Langovoy},\ and\ \citenamefont
  {Ceriotti}}]{musi+19jctc}%
  \BibitemOpen
  \bibfield  {author} {\bibinfo {author} {\bibfnamefont {F.}~\bibnamefont
  {Musil}}, \bibinfo {author} {\bibfnamefont {M.~J.}\ \bibnamefont {Willatt}},
  \bibinfo {author} {\bibfnamefont {M.~A.}\ \bibnamefont {Langovoy}},\ and\
  \bibinfo {author} {\bibfnamefont {M.}~\bibnamefont {Ceriotti}},\ }\bibfield
  {title} {\enquote {\bibinfo {title} {Fast and {{Accurate Uncertainty
  Estimation}} in {{Chemical Machine Learning}}},}\ }\href
  {https://doi.org/10.1021/acs.jctc.8b00959} {\bibfield  {journal} {\bibinfo
  {journal} {J. Chem. Theory Comput.}\ }\textbf {\bibinfo {volume} {15}},\
  \bibinfo {pages} {906--915} (\bibinfo {year} {2019})}\BibitemShut {NoStop}%
\bibitem [{\citenamefont {Raimbault}\ \emph {et~al.}(2019)\citenamefont
  {Raimbault}, \citenamefont {Grisafi}, \citenamefont {Ceriotti},\ and\
  \citenamefont {Rossi}}]{raim+19njp}%
  \BibitemOpen
  \bibfield  {author} {\bibinfo {author} {\bibfnamefont {N.}~\bibnamefont
  {Raimbault}}, \bibinfo {author} {\bibfnamefont {A.}~\bibnamefont {Grisafi}},
  \bibinfo {author} {\bibfnamefont {M.}~\bibnamefont {Ceriotti}},\ and\
  \bibinfo {author} {\bibfnamefont {M.}~\bibnamefont {Rossi}},\ }\bibfield
  {title} {\enquote {\bibinfo {title} {Using {{Gaussian}} process regression to
  simulate the vibrational {{Raman}} spectra of molecular crystals},}\ }\href
  {https://doi.org/10.1088/1367-2630/ab4509} {\bibfield  {journal} {\bibinfo
  {journal} {New J. Phys.}\ }\textbf {\bibinfo {volume} {21}},\ \bibinfo
  {pages} {105001} (\bibinfo {year} {2019})}\BibitemShut {NoStop}%
\bibitem [{\citenamefont {Tuckerman}, \citenamefont {Berne},\ and\
  \citenamefont {Martyna}(1992)}]{tuck+92jcp}%
  \BibitemOpen
  \bibfield  {author} {\bibinfo {author} {\bibfnamefont {M.}~\bibnamefont
  {Tuckerman}}, \bibinfo {author} {\bibfnamefont {B.~J.}\ \bibnamefont
  {Berne}},\ and\ \bibinfo {author} {\bibfnamefont {G.~J.}\ \bibnamefont
  {Martyna}},\ }\bibfield  {title} {\enquote {\bibinfo {title} {Reversible
  multiple time scale molecular dynamics},}\ }\href
  {https://doi.org/10.1063/1.463137} {\bibfield  {journal} {\bibinfo  {journal}
  {J. Chem. Phys.}\ }\textbf {\bibinfo {volume} {97}},\ \bibinfo {pages} {1990}
  (\bibinfo {year} {1992})}\BibitemShut {NoStop}%
\bibitem [{\citenamefont {Markland}\ and\ \citenamefont
  {Manolopoulos}(2008)}]{mark-mano08cpl}%
  \BibitemOpen
  \bibfield  {author} {\bibinfo {author} {\bibfnamefont {T.~E.}\ \bibnamefont
  {Markland}}\ and\ \bibinfo {author} {\bibfnamefont {D.~E.}\ \bibnamefont
  {Manolopoulos}},\ }\bibfield  {title} {\enquote {\bibinfo {title} {A refined
  ring polymer contraction scheme for systems with electrostatic
  interactions},}\ }\href@noop {} {\bibfield  {journal} {\bibinfo  {journal}
  {Chem. Phys. Lett.}\ }\textbf {\bibinfo {volume} {464}},\ \bibinfo {pages}
  {256} (\bibinfo {year} {2008})}\BibitemShut {NoStop}%
\bibitem [{\citenamefont {Kapil}, \citenamefont {VandeVondele},\ and\
  \citenamefont {Ceriotti}(2016)}]{kapi+16jcp}%
  \BibitemOpen
  \bibfield  {author} {\bibinfo {author} {\bibfnamefont {V.}~\bibnamefont
  {Kapil}}, \bibinfo {author} {\bibfnamefont {J.}~\bibnamefont
  {VandeVondele}},\ and\ \bibinfo {author} {\bibfnamefont {M.}~\bibnamefont
  {Ceriotti}},\ }\bibfield  {title} {\enquote {\bibinfo {title} {Accurate
  molecular dynamics and nuclear quantum effects at low cost by multiple steps
  in real and imaginary time: {{Using}} density functional theory to accelerate
  wavefunction methods},}\ }\href {https://doi.org/10.1063/1.4941091}
  {\bibfield  {journal} {\bibinfo  {journal} {J. Chem. Phys.}\ }\textbf
  {\bibinfo {volume} {144}},\ \bibinfo {pages} {054111} (\bibinfo {year}
  {2016})}\BibitemShut {NoStop}%
\bibitem [{\citenamefont {Ramakrishnan}\ \emph {et~al.}(2015)\citenamefont
  {Ramakrishnan}, \citenamefont {Dral}, \citenamefont {Rupp},\ and\
  \citenamefont {Von~Lilienfeld}}]{rama+15jctc}%
  \BibitemOpen
  \bibfield  {author} {\bibinfo {author} {\bibfnamefont {R.}~\bibnamefont
  {Ramakrishnan}}, \bibinfo {author} {\bibfnamefont {P.~O.}\ \bibnamefont
  {Dral}}, \bibinfo {author} {\bibfnamefont {M.}~\bibnamefont {Rupp}},\ and\
  \bibinfo {author} {\bibfnamefont {O.~A.}\ \bibnamefont {Von~Lilienfeld}},\
  }\bibfield  {title} {\enquote {\bibinfo {title} {Big data meets quantum
  chemistry approximations: {{The $\Delta$}}-machine learning approach},}\
  }\href {https://doi.org/10.1021/acs.jctc.5b00099} {\bibfield  {journal}
  {\bibinfo  {journal} {J. Chem. Theory Comput.}\ }\textbf {\bibinfo {volume}
  {11}},\ \bibinfo {pages} {2087--2096} (\bibinfo {year} {2015})}\BibitemShut
  {NoStop}%
\bibitem [{\citenamefont {Bart{\'o}k}\ \emph {et~al.}(2017)\citenamefont
  {Bart{\'o}k}, \citenamefont {De}, \citenamefont {Poelking}, \citenamefont
  {Bernstein}, \citenamefont {Kermode}, \citenamefont {Cs{\'a}nyi},\ and\
  \citenamefont {Ceriotti}}]{bart+17sa}%
  \BibitemOpen
  \bibfield  {author} {\bibinfo {author} {\bibfnamefont {A.~P.}\ \bibnamefont
  {Bart{\'o}k}}, \bibinfo {author} {\bibfnamefont {S.}~\bibnamefont {De}},
  \bibinfo {author} {\bibfnamefont {C.}~\bibnamefont {Poelking}}, \bibinfo
  {author} {\bibfnamefont {N.}~\bibnamefont {Bernstein}}, \bibinfo {author}
  {\bibfnamefont {J.~R.}\ \bibnamefont {Kermode}}, \bibinfo {author}
  {\bibfnamefont {G.}~\bibnamefont {Cs{\'a}nyi}},\ and\ \bibinfo {author}
  {\bibfnamefont {M.}~\bibnamefont {Ceriotti}},\ }\bibfield  {title} {\enquote
  {\bibinfo {title} {Machine learning unifies the modeling of materials and
  molecules},}\ }\href {https://doi.org/10.1126/sciadv.1701816} {\bibfield
  {journal} {\bibinfo  {journal} {Sci. Adv.}\ }\textbf {\bibinfo {volume}
  {3}},\ \bibinfo {pages} {e1701816} (\bibinfo {year} {2017})}\BibitemShut
  {NoStop}%
\bibitem [{\citenamefont {Li}, \citenamefont {Kermode},\ and\ \citenamefont
  {De~Vita}(2015{\natexlab{a}})}]{li+15prl}%
  \BibitemOpen
  \bibfield  {author} {\bibinfo {author} {\bibfnamefont {Z.}~\bibnamefont
  {Li}}, \bibinfo {author} {\bibfnamefont {J.~R.}\ \bibnamefont {Kermode}},\
  and\ \bibinfo {author} {\bibfnamefont {A.}~\bibnamefont {De~Vita}},\
  }\bibfield  {title} {\enquote {\bibinfo {title} {Molecular dynamics with
  on-the-fly machine learning of quantum-mechanical forces},}\ }\href
  {https://doi.org/10.1103/PhysRevLett.114.096405} {\bibfield  {journal}
  {\bibinfo  {journal} {Phys. Rev. Lett.}\ }\textbf {\bibinfo {volume} {114}},\
  \bibinfo {pages} {096405} (\bibinfo {year} {2015}{\natexlab{a}})}\BibitemShut
  {NoStop}%
\bibitem [{\citenamefont {Smith}\ \emph {et~al.}(2018)\citenamefont {Smith},
  \citenamefont {Nebgen}, \citenamefont {Lubbers}, \citenamefont {Isayev},\
  and\ \citenamefont {Roitberg}}]{smit+18jcp}%
  \BibitemOpen
  \bibfield  {author} {\bibinfo {author} {\bibfnamefont {J.~S.}\ \bibnamefont
  {Smith}}, \bibinfo {author} {\bibfnamefont {B.}~\bibnamefont {Nebgen}},
  \bibinfo {author} {\bibfnamefont {N.}~\bibnamefont {Lubbers}}, \bibinfo
  {author} {\bibfnamefont {O.}~\bibnamefont {Isayev}},\ and\ \bibinfo {author}
  {\bibfnamefont {A.~E.}\ \bibnamefont {Roitberg}},\ }\bibfield  {title}
  {\enquote {\bibinfo {title} {Less is more: {{Sampling}} chemical space with
  active learning},}\ }\href {https://doi.org/10.1063/1.5023802} {\bibfield
  {journal} {\bibinfo  {journal} {The Journal of Chemical Physics}\ }\textbf
  {\bibinfo {volume} {148}},\ \bibinfo {pages} {241733} (\bibinfo {year}
  {2018})}\BibitemShut {NoStop}%
\bibitem [{\citenamefont {Janet}\ \emph {et~al.}(2019)\citenamefont {Janet},
  \citenamefont {Duan}, \citenamefont {Yang}, \citenamefont {Nandy},\ and\
  \citenamefont {Kulik}}]{jane+19cs}%
  \BibitemOpen
  \bibfield  {author} {\bibinfo {author} {\bibfnamefont {J.~P.}\ \bibnamefont
  {Janet}}, \bibinfo {author} {\bibfnamefont {C.}~\bibnamefont {Duan}},
  \bibinfo {author} {\bibfnamefont {T.}~\bibnamefont {Yang}}, \bibinfo {author}
  {\bibfnamefont {A.}~\bibnamefont {Nandy}},\ and\ \bibinfo {author}
  {\bibfnamefont {H.~J.}\ \bibnamefont {Kulik}},\ }\bibfield  {title} {\enquote
  {\bibinfo {title} {A quantitative uncertainty metric controls error in neural
  network-driven chemical discovery},}\ }\href
  {https://doi.org/10.1039/C9SC02298H} {\bibfield  {journal} {\bibinfo
  {journal} {Chem. Sci.}\ }\textbf {\bibinfo {volume} {10}},\ \bibinfo {pages}
  {7913--7922} (\bibinfo {year} {2019})}\BibitemShut {NoStop}%
\bibitem [{\citenamefont {Ben~Mahmoud}\ \emph {et~al.}(2020)\citenamefont
  {Ben~Mahmoud}, \citenamefont {Anelli}, \citenamefont {Cs{\'a}nyi},\ and\
  \citenamefont {Ceriotti}}]{chiheb2020}%
  \BibitemOpen
  \bibfield  {author} {\bibinfo {author} {\bibfnamefont {C.}~\bibnamefont
  {Ben~Mahmoud}}, \bibinfo {author} {\bibfnamefont {A.}~\bibnamefont {Anelli}},
  \bibinfo {author} {\bibfnamefont {G.}~\bibnamefont {Cs{\'a}nyi}},\ and\
  \bibinfo {author} {\bibfnamefont {M.}~\bibnamefont {Ceriotti}},\ }\bibfield
  {title} {\enquote {\bibinfo {title} {Learning the electronic density of
  states in condensed matter},}\ }\href@noop {} {\bibfield  {journal} {\bibinfo
   {journal} {arXiv preprint arXiv:2006.11803}\ } (\bibinfo {year}
  {2020})}\BibitemShut {NoStop}%
\bibitem [{\citenamefont {Torrie}\ and\ \citenamefont
  {Valleau}(1977)}]{torr-vall99jcp}%
  \BibitemOpen
  \bibfield  {author} {\bibinfo {author} {\bibfnamefont {G.~M.}\ \bibnamefont
  {Torrie}}\ and\ \bibinfo {author} {\bibfnamefont {J.~P.}\ \bibnamefont
  {Valleau}},\ }\bibfield  {title} {\enquote {\bibinfo {title} {Nonphysical
  sampling distributions in {{Monte Carlo}} free-energy estimation:
  {{Umbrella}} sampling},}\ }\href
  {https://doi.org/10.1016/0021-9991(77)90121-8} {\bibfield  {journal}
  {\bibinfo  {journal} {J. Comput. Phys.}\ }\textbf {\bibinfo {volume} {23}},\
  \bibinfo {pages} {187--199} (\bibinfo {year} {1977})}\BibitemShut {NoStop}%
\bibitem [{\citenamefont {Cs\'anyi}\ \emph {et~al.}(2004)\citenamefont
  {Cs\'anyi}, \citenamefont {Albaret}, \citenamefont {Payne},\ and\
  \citenamefont {De~Vita}}]{csanyiPRL2003}%
  \BibitemOpen
  \bibfield  {author} {\bibinfo {author} {\bibfnamefont {G.}~\bibnamefont
  {Cs\'anyi}}, \bibinfo {author} {\bibfnamefont {T.}~\bibnamefont {Albaret}},
  \bibinfo {author} {\bibfnamefont {M.~C.}\ \bibnamefont {Payne}},\ and\
  \bibinfo {author} {\bibfnamefont {A.}~\bibnamefont {De~Vita}},\ }\bibfield
  {title} {\enquote {\bibinfo {title} {``learn on the fly'': A hybrid classical
  and quantum-mechanical molecular dynamics simulation},}\ }\href
  {https://doi.org/10.1103/PhysRevLett.93.175503} {\bibfield  {journal}
  {\bibinfo  {journal} {Phys. Rev. Lett.}\ }\textbf {\bibinfo {volume} {93}},\
  \bibinfo {pages} {175503} (\bibinfo {year} {2004})}\BibitemShut {NoStop}%
\bibitem [{\citenamefont {Li}, \citenamefont {Kermode},\ and\ \citenamefont
  {De~Vita}(2015{\natexlab{b}})}]{li2015molecular}%
  \BibitemOpen
  \bibfield  {author} {\bibinfo {author} {\bibfnamefont {Z.}~\bibnamefont
  {Li}}, \bibinfo {author} {\bibfnamefont {J.~R.}\ \bibnamefont {Kermode}},\
  and\ \bibinfo {author} {\bibfnamefont {A.}~\bibnamefont {De~Vita}},\
  }\bibfield  {title} {\enquote {\bibinfo {title} {Molecular dynamics with
  on-the-fly machine learning of quantum-mechanical forces},}\ }\href@noop {}
  {\bibfield  {journal} {\bibinfo  {journal} {Physical review letters}\
  }\textbf {\bibinfo {volume} {114}},\ \bibinfo {pages} {096405} (\bibinfo
  {year} {2015}{\natexlab{b}})}\BibitemShut {NoStop}%
\bibitem [{\citenamefont {Ceriotti}\ \emph {et~al.}(2012)\citenamefont
  {Ceriotti}, \citenamefont {Brain}, \citenamefont {Riordan},\ and\
  \citenamefont {Manolopoulos}}]{ceri+12prsa}%
  \BibitemOpen
  \bibfield  {author} {\bibinfo {author} {\bibfnamefont {M.}~\bibnamefont
  {Ceriotti}}, \bibinfo {author} {\bibfnamefont {G.~A.}\ \bibnamefont {Brain}},
  \bibinfo {author} {\bibfnamefont {O.}~\bibnamefont {Riordan}},\ and\ \bibinfo
  {author} {\bibfnamefont {D.~E.}\ \bibnamefont {Manolopoulos}},\ }\bibfield
  {title} {\enquote {\bibinfo {title} {The inefficiency of re-weighted sampling
  and the curse of system size in high-order path integration},}\ }\href
  {https://doi.org/10.1098/rspa.2011.0413} {\bibfield  {journal} {\bibinfo
  {journal} {Proc. R. Soc. Math. Phys. Eng. Sci.}\ }\textbf {\bibinfo {volume}
  {468}},\ \bibinfo {pages} {2--17} (\bibinfo {year} {2012})}\BibitemShut
  {NoStop}%
\bibitem [{\citenamefont {Kapil}\ \emph
  {et~al.}(2019{\natexlab{b}})\citenamefont {Kapil}, \citenamefont {Rossi},
  \citenamefont {Marsalek}, \citenamefont {Petraglia}, \citenamefont {Litman},
  \citenamefont {Spura}, \citenamefont {Cheng}, \citenamefont {Cuzzocrea},
  \citenamefont {Mei{\ss}ner}, \citenamefont {Wilkins}, \citenamefont
  {Helfrecht}, \citenamefont {Juda}, \citenamefont {Bienvenue}, \citenamefont
  {Fang}, \citenamefont {Kessler}, \citenamefont {Poltavsky}, \citenamefont
  {Vandenbrande}, \citenamefont {Wieme}, \citenamefont {Corminboeuf},
  \citenamefont {K{\"u}hne}, \citenamefont {Manolopoulos}, \citenamefont
  {Markland}, \citenamefont {Richardson}, \citenamefont {Tkatchenko},
  \citenamefont {Tribello}, \citenamefont {Van~Speybroeck},\ and\ \citenamefont
  {Ceriotti}}]{kapi+19cpc}%
  \BibitemOpen
  \bibfield  {author} {\bibinfo {author} {\bibfnamefont {V.}~\bibnamefont
  {Kapil}}, \bibinfo {author} {\bibfnamefont {M.}~\bibnamefont {Rossi}},
  \bibinfo {author} {\bibfnamefont {O.}~\bibnamefont {Marsalek}}, \bibinfo
  {author} {\bibfnamefont {R.}~\bibnamefont {Petraglia}}, \bibinfo {author}
  {\bibfnamefont {Y.}~\bibnamefont {Litman}}, \bibinfo {author} {\bibfnamefont
  {T.}~\bibnamefont {Spura}}, \bibinfo {author} {\bibfnamefont
  {B.}~\bibnamefont {Cheng}}, \bibinfo {author} {\bibfnamefont
  {A.}~\bibnamefont {Cuzzocrea}}, \bibinfo {author} {\bibfnamefont {R.~H.}\
  \bibnamefont {Mei{\ss}ner}}, \bibinfo {author} {\bibfnamefont {D.~M.}\
  \bibnamefont {Wilkins}}, \bibinfo {author} {\bibfnamefont {B.~A.}\
  \bibnamefont {Helfrecht}}, \bibinfo {author} {\bibfnamefont {P.}~\bibnamefont
  {Juda}}, \bibinfo {author} {\bibfnamefont {S.~P.}\ \bibnamefont {Bienvenue}},
  \bibinfo {author} {\bibfnamefont {W.}~\bibnamefont {Fang}}, \bibinfo {author}
  {\bibfnamefont {J.}~\bibnamefont {Kessler}}, \bibinfo {author} {\bibfnamefont
  {I.}~\bibnamefont {Poltavsky}}, \bibinfo {author} {\bibfnamefont
  {S.}~\bibnamefont {Vandenbrande}}, \bibinfo {author} {\bibfnamefont
  {J.}~\bibnamefont {Wieme}}, \bibinfo {author} {\bibfnamefont
  {C.}~\bibnamefont {Corminboeuf}}, \bibinfo {author} {\bibfnamefont {T.~D.}\
  \bibnamefont {K{\"u}hne}}, \bibinfo {author} {\bibfnamefont {D.~E.}\
  \bibnamefont {Manolopoulos}}, \bibinfo {author} {\bibfnamefont {T.~E.}\
  \bibnamefont {Markland}}, \bibinfo {author} {\bibfnamefont {J.~O.}\
  \bibnamefont {Richardson}}, \bibinfo {author} {\bibfnamefont
  {A.}~\bibnamefont {Tkatchenko}}, \bibinfo {author} {\bibfnamefont {G.~A.}\
  \bibnamefont {Tribello}}, \bibinfo {author} {\bibfnamefont {V.}~\bibnamefont
  {Van~Speybroeck}},\ and\ \bibinfo {author} {\bibfnamefont {M.}~\bibnamefont
  {Ceriotti}},\ }\bibfield  {title} {\enquote {\bibinfo {title} {I-{{PI}} 2.0:
  {{A}} universal force engine for advanced molecular simulations},}\ }\href
  {https://doi.org/10.1016/j.cpc.2018.09.020} {\bibfield  {journal} {\bibinfo
  {journal} {Comput. Phys. Commun.}\ }\textbf {\bibinfo {volume} {236}},\
  \bibinfo {pages} {214--223} (\bibinfo {year}
  {2019}{\natexlab{b}})}\BibitemShut {NoStop}%
\bibitem [{\citenamefont {Plimpton}(1995)}]{plim95jcp}%
  \BibitemOpen
  \bibfield  {author} {\bibinfo {author} {\bibfnamefont {S.}~\bibnamefont
  {Plimpton}},\ }\bibfield  {title} {\enquote {\bibinfo {title} {Fast
  {{Parallel Algorithms}} for {{Short}}-{{Range Molecular Dynamics}}},}\ }\href
  {https://doi.org/10.1006/jcph.1995.1039} {\bibfield  {journal} {\bibinfo
  {journal} {J. Comput. Phys.}\ }\textbf {\bibinfo {volume} {117}},\ \bibinfo
  {pages} {1--19} (\bibinfo {year} {1995})}\BibitemShut {NoStop}%
\bibitem [{\citenamefont {Singraber}, \citenamefont {Behler},\ and\
  \citenamefont {Dellago}(2019)}]{singraber2019library}%
  \BibitemOpen
  \bibfield  {author} {\bibinfo {author} {\bibfnamefont {A.}~\bibnamefont
  {Singraber}}, \bibinfo {author} {\bibfnamefont {J.}~\bibnamefont {Behler}},\
  and\ \bibinfo {author} {\bibfnamefont {C.}~\bibnamefont {Dellago}},\
  }\bibfield  {title} {\enquote {\bibinfo {title} {Library-based lammps
  implementation of high-dimensional neural network potentials},}\ }\href@noop
  {} {\bibfield  {journal} {\bibinfo  {journal} {Journal of chemical theory and
  computation}\ }\textbf {\bibinfo {volume} {15}},\ \bibinfo {pages}
  {1827--1840} (\bibinfo {year} {2019})}\BibitemShut {NoStop}%
\bibitem [{\citenamefont {Petraglia}\ \emph {et~al.}(2015)\citenamefont
  {Petraglia}, \citenamefont {Nicolaï}, \citenamefont {Wodrich}, \citenamefont
  {Ceriotti},\ and\ \citenamefont {Corminboeuf}}]{petraglia2016beyond}%
  \BibitemOpen
  \bibfield  {author} {\bibinfo {author} {\bibfnamefont {R.}~\bibnamefont
  {Petraglia}}, \bibinfo {author} {\bibfnamefont {A.}~\bibnamefont {Nicolaï}},
  \bibinfo {author} {\bibfnamefont {M.~D.}\ \bibnamefont {Wodrich}}, \bibinfo
  {author} {\bibfnamefont {M.}~\bibnamefont {Ceriotti}},\ and\ \bibinfo
  {author} {\bibfnamefont {C.}~\bibnamefont {Corminboeuf}},\ }\bibfield
  {title} {\enquote {\bibinfo {title} {Beyond static structures: {Putting}
  forth {REMD} as a tool to solve problems in computational organic
  chemistry},}\ }\href {https://doi.org/10.1002/jcc.24025} {\bibfield
  {journal} {\bibinfo  {journal} {J. Comput. Chem.}\ }\textbf {\bibinfo
  {volume} {37}},\ \bibinfo {pages} {83--92} (\bibinfo {year}
  {2015})}\BibitemShut {NoStop}%
\bibitem [{\citenamefont {Aradi}, \citenamefont {Hourahine},\ and\
  \citenamefont {Frauenheim}(2007)}]{aradi2007dftb+}%
  \BibitemOpen
  \bibfield  {author} {\bibinfo {author} {\bibfnamefont {B.}~\bibnamefont
  {Aradi}}, \bibinfo {author} {\bibfnamefont {B.}~\bibnamefont {Hourahine}},\
  and\ \bibinfo {author} {\bibfnamefont {T.}~\bibnamefont {Frauenheim}},\
  }\bibfield  {title} {\enquote {\bibinfo {title} {{DFTB+,} a sparse
  matrix-based implementation of the {DFTB} method\dag},}\ }\href
  {https://doi.org/10.1021/jp070186p} {\bibfield  {journal} {\bibinfo
  {journal} {J. Phys. Chem. A}\ }\textbf {\bibinfo {volume} {111}},\ \bibinfo
  {pages} {5678--5684} (\bibinfo {year} {2007})}\BibitemShut {NoStop}%
\bibitem [{\citenamefont {Gaus}, \citenamefont {Goez},\ and\ \citenamefont
  {Elstner}(2012)}]{gaus2012parametrization}%
  \BibitemOpen
  \bibfield  {author} {\bibinfo {author} {\bibfnamefont {M.}~\bibnamefont
  {Gaus}}, \bibinfo {author} {\bibfnamefont {A.}~\bibnamefont {Goez}},\ and\
  \bibinfo {author} {\bibfnamefont {M.}~\bibnamefont {Elstner}},\ }\bibfield
  {title} {\enquote {\bibinfo {title} {Parametrization and benchmark of {DFTB3}
  for organic molecules},}\ }\href {https://doi.org/10.1021/ct300849w}
  {\bibfield  {journal} {\bibinfo  {journal} {J. Chem. Theory Comput.}\
  }\textbf {\bibinfo {volume} {9}},\ \bibinfo {pages} {338--354} (\bibinfo
  {year} {2012})}\BibitemShut {NoStop}%
\bibitem [{\citenamefont {Gaus}\ \emph {et~al.}(2014)\citenamefont {Gaus},
  \citenamefont {Lu}, \citenamefont {Elstner},\ and\ \citenamefont
  {Cui}}]{gaus2014parameterization}%
  \BibitemOpen
  \bibfield  {author} {\bibinfo {author} {\bibfnamefont {M.}~\bibnamefont
  {Gaus}}, \bibinfo {author} {\bibfnamefont {X.}~\bibnamefont {Lu}}, \bibinfo
  {author} {\bibfnamefont {M.}~\bibnamefont {Elstner}},\ and\ \bibinfo {author}
  {\bibfnamefont {Q.}~\bibnamefont {Cui}},\ }\bibfield  {title} {\enquote
  {\bibinfo {title} {Parameterization of {DFTB3/3OB} for sulfur and phosphorus
  for chemical and biological applications},}\ }\href
  {https://doi.org/10.1021/ct401002w} {\bibfield  {journal} {\bibinfo
  {journal} {J. Chem. Theory Comput.}\ }\textbf {\bibinfo {volume} {10}},\
  \bibinfo {pages} {1518--1537} (\bibinfo {year} {2014})}\BibitemShut {NoStop}%
\bibitem [{\citenamefont {Grimme}, \citenamefont {Ehrlich},\ and\ \citenamefont
  {Goerigk}(2011)}]{grimme2011effect}%
  \BibitemOpen
  \bibfield  {author} {\bibinfo {author} {\bibfnamefont {S.}~\bibnamefont
  {Grimme}}, \bibinfo {author} {\bibfnamefont {S.}~\bibnamefont {Ehrlich}},\
  and\ \bibinfo {author} {\bibfnamefont {L.}~\bibnamefont {Goerigk}},\
  }\bibfield  {title} {\enquote {\bibinfo {title} {Effect of the damping
  function in dispersion corrected density functional theory},}\ }\href
  {https://doi.org/10.1002/jcc.21759} {\bibfield  {journal} {\bibinfo
  {journal} {J. Comput. Chem.}\ }\textbf {\bibinfo {volume} {32}},\ \bibinfo
  {pages} {1456--1465} (\bibinfo {year} {2011})}\BibitemShut {NoStop}%
\bibitem [{\citenamefont {Schmidt}\ \emph {et~al.}(1993)\citenamefont
  {Schmidt}, \citenamefont {Baldridge}, \citenamefont {Boatz}, \citenamefont
  {Elbert}, \citenamefont {Gordon}, \citenamefont {Jensen}, \citenamefont
  {Koseki}, \citenamefont {Matsunaga}, \citenamefont {Nguyen}, \citenamefont
  {Su} \emph {et~al.}}]{schmidt1993general}%
  \BibitemOpen
  \bibfield  {author} {\bibinfo {author} {\bibfnamefont {M.~W.}\ \bibnamefont
  {Schmidt}}, \bibinfo {author} {\bibfnamefont {K.~K.}\ \bibnamefont
  {Baldridge}}, \bibinfo {author} {\bibfnamefont {J.~A.}\ \bibnamefont
  {Boatz}}, \bibinfo {author} {\bibfnamefont {S.~T.}\ \bibnamefont {Elbert}},
  \bibinfo {author} {\bibfnamefont {M.~S.}\ \bibnamefont {Gordon}}, \bibinfo
  {author} {\bibfnamefont {J.~H.}\ \bibnamefont {Jensen}}, \bibinfo {author}
  {\bibfnamefont {S.}~\bibnamefont {Koseki}}, \bibinfo {author} {\bibfnamefont
  {N.}~\bibnamefont {Matsunaga}}, \bibinfo {author} {\bibfnamefont {K.~A.}\
  \bibnamefont {Nguyen}}, \bibinfo {author} {\bibfnamefont {S.}~\bibnamefont
  {Su}}, \emph {et~al.},\ }\bibfield  {title} {\enquote {\bibinfo {title}
  {General atomic and molecular electronic structure system},}\ }\href@noop {}
  {\bibfield  {journal} {\bibinfo  {journal} {J. Comput. Chem.}\ }\textbf
  {\bibinfo {volume} {14}},\ \bibinfo {pages} {1347--1363} (\bibinfo {year}
  {1993})}\BibitemShut {NoStop}%
\bibitem [{\citenamefont {Gordon}\ and\ \citenamefont
  {Schmidt}(2005)}]{gordon2005advances}%
  \BibitemOpen
  \bibfield  {author} {\bibinfo {author} {\bibfnamefont {M.~S.}\ \bibnamefont
  {Gordon}}\ and\ \bibinfo {author} {\bibfnamefont {M.~W.}\ \bibnamefont
  {Schmidt}},\ }\bibfield  {title} {\enquote {\bibinfo {title} {Advances in
  electronic structure theory: Gamess a decade later},}\ }in\ \href@noop {}
  {\emph {\bibinfo {booktitle} {Theory and applications of computational
  chemistry}}}\ (\bibinfo  {publisher} {Elsevier},\ \bibinfo {year} {2005})\
  pp.\ \bibinfo {pages} {1167--1189}\BibitemShut {NoStop}%
\bibitem [{\citenamefont {Perdew}, \citenamefont {Burke},\ and\ \citenamefont
  {Ernzerhof}(1996{\natexlab{a}})}]{perdew1996generalized}%
  \BibitemOpen
  \bibfield  {author} {\bibinfo {author} {\bibfnamefont {J.~P.}\ \bibnamefont
  {Perdew}}, \bibinfo {author} {\bibfnamefont {K.}~\bibnamefont {Burke}},\ and\
  \bibinfo {author} {\bibfnamefont {M.}~\bibnamefont {Ernzerhof}},\ }\bibfield
  {title} {\enquote {\bibinfo {title} {Generalized gradient approximation made
  simple},}\ }\href {https://doi.org/10.1103/physrevlett.77.3865} {\bibfield
  {journal} {\bibinfo  {journal} {Phys. Rev. Lett.}\ }\textbf {\bibinfo
  {volume} {77}},\ \bibinfo {pages} {3865--3868} (\bibinfo {year}
  {1996}{\natexlab{a}})}\BibitemShut {NoStop}%
\bibitem [{\citenamefont {Steinmann}\ and\ \citenamefont
  {Corminboeuf}(2010)}]{steinmann2010system}%
  \BibitemOpen
  \bibfield  {author} {\bibinfo {author} {\bibfnamefont {S.~N.}\ \bibnamefont
  {Steinmann}}\ and\ \bibinfo {author} {\bibfnamefont {C.}~\bibnamefont
  {Corminboeuf}},\ }\bibfield  {title} {\enquote {\bibinfo {title} {A
  system-dependent density-based dispersion correction},}\ }\href
  {https://doi.org/10.1021/ct1001494} {\bibfield  {journal} {\bibinfo
  {journal} {J. Chem. Theory Comput.}\ }\textbf {\bibinfo {volume} {6}},\
  \bibinfo {pages} {1990--2001} (\bibinfo {year} {2010})}\BibitemShut {NoStop}%
\bibitem [{\citenamefont {Steinmann}\ and\ \citenamefont
  {Corminboeuf}(2011{\natexlab{a}})}]{steinmann2011comprehensive}%
  \BibitemOpen
  \bibfield  {author} {\bibinfo {author} {\bibfnamefont {S.~N.}\ \bibnamefont
  {Steinmann}}\ and\ \bibinfo {author} {\bibfnamefont {C.}~\bibnamefont
  {Corminboeuf}},\ }\bibfield  {title} {\enquote {\bibinfo {title}
  {Comprehensive benchmarking of a density-dependent dispersion correction},}\
  }\href {https://doi.org/10.1021/ct200602x} {\bibfield  {journal} {\bibinfo
  {journal} {J. Chem. Theory Comput.}\ }\textbf {\bibinfo {volume} {7}},\
  \bibinfo {pages} {3567--3577} (\bibinfo {year}
  {2011}{\natexlab{a}})}\BibitemShut {NoStop}%
\bibitem [{\citenamefont {Steinmann}\ and\ \citenamefont
  {Corminboeuf}(2011{\natexlab{b}})}]{steinmann2011generalized}%
  \BibitemOpen
  \bibfield  {author} {\bibinfo {author} {\bibfnamefont {S.~N.}\ \bibnamefont
  {Steinmann}}\ and\ \bibinfo {author} {\bibfnamefont {C.}~\bibnamefont
  {Corminboeuf}},\ }\bibfield  {title} {\enquote {\bibinfo {title} {A
  generalized-gradient approximation exchange hole model for dispersion
  coefficients},}\ }\href {https://doi.org/10.1063/1.3545985} {\bibfield
  {journal} {\bibinfo  {journal} {J. Chem. Phys.}\ }\textbf {\bibinfo {volume}
  {134}},\ \bibinfo {pages} {044117} (\bibinfo {year}
  {2011}{\natexlab{b}})}\BibitemShut {NoStop}%
\bibitem [{\citenamefont {Sch{\"{a}}fer}, \citenamefont {Horn},\ and\
  \citenamefont {Ahlrichs}(1992)}]{Schafer1992}%
  \BibitemOpen
  \bibfield  {author} {\bibinfo {author} {\bibfnamefont {A.}~\bibnamefont
  {Sch{\"{a}}fer}}, \bibinfo {author} {\bibfnamefont {H.}~\bibnamefont
  {Horn}},\ and\ \bibinfo {author} {\bibfnamefont {R.}~\bibnamefont
  {Ahlrichs}},\ }\bibfield  {title} {\enquote {\bibinfo {title} {{Fully
  optimized contracted Gaussian basis sets for atoms Li to Kr}},}\ }\href
  {https://doi.org/10.1063/1.463096} {\bibfield  {journal} {\bibinfo  {journal}
  {The Journal of Chemical Physics}\ } (\bibinfo {year} {1992}),\
  10.1063/1.463096}\BibitemShut {NoStop}%
\bibitem [{\citenamefont {Burns}\ \emph {et~al.}(2017)\citenamefont {Burns},
  \citenamefont {Faver}, \citenamefont {Zheng}, \citenamefont {Marshall},
  \citenamefont {Smith}, \citenamefont {Vanommeslaeghe}, \citenamefont
  {MacKerell}, \citenamefont {Merz},\ and\ \citenamefont
  {Sherrill}}]{burns_biofragment_2017}%
  \BibitemOpen
  \bibfield  {author} {\bibinfo {author} {\bibfnamefont {L.~A.}\ \bibnamefont
  {Burns}}, \bibinfo {author} {\bibfnamefont {J.~C.}\ \bibnamefont {Faver}},
  \bibinfo {author} {\bibfnamefont {Z.}~\bibnamefont {Zheng}}, \bibinfo
  {author} {\bibfnamefont {M.~S.}\ \bibnamefont {Marshall}}, \bibinfo {author}
  {\bibfnamefont {D.~G.~A.}\ \bibnamefont {Smith}}, \bibinfo {author}
  {\bibfnamefont {K.}~\bibnamefont {Vanommeslaeghe}}, \bibinfo {author}
  {\bibfnamefont {A.~D.}\ \bibnamefont {MacKerell}}, \bibinfo {author}
  {\bibfnamefont {K.~M.}\ \bibnamefont {Merz}},\ and\ \bibinfo {author}
  {\bibfnamefont {C.~D.}\ \bibnamefont {Sherrill}},\ }\bibfield  {title}
  {\enquote {\bibinfo {title} {The {BioFragment} {Database} ({BFDb}): {An}
  open-data platform for computational chemistry analysis of noncovalent
  interactions},}\ }\href {https://doi.org/10.1063/1.5001028} {\bibfield
  {journal} {\bibinfo  {journal} {J. Chem. Phys.}\ }\textbf {\bibinfo {volume}
  {147}},\ \bibinfo {pages} {161727} (\bibinfo {year} {2017})}\BibitemShut
  {NoStop}%
\bibitem [{\citenamefont {Singraber}\ \emph {et~al.}(2019)\citenamefont
  {Singraber}, \citenamefont {Morawietz}, \citenamefont {Behler},\ and\
  \citenamefont {Dellago}}]{singraber2019parallel}%
  \BibitemOpen
  \bibfield  {author} {\bibinfo {author} {\bibfnamefont {A.}~\bibnamefont
  {Singraber}}, \bibinfo {author} {\bibfnamefont {T.}~\bibnamefont
  {Morawietz}}, \bibinfo {author} {\bibfnamefont {J.}~\bibnamefont {Behler}},\
  and\ \bibinfo {author} {\bibfnamefont {C.}~\bibnamefont {Dellago}},\
  }\bibfield  {title} {\enquote {\bibinfo {title} {Parallel multistream
  training of high-dimensional neural network potentials},}\ }\href@noop {}
  {\bibfield  {journal} {\bibinfo  {journal} {Journal of chemical theory and
  computation}\ }\textbf {\bibinfo {volume} {15}},\ \bibinfo {pages}
  {3075--3092} (\bibinfo {year} {2019})}\BibitemShut {NoStop}%
\bibitem [{\citenamefont {Cheng}\ \emph {et~al.}(2018)\citenamefont {Cheng},
  \citenamefont {Engel}, \citenamefont {Behler}, \citenamefont {Dellago},\ and\
  \citenamefont {Ceriotti}}]{matcloud18a}%
  \BibitemOpen
  \bibfield  {author} {\bibinfo {author} {\bibfnamefont {B.}~\bibnamefont
  {Cheng}}, \bibinfo {author} {\bibfnamefont {E.}~\bibnamefont {Engel}},
  \bibinfo {author} {\bibfnamefont {J.}~\bibnamefont {Behler}}, \bibinfo
  {author} {\bibfnamefont {C.}~\bibnamefont {Dellago}},\ and\ \bibinfo {author}
  {\bibfnamefont {M.}~\bibnamefont {Ceriotti}},\ }\href
  {https://doi.org/10.24435/materialscloud:2018.0020/v1} {\enquote {\bibinfo
  {title} {Dataset: {{Ab}} initio thermodynamics of liquid and solid water},}\
  } (\bibinfo {year} {2018})\BibitemShut {NoStop}%
\bibitem [{\citenamefont {Morawietz}\ \emph {et~al.}(2016)\citenamefont
  {Morawietz}, \citenamefont {Singraber}, \citenamefont {Dellago},\ and\
  \citenamefont {Behler}}]{mora+16pnas}%
  \BibitemOpen
  \bibfield  {author} {\bibinfo {author} {\bibfnamefont {T.}~\bibnamefont
  {Morawietz}}, \bibinfo {author} {\bibfnamefont {A.}~\bibnamefont
  {Singraber}}, \bibinfo {author} {\bibfnamefont {C.}~\bibnamefont {Dellago}},\
  and\ \bibinfo {author} {\bibfnamefont {J.}~\bibnamefont {Behler}},\
  }\bibfield  {title} {\enquote {\bibinfo {title} {How van der waals
  interactions determine the unique properties of water},}\ }\href
  {https://doi.org/10.1073/pnas.1602375113} {\bibfield  {journal} {\bibinfo
  {journal} {Proc. Natl. Acad. Sci. U. S. A.}\ }\textbf {\bibinfo {volume}
  {113}},\ \bibinfo {pages} {8368--8373} (\bibinfo {year} {2016})}\BibitemShut
  {NoStop}%
\bibitem [{\citenamefont {Cheng}\ \emph {et~al.}(2019)\citenamefont {Cheng},
  \citenamefont {Engel}, \citenamefont {Behler}, \citenamefont {Dellago},\ and\
  \citenamefont {Ceriotti}}]{chen+19pnas}%
  \BibitemOpen
  \bibfield  {author} {\bibinfo {author} {\bibfnamefont {B.}~\bibnamefont
  {Cheng}}, \bibinfo {author} {\bibfnamefont {E.~A.}\ \bibnamefont {Engel}},
  \bibinfo {author} {\bibfnamefont {J.}~\bibnamefont {Behler}}, \bibinfo
  {author} {\bibfnamefont {C.}~\bibnamefont {Dellago}},\ and\ \bibinfo {author}
  {\bibfnamefont {M.}~\bibnamefont {Ceriotti}},\ }\bibfield  {title} {\enquote
  {\bibinfo {title} {Ab initio thermodynamics of liquid and solid water},}\
  }\href {https://doi.org/10.1073/pnas.1815117116} {\bibfield  {journal}
  {\bibinfo  {journal} {Proc. Natl. Acad. Sci. U. S. A.}\ }\textbf {\bibinfo
  {volume} {116}},\ \bibinfo {pages} {1110--1115} (\bibinfo {year}
  {2019})}\BibitemShut {NoStop}%
\bibitem [{\citenamefont {Rossi}\ \emph
  {et~al.}(2020{\natexlab{b}})\citenamefont {Rossi}, \citenamefont {Juraskova},
  \citenamefont {Wischert}, \citenamefont {Garel}, \citenamefont
  {Corminboeuf},\ and\ \citenamefont {Ceriotti}}]{matcloud20d}%
  \BibitemOpen
  \bibfield  {author} {\bibinfo {author} {\bibfnamefont {K.}~\bibnamefont
  {Rossi}}, \bibinfo {author} {\bibfnamefont {V.}~\bibnamefont {Juraskova}},
  \bibinfo {author} {\bibfnamefont {R.}~\bibnamefont {Wischert}}, \bibinfo
  {author} {\bibfnamefont {L.}~\bibnamefont {Garel}}, \bibinfo {author}
  {\bibfnamefont {C.}~\bibnamefont {Corminboeuf}},\ and\ \bibinfo {author}
  {\bibfnamefont {M.}~\bibnamefont {Ceriotti}},\ }\href
  {https://doi.org/10.24435/MATERIALSCLOUD:Z9-ZR} {\enquote {\bibinfo {title}
  {Dataset: {{Simulating}} solvation and acidity in complex mixtures with
  first-principles accuracy: The case of {{CH}}{$_{3}$}{{SO}}{$_{3}$}{{H}} and
  {{H}}{$_{2}$}{{O}}{$_2$} in phenol},}\ } (\bibinfo {year}
  {2020}{\natexlab{b}})\BibitemShut {NoStop}%
\bibitem [{\citenamefont {Pedersen}\ \emph {et~al.}(2013)\citenamefont
  {Pedersen}, \citenamefont {Hummel}, \citenamefont {Kresse}, \citenamefont
  {Kahl},\ and\ \citenamefont
  {Dellago}}]{Pedersen_Hummel_Kresse_Kahl_Dellago_2013}%
  \BibitemOpen
  \bibfield  {author} {\bibinfo {author} {\bibfnamefont {U.~R.}\ \bibnamefont
  {Pedersen}}, \bibinfo {author} {\bibfnamefont {F.}~\bibnamefont {Hummel}},
  \bibinfo {author} {\bibfnamefont {G.}~\bibnamefont {Kresse}}, \bibinfo
  {author} {\bibfnamefont {G.}~\bibnamefont {Kahl}},\ and\ \bibinfo {author}
  {\bibfnamefont {C.}~\bibnamefont {Dellago}},\ }\bibfield  {title} {\enquote
  {\bibinfo {title} {Computing gibbs free energy differences by interface
  pinning},}\ }\href {https://doi.org/10.1103/PhysRevB.88.094101} {\bibfield
  {journal} {\bibinfo  {journal} {Physical Review B}\ }\textbf {\bibinfo
  {volume} {88}},\ \bibinfo {pages} {094101} (\bibinfo {year}
  {2013})}\BibitemShut {NoStop}%
\bibitem [{\citenamefont {Lechner}\ and\ \citenamefont
  {Dellago}(2008)}]{lechner2008accurate}%
  \BibitemOpen
  \bibfield  {author} {\bibinfo {author} {\bibfnamefont {W.}~\bibnamefont
  {Lechner}}\ and\ \bibinfo {author} {\bibfnamefont {C.}~\bibnamefont
  {Dellago}},\ }\bibfield  {title} {\enquote {\bibinfo {title} {Accurate
  determination of crystal structures based on averaged local bond order
  parameters},}\ }\href@noop {} {\bibfield  {journal} {\bibinfo  {journal} {The
  Journal of chemical physics}\ }\textbf {\bibinfo {volume} {129}},\ \bibinfo
  {pages} {114707} (\bibinfo {year} {2008})}\BibitemShut {NoStop}%
\bibitem [{\citenamefont {Steinhardt}, \citenamefont {Nelson},\ and\
  \citenamefont {Ronchetti}(1983)}]{stei+83prb}%
  \BibitemOpen
  \bibfield  {author} {\bibinfo {author} {\bibfnamefont {P.~J.}\ \bibnamefont
  {Steinhardt}}, \bibinfo {author} {\bibfnamefont {D.~R.}\ \bibnamefont
  {Nelson}},\ and\ \bibinfo {author} {\bibfnamefont {M.}~\bibnamefont
  {Ronchetti}},\ }\bibfield  {title} {\enquote {\bibinfo {title}
  {Bond-orientational order in liquids and glasses},}\ }\href
  {https://doi.org/10.1103/PhysRevB.28.784} {\bibfield  {journal} {\bibinfo
  {journal} {Phys. Rev. B}\ }\textbf {\bibinfo {volume} {28}},\ \bibinfo
  {pages} {784--805} (\bibinfo {year} {1983})}\BibitemShut {NoStop}%
\bibitem [{\citenamefont {Bonomi}\ \emph {et~al.}(2009)\citenamefont {Bonomi},
  \citenamefont {Branduardi}, \citenamefont {Bussi}, \citenamefont {Camilloni},
  \citenamefont {Provasi}, \citenamefont {Raiteri}, \citenamefont {Donadio},
  \citenamefont {Marinelli}, \citenamefont {Pietrucci}, \citenamefont
  {Broglia},\ and\ \citenamefont {Parrinello}}]{PLUMED}%
  \BibitemOpen
  \bibfield  {author} {\bibinfo {author} {\bibfnamefont {M.}~\bibnamefont
  {Bonomi}}, \bibinfo {author} {\bibfnamefont {D.}~\bibnamefont {Branduardi}},
  \bibinfo {author} {\bibfnamefont {G.}~\bibnamefont {Bussi}}, \bibinfo
  {author} {\bibfnamefont {C.}~\bibnamefont {Camilloni}}, \bibinfo {author}
  {\bibfnamefont {D.}~\bibnamefont {Provasi}}, \bibinfo {author} {\bibfnamefont
  {P.}~\bibnamefont {Raiteri}}, \bibinfo {author} {\bibfnamefont
  {D.}~\bibnamefont {Donadio}}, \bibinfo {author} {\bibfnamefont
  {F.}~\bibnamefont {Marinelli}}, \bibinfo {author} {\bibfnamefont
  {F.}~\bibnamefont {Pietrucci}}, \bibinfo {author} {\bibfnamefont {R.~A.}\
  \bibnamefont {Broglia}},\ and\ \bibinfo {author} {\bibfnamefont
  {M.}~\bibnamefont {Parrinello}},\ }\bibfield  {title} {\enquote {\bibinfo
  {title} {{{PLUMED}}: {{A}} portable plugin for free-energy calculations with
  molecular dynamics},}\ }\href {https://doi.org/10.1016/j.cpc.2009.05.011}
  {\bibfield  {journal} {\bibinfo  {journal} {Comput. Phys. Commun.}\ }\textbf
  {\bibinfo {volume} {180}},\ \bibinfo {pages} {1961--1972} (\bibinfo {year}
  {2009})}\BibitemShut {NoStop}%
\bibitem [{\citenamefont {Giberti}\ \emph {et~al.}(2020)\citenamefont
  {Giberti}, \citenamefont {Cheng}, \citenamefont {Tribello},\ and\
  \citenamefont {Ceriotti}}]{Giberti_Cheng_Tribello_Ceriotti_2020}%
  \BibitemOpen
  \bibfield  {author} {\bibinfo {author} {\bibfnamefont {F.}~\bibnamefont
  {Giberti}}, \bibinfo {author} {\bibfnamefont {B.}~\bibnamefont {Cheng}},
  \bibinfo {author} {\bibfnamefont {G.~A.}\ \bibnamefont {Tribello}},\ and\
  \bibinfo {author} {\bibfnamefont {M.}~\bibnamefont {Ceriotti}},\ }\bibfield
  {title} {\enquote {\bibinfo {title} {Iterative unbiasing of quasi-equilibrium
  sampling},}\ }\href {https://doi.org/10.1021/acs.jctc.9b00907} {\bibfield
  {journal} {\bibinfo  {journal} {Journal of Chemical Theory and Computation}\
  }\textbf {\bibinfo {volume} {16}},\ \bibinfo {pages} {100–107} (\bibinfo
  {year} {2020})}\BibitemShut {NoStop}%
\bibitem [{\citenamefont {Zamani}\ \emph
  {et~al.}(2020{\natexlab{b}})\citenamefont {Zamani}, \citenamefont
  {Imbalzano}, \citenamefont {Tappy}, \citenamefont {Alexander}, \citenamefont
  {Mart{\'i}-S{\'a}nchez}, \citenamefont {Ghisalberti}, \citenamefont
  {Ramasse}, \citenamefont {Friedl}, \citenamefont {T{\"u}t{\"u}nc{\"u}oglu},
  \citenamefont {Francaviglia}, \citenamefont {Bienvenue}, \citenamefont
  {H{\'e}bert}, \citenamefont {Arbiol}, \citenamefont {Ceriotti},\ and\
  \citenamefont {{Fontcuberta i Morral}}}]{zama+20am}%
  \BibitemOpen
  \bibfield  {author} {\bibinfo {author} {\bibfnamefont {M.}~\bibnamefont
  {Zamani}}, \bibinfo {author} {\bibfnamefont {G.}~\bibnamefont {Imbalzano}},
  \bibinfo {author} {\bibfnamefont {N.}~\bibnamefont {Tappy}}, \bibinfo
  {author} {\bibfnamefont {D.~T.~L.}\ \bibnamefont {Alexander}}, \bibinfo
  {author} {\bibfnamefont {S.}~\bibnamefont {Mart{\'i}-S{\'a}nchez}}, \bibinfo
  {author} {\bibfnamefont {L.}~\bibnamefont {Ghisalberti}}, \bibinfo {author}
  {\bibfnamefont {Q.~M.}\ \bibnamefont {Ramasse}}, \bibinfo {author}
  {\bibfnamefont {M.}~\bibnamefont {Friedl}}, \bibinfo {author} {\bibfnamefont
  {G.}~\bibnamefont {T{\"u}t{\"u}nc{\"u}oglu}}, \bibinfo {author}
  {\bibfnamefont {L.}~\bibnamefont {Francaviglia}}, \bibinfo {author}
  {\bibfnamefont {S.}~\bibnamefont {Bienvenue}}, \bibinfo {author}
  {\bibfnamefont {C.}~\bibnamefont {H{\'e}bert}}, \bibinfo {author}
  {\bibfnamefont {J.}~\bibnamefont {Arbiol}}, \bibinfo {author} {\bibfnamefont
  {M.}~\bibnamefont {Ceriotti}},\ and\ \bibinfo {author} {\bibfnamefont
  {A.}~\bibnamefont {{Fontcuberta i Morral}}},\ }\bibfield  {title} {\enquote
  {\bibinfo {title} {{{3D Ordering}} at the {{Liquid}}\textendash{{Solid Polar
  Interface}} of {{Nanowires}}},}\ }\href
  {https://doi.org/10.1002/adma.202001030} {\bibfield  {journal} {\bibinfo
  {journal} {Adv. Mater.}\ }\textbf {\bibinfo {volume} {32}},\ \bibinfo {pages}
  {2001030} (\bibinfo {year} {2020}{\natexlab{b}})}\BibitemShut {NoStop}%
\bibitem [{\citenamefont {Pozdnyakov}, \citenamefont {Willatt},\ and\
  \citenamefont {Ceriotti}(2020)}]{matcloud20a}%
  \BibitemOpen
  \bibfield  {author} {\bibinfo {author} {\bibfnamefont {S.}~\bibnamefont
  {Pozdnyakov}}, \bibinfo {author} {\bibfnamefont {M.}~\bibnamefont
  {Willatt}},\ and\ \bibinfo {author} {\bibfnamefont {M.}~\bibnamefont
  {Ceriotti}},\ }\href {https://doi.org/10.24435/MATERIALSCLOUD:QY-DP}
  {\enquote {\bibinfo {title} {Dataset: {{Randomly}}-displaced methane
  configurations},}\ } (\bibinfo {year} {2020})\BibitemShut {NoStop}%
\bibitem [{\citenamefont {Ceriotti}, \citenamefont {Bussi},\ and\ \citenamefont
  {Parrinello}(2010)}]{ceri+10jctc}%
  \BibitemOpen
  \bibfield  {author} {\bibinfo {author} {\bibfnamefont {M.}~\bibnamefont
  {Ceriotti}}, \bibinfo {author} {\bibfnamefont {G.}~\bibnamefont {Bussi}},\
  and\ \bibinfo {author} {\bibfnamefont {M.}~\bibnamefont {Parrinello}},\
  }\bibfield  {title} {\enquote {\bibinfo {title} {Colored-noise thermostats
  \`a la {{Carte}}},}\ }\href {https://doi.org/10.1021/ct900563s} {\bibfield
  {journal} {\bibinfo  {journal} {J. Chem. Theory Comput.}\ }\textbf {\bibinfo
  {volume} {6}},\ \bibinfo {pages} {1170--1180} (\bibinfo {year}
  {2010})}\BibitemShut {NoStop}%
\bibitem [{\citenamefont {Bussi}, \citenamefont {Donadio},\ and\ \citenamefont
  {Parrinello}(2007)}]{bussi2007canonical}%
  \BibitemOpen
  \bibfield  {author} {\bibinfo {author} {\bibfnamefont {G.}~\bibnamefont
  {Bussi}}, \bibinfo {author} {\bibfnamefont {D.}~\bibnamefont {Donadio}},\
  and\ \bibinfo {author} {\bibfnamefont {M.}~\bibnamefont {Parrinello}},\
  }\bibfield  {title} {\enquote {\bibinfo {title} {Canonical sampling through
  velocity rescaling},}\ }\href@noop {} {\bibfield  {journal} {\bibinfo
  {journal} {The Journal of chemical physics}\ }\textbf {\bibinfo {volume}
  {126}},\ \bibinfo {pages} {014101} (\bibinfo {year} {2007})}\BibitemShut
  {NoStop}%
\bibitem [{\citenamefont {Perdew}, \citenamefont {Burke},\ and\ \citenamefont
  {Ernzerhof}(1996{\natexlab{b}})}]{perd+96prl}%
  \BibitemOpen
  \bibfield  {author} {\bibinfo {author} {\bibfnamefont {J.~P.}\ \bibnamefont
  {Perdew}}, \bibinfo {author} {\bibfnamefont {K.}~\bibnamefont {Burke}},\ and\
  \bibinfo {author} {\bibfnamefont {M.}~\bibnamefont {Ernzerhof}},\ }\bibfield
  {title} {\enquote {\bibinfo {title} {Generalized {{Gradient Approximation}}
  made simple},}\ }\href {http://www.ncbi.nlm.nih.gov/pubmed/10062328}
  {\bibfield  {journal} {\bibinfo  {journal} {Phys. Rev. Lett.}\ }\textbf
  {\bibinfo {volume} {77}},\ \bibinfo {pages} {3865} (\bibinfo {year}
  {1996}{\natexlab{b}})}\BibitemShut {NoStop}%
\bibitem [{\citenamefont {Giannozzi}\ \emph {et~al.}(2009)\citenamefont
  {Giannozzi}, \citenamefont {Baroni}, \citenamefont {Bonini}, \citenamefont
  {Calandra}, \citenamefont {Car}, \citenamefont {Cavazzoni}, \citenamefont
  {Ceresoli}, \citenamefont {Chiarotti}, \citenamefont {Cococcioni},
  \citenamefont {Dabo} \emph {et~al.}}]{quantum-espresso-1}%
  \BibitemOpen
  \bibfield  {author} {\bibinfo {author} {\bibfnamefont {P.}~\bibnamefont
  {Giannozzi}}, \bibinfo {author} {\bibfnamefont {S.}~\bibnamefont {Baroni}},
  \bibinfo {author} {\bibfnamefont {N.}~\bibnamefont {Bonini}}, \bibinfo
  {author} {\bibfnamefont {M.}~\bibnamefont {Calandra}}, \bibinfo {author}
  {\bibfnamefont {R.}~\bibnamefont {Car}}, \bibinfo {author} {\bibfnamefont
  {C.}~\bibnamefont {Cavazzoni}}, \bibinfo {author} {\bibfnamefont
  {D.}~\bibnamefont {Ceresoli}}, \bibinfo {author} {\bibfnamefont {G.~L.}\
  \bibnamefont {Chiarotti}}, \bibinfo {author} {\bibfnamefont {M.}~\bibnamefont
  {Cococcioni}}, \bibinfo {author} {\bibfnamefont {I.}~\bibnamefont {Dabo}},
  \emph {et~al.},\ }\bibfield  {title} {\enquote {\bibinfo {title} {{QUANTUM
  ESPRESSO}: a modular and open-source software project for quantum simulations
  of materials},}\ }\href@noop {} {\bibfield  {journal} {\bibinfo  {journal}
  {Journal of physics: Condensed matter}\ }\textbf {\bibinfo {volume} {21}},\
  \bibinfo {pages} {395502} (\bibinfo {year} {2009})}\BibitemShut {NoStop}%
\bibitem [{\citenamefont {Giannozzi}\ \emph {et~al.}(2017)\citenamefont
  {Giannozzi}, \citenamefont {Andreussi}, \citenamefont {Brumme}, \citenamefont
  {Bunau}, \citenamefont {Nardelli}, \citenamefont {Calandra}, \citenamefont
  {Car}, \citenamefont {Cavazzoni}, \citenamefont {Ceresoli}, \citenamefont
  {Cococcioni} \emph {et~al.}}]{quantum-espresso-2}%
  \BibitemOpen
  \bibfield  {author} {\bibinfo {author} {\bibfnamefont {P.}~\bibnamefont
  {Giannozzi}}, \bibinfo {author} {\bibfnamefont {O.}~\bibnamefont
  {Andreussi}}, \bibinfo {author} {\bibfnamefont {T.}~\bibnamefont {Brumme}},
  \bibinfo {author} {\bibfnamefont {O.}~\bibnamefont {Bunau}}, \bibinfo
  {author} {\bibfnamefont {M.~B.}\ \bibnamefont {Nardelli}}, \bibinfo {author}
  {\bibfnamefont {M.}~\bibnamefont {Calandra}}, \bibinfo {author}
  {\bibfnamefont {R.}~\bibnamefont {Car}}, \bibinfo {author} {\bibfnamefont
  {C.}~\bibnamefont {Cavazzoni}}, \bibinfo {author} {\bibfnamefont
  {D.}~\bibnamefont {Ceresoli}}, \bibinfo {author} {\bibfnamefont
  {M.}~\bibnamefont {Cococcioni}}, \emph {et~al.},\ }\bibfield  {title}
  {\enquote {\bibinfo {title} {Advanced capabilities for materials modelling
  with {Quantum ESPRESSO}},}\ }\href@noop {} {\bibfield  {journal} {\bibinfo
  {journal} {Journal of physics: Condensed matter}\ }\textbf {\bibinfo {volume}
  {29}},\ \bibinfo {pages} {465901} (\bibinfo {year} {2017})}\BibitemShut
  {NoStop}%
\bibitem [{\citenamefont {Kapil}\ \emph {et~al.}(2018)\citenamefont {Kapil},
  \citenamefont {Rossi}, \citenamefont {Marsalek}, \citenamefont {Petraglia},
  \citenamefont {Litman}, \citenamefont {Spura}, \citenamefont {Cheng},
  \citenamefont {Cuzzocrea}, \citenamefont {Mei{\ss}ner}, \citenamefont
  {Wilkins}, \citenamefont {Juda}, \citenamefont {Bienvenue}, \citenamefont
  {Fang}, \citenamefont {Kessler}, \citenamefont {Poltavsky}, \citenamefont
  {Vandenbrande}, \citenamefont {Wieme}, \citenamefont {Corminboeuf},
  \citenamefont {K{\"u}hne}, \citenamefont {Manolopoulos}, \citenamefont
  {Markland}, \citenamefont {Richardson}, \citenamefont {Tkatchenko},
  \citenamefont {Tribello}, \citenamefont {Van~Speybroeck},\ and\ \citenamefont
  {Ceriotti}}]{ipicode}%
  \BibitemOpen
  \bibfield  {author} {\bibinfo {author} {\bibfnamefont {V.}~\bibnamefont
  {Kapil}}, \bibinfo {author} {\bibfnamefont {M.}~\bibnamefont {Rossi}},
  \bibinfo {author} {\bibfnamefont {O.}~\bibnamefont {Marsalek}}, \bibinfo
  {author} {\bibfnamefont {R.}~\bibnamefont {Petraglia}}, \bibinfo {author}
  {\bibfnamefont {Y.}~\bibnamefont {Litman}}, \bibinfo {author} {\bibfnamefont
  {T.}~\bibnamefont {Spura}}, \bibinfo {author} {\bibfnamefont
  {B.}~\bibnamefont {Cheng}}, \bibinfo {author} {\bibfnamefont
  {A.}~\bibnamefont {Cuzzocrea}}, \bibinfo {author} {\bibfnamefont {R.~H.}\
  \bibnamefont {Mei{\ss}ner}}, \bibinfo {author} {\bibfnamefont {D.~M.}\
  \bibnamefont {Wilkins}}, \bibinfo {author} {\bibfnamefont {P.}~\bibnamefont
  {Juda}}, \bibinfo {author} {\bibfnamefont {S.~P.}\ \bibnamefont {Bienvenue}},
  \bibinfo {author} {\bibfnamefont {W.}~\bibnamefont {Fang}}, \bibinfo {author}
  {\bibfnamefont {J.}~\bibnamefont {Kessler}}, \bibinfo {author} {\bibfnamefont
  {I.}~\bibnamefont {Poltavsky}}, \bibinfo {author} {\bibfnamefont
  {S.}~\bibnamefont {Vandenbrande}}, \bibinfo {author} {\bibfnamefont
  {J.}~\bibnamefont {Wieme}}, \bibinfo {author} {\bibfnamefont
  {C.}~\bibnamefont {Corminboeuf}}, \bibinfo {author} {\bibfnamefont {T.~D.}\
  \bibnamefont {K{\"u}hne}}, \bibinfo {author} {\bibfnamefont {D.~E.}\
  \bibnamefont {Manolopoulos}}, \bibinfo {author} {\bibfnamefont {T.~E.}\
  \bibnamefont {Markland}}, \bibinfo {author} {\bibfnamefont {J.~O.}\
  \bibnamefont {Richardson}}, \bibinfo {author} {\bibfnamefont
  {A.}~\bibnamefont {Tkatchenko}}, \bibinfo {author} {\bibfnamefont {G.~A.}\
  \bibnamefont {Tribello}}, \bibinfo {author} {\bibfnamefont {V.}~\bibnamefont
  {Van~Speybroeck}},\ and\ \bibinfo {author} {\bibfnamefont {M.}~\bibnamefont
  {Ceriotti}},\ }\href {http://ipi-code.org} {\enquote {\bibinfo {title}
  {I-{{PI Software}}},}\ } (\bibinfo {year} {2018})\BibitemShut {NoStop}%
\bibitem [{\citenamefont {Boos}\ and\ \citenamefont
  {Hughes-Oliver}(1998)}]{boos1998Basu}%
  \BibitemOpen
  \bibfield  {author} {\bibinfo {author} {\bibfnamefont {D.~D.}\ \bibnamefont
  {Boos}}\ and\ \bibinfo {author} {\bibfnamefont {J.~M.}\ \bibnamefont
  {Hughes-Oliver}},\ }\bibfield  {title} {\enquote {\bibinfo {title}
  {Applications of basu's theorem},}\ }\href@noop {} {\bibfield  {journal}
  {\bibinfo  {journal} {The American Statistician}\ }\textbf {\bibinfo {volume}
  {52}},\ \bibinfo {pages} {218--221} (\bibinfo {year} {1998})}\BibitemShut
  {NoStop}%
\end{thebibliography}
\end{document}